\documentclass[twocolumn,traditabstract]{aa}
\usepackage{fixltx2e}
\usepackage[english]{babel}
\usepackage{graphicx,amsmath}
\usepackage{epsf,color}
\usepackage[mathscr]{eucal}
\usepackage{amsmath}
\usepackage{amssymb,amsfonts}
\usepackage{natbib}
\usepackage{txfonts}
\usepackage{dsfont}
\definecolor{Mygreen}{rgb}{0.00, 0.72, 0.0}
\definecolor{Mypink}{rgb}{1.0, 0.0, 0.5}
\usepackage[breaklinks, citecolor=blue, linkcolor=Mygreen, urlcolor=Mypink, colorlinks=true, debug, baseurl=' ']{hyperref}
\usepackage{float}
\usepackage{color}
\usepackage{widetext}

\newcommand{\nika}{{\it NIKA}}
\newcommand{\nikad}{{\it NIKA2}}

\def\simlt{\lower.5ex\hbox{$\; \buildrel < \over \sim \;$}}
\def\simgt{\lower.5ex\hbox{$\; \buildrel > \over \sim \;$}}
\def\NIKA{\textit{NIKA}}

\bibpunct{(}{)}{;}{a}{}{,}
\begin{document}
\title{Polarimetry at millimeter wavelengths with the \nika\ camera: calibration  and performance}
\author{A.~Ritacco \inst{\ref{LPSC}},\inst{\ref{IRAME}}\thanks{Corresponding author: Alessia Ritacco, \url{ritaccoa@iram.es}}
\and  N.~Ponthieu \inst{\ref{IPAG}}
\and  A.~Catalano \inst{\ref{LPSC}}
\and R.~Adam \inst{\ref{LPSC},\ref{OCA}}
\and  P.~Ade \inst{\ref{Cardiff}}
\and  P.~Andr\'e \inst{\ref{CEA}}
\and  A.~Beelen \inst{\ref{IAS}}
\and  A.~Beno\^it \inst{\ref{Neel}}
\and  A.~Bideaud \inst{\ref{Neel}}
\and  N.~Billot \inst{\ref{IRAME}}
\and  O.~Bourrion \inst{\ref{LPSC}}
\and  M.~Calvo \inst{\ref{Neel}}
\and  G.~Coiffard \inst{\ref{IRAMF}}
\and  B.~Comis \inst{\ref{LPSC}}
\and  F.-X.~D\'esert \inst{\ref{IPAG}}
\and  S.~Doyle \inst{\ref{Cardiff}}
\and  J.~Goupy \inst{\ref{Neel}}
\and  C.~Kramer \inst{\ref{IRAME}}
\and  S.~Leclercq \inst{\ref{IRAMF}}
\and  J.F.~Mac\'ias-P\'erez \inst{\ref{LPSC}}
\and  P.~Mauskopf \inst{\ref{Cardiff},\ref{Arizona}}
\and A. Maury \inst{\ref{CEA}}
\and  F.~Mayet \inst{\ref{LPSC}}
\and  A.~Monfardini \inst{\ref{Neel}}
\and  F.~Pajot \inst{\ref{IAS}}
\and  E.~Pascale \inst{\ref{Cardiff}}
\and  L.~Perotto \inst{\ref{LPSC}}
\and  G.~Pisano \inst{\ref{Cardiff}}
\and M.~Rebolo-Iglesias  \inst{\ref{LPSC}}
\and  V.~Rev\'eret \inst{\ref{CEA}}
\and  L.~Rodriguez \inst{\ref{CEA}}
\and  C.~Romero \inst{\ref{IRAMF}}
\and F.~Ruppin \inst{\ref{LPSC}}
\and G.~Savini \inst{\ref{London_college}}
\and  K.~Schuster \inst{\ref{IRAMF}}
\and  A.~Sievers \inst{\ref{IRAME}}
\and C. Thum \inst{\ref{IRAME}}
\and  S.~Triqueneaux \inst{\ref{Neel}}
\and  C.~Tucker \inst{\ref{Cardiff}}
\and  R.~Zylka \inst{\ref{IRAMF}}}

\institute{
Laboratoire de Physique Subatomique et de Cosmologie, Universit\'e Grenoble Alpes, CNRS/IN2P3, 53, avenue des Martyrs, Grenoble, France
  \label{LPSC}
  \and
  Laboratoire Lagrange, Universit\'e C\^ote d'Azur, Observatoire de la C\^ote d'Azur, CNRS, Blvd de l'Observatoire, CS 34229, 06304 Nice cedex 4, France
  \label{OCA}
  \and
Astronomy Instrumentation Group, University of Cardiff, UK
  \label{Cardiff}
\and
Laboratoire AIM, CEA/IRFU, CNRS/INSU, Universit\'e Paris Diderot, CEA-Saclay, 91191 Gif-Sur-Yvette, France 
  \label{CEA}
\and
Institut d'Astrophysique Spatiale (IAS), CNRS and Universit\'e Paris Sud, Orsay, France
  \label{IAS}
\and
Institut N\'eel, CNRS and Universit\'e Grenoble Alpes, France
  \label{Neel}
\and
Institut de RadioAstronomie Millim\'etrique (IRAM), Granada, Spain
  \label{IRAME}
\and
Institut de RadioAstronomie Millim\'etrique (IRAM), Grenoble, France
  \label{IRAMF}
\and
Univ. Grenoble Alpes, CNRS, IPAG, F-38000 Grenoble, France 
  \label{IPAG}
    \and
School of Earth and Space Exploration and Department of Physics, Arizona State University, Tempe, AZ 85287
  \label{Arizona}
 \and 
 University College London, Department 
of Physics and Astronomy, Gower Street, London WC1E 6BT, UK
\label{London_college}
}

\date{Received 7 September 2016 \ / Accepted 3 December 2016}
	
\abstract{
  Magnetic fields, which play a major role in a large number of astrophysical processes can be traced via observations of dust polarization.
  In particular, {\it Planck} low-resolution observations of dust polarization have demonstrated that Galactic filamentary structures, where star formation takes place, are associated to well organized magnetic fields. 
  A better understanding of this process requires detailed
  observations of galactic dust polarization on scales of 0.01 to 0.1
  pc.  Such high-resolution polarization observations can be carried out at the
  IRAM 30 m telescope using the recently installed \nikad\ camera, which features
  two frequency bands at 260 and 150 GHz (respectively 1.15 and 2.05 mm),
   the 260~GHz band being polarization sensitive. \nikad\ so far in commissioning phase, has its focal plane
  filled with $\sim$ 3300 detectors to cover a Field of View (FoV) of 6.5 arcminutes diameter. 
   The \nika\ camera, which consisted of two arrays of 132
  and 224 Lumped Element Kinetic Inductance Detectors (LEKIDs) and a FWHM (Full-Width-Half-Maximum) of 12 and 18.2 arcsecond at 1.15 and
  2.05~mm respectively, has been operated at the IRAM 30 m telescope from 2012 to 2015 as a
  test-bench for \nikad. \nika\ was equipped of a room temperature polarization
  system (a half wave plate (HWP) and a grid polarizer facing the
  \nika\ cryostat window). The fast and continuous rotation of the HWP
  permits the quasi simultaneous reconstruction of the three Stokes
  parameters, $I$, $Q$, and $U$ at 150 and 260 GHz. 
  This paper presents the first polarization
  measurements with KIDs and reports the
  polarization performance of the \nika\ camera and the pertinence of the choice
  of the polarization setup in the perspective of \nikad.
  We describe the polarized data reduction pipeline, 
  specifically developed for this project and how the continuous rotation of the HWP permits 
  to shift the polarized signal far from any low frequency noise. We also present the dedicated 
  algorithm developed to correct systematic leakage effects. We report results on compact and extended sources obtained during the February 2015 technical
  campaign. These results demonstrate a good understanding of polarization
  systematics and state-of-the-art performance in terms of photometry,
  polarization degree and polarization angle reconstruction.  }
\titlerunning{\NIKA\ polarization performance}	
\authorrunning{A. Ritacco, N. Ponthieu, A. Catalano et al.}
\maketitle
\section{Introduction}\label{sec:introduction}
\vspace{0.2cm}
Magnetic fields have been proven to play a predominant role in a large number of
astrophysical processes from galactic to cosmological scales. In particular, recent observations obtained with {\it Herschel} and {\it Planck} \citep{planck2013mission}
satellites have provided us with sensitive maps of the star-forming
complexes in the galaxy. These maps reveal large-scale filamentary structures as
the preferential sites of star formation
\citep{2010A&A...518L.100M,arzoumianian}. These filamentary structures
are associated with organized magnetic field topology at scales larger than 0.5 pc
\citep{2014prpl.conf...27A} and indicate that magnetic field must be explored on scales of
0.01 to 0.1 pc \citep{2004ApJ...603..584P,planckXXXIII}.
At millimeter and sub-millimeter wavelengths, the magnetic field orientation can
be explored using the polarized thermal dust emission
\citep{2015A&A...576A.104P,2016arXiv160100546P}. Dust grains are generally prolate. The polarization emission of a grain
depends on the orientation and acceleration of its magnetic dipole moment and is stronger along the major axis of the grain that aligns 
orthogonally to the magnetic field \citep{2009ASPC..414..482L}. This results in coherently
polarized dust emission in the plane perpendicular to the magnetic field
lines. Thus, polarized dust emission permits us to recover the direction of the
magnetic field lines projected on the plane of the sky
\citep[{\it e.g.,}][]{2015A&A...576A.106P}. The {\it Planck} satellite has mapped the polarized
dust emission at 353 GHz on large angular scales over the entire sky
\citep{2014A&A...571A...8P,2015arXiv150201587P} and suggests a high degree
of polarization, up to 15 \% \citep{planckdust}, confirming previous {\it Archeops}
results \citep{2004A&A...424..571B}. This opens a new window on the
understanding of galactic magnetic fields.

Unfortunately, the 5 arcminutes resolution of the {\it Planck} 353~GHz data limits the study of the galactic magnetic field at scales
of 0.2 to 0.5 pc even for the closest clouds. For a detailed exploration of the
magnetic field lines in the star-forming filamentary structures we need to perform high-resolution observations (10-20 arcsec
resolution) of the polarized dust emission \citep{2014ApJ...792..116Z}.

The \nikad\ dual-band millimeter camera \citep{Calvo2016,2016arXiv160508628C}, recently
(October 2015) installed at the IRAM 30 m telescope in Pico Veleta (Spain), is
particularly well adapted to such high-resolution observations of the
polarized thermal dust emission.  \nikad\ features two frequency bands at 260
(polarized) and 150 (non polarized) GHz for a total of 3300 Lumped
Element Kinetic Inductance Detectors (LEKIDs). \nikad\ has expected to have 12
(resp.~18.2) arcsec Full Width Half Maximum (FWHM) resolution at 260~GHz
  (resp.~150~GHz) and a 6.5 arcmin diameter Field of View (FoV) at both
  frequencies. Between 2012 and 2015, a prototype version of \nikad\ named
\nika\ \citep{monfardini2010,catalano2014} was operated at the IRAM 30 m
telescope as a test-bench. \nika\ was also a dual-band camera at 150 and 260 GHz
with a total of 356 LEKIDs, 12 and 18.2 arcsec resolution, but a 1.8~arcmin diameter
FoV. Thanks to a specifically designed polarization setup \nika\ has provided
polarized observations at both frequency bands \citep{Ritacco2015}. This
polarization setup includes an analyzer and a half-wave plate (HWP).

Experiments such as {\it Planck} \citep{planck_mission}, BICEP \citep{bicep}, ACTPol
\citep{ACTPOL}, QUaD \citep{QUAD}, QUIET \citep{QUIET} and QUIJOTE \citep{QUIJOTE}
rotate the instrument with respect to the sky. This modulates the input polarization
signal providing the required angular coverage to reconstruct the $I$, $Q$, and $U$
Stokes parameters. By contrast, other experiments rotate a HWP in front of an
analyzer to modulate the incoming sky polarization. HERTZ \citep{1997PASP..109..307S}, SCUPOL \citep{2003MNRAS.340..353G}, SHARP polarimeter \citep{2008ApOpt..47..422L}, PILOT \citep{PILOT},
BLASTPol \citep{BLASTPol}, SPIDER \citep{SPIDER}, POLARBEAR \citep{polarbear},
and SMA \citep{2008SPIE.7020E..2BM} change the HWP orientation step by step and maintain it fixed during some
periods of observation. EBEX \citep{ebex}, POLKA \citep{polka_apex}, ABS \citep{2016arXiv160105901Er}, \nika, and \nikad\ take another option,
that is, to rotate the HWP continuously. 

We discuss in this paper the
polarization performance of the \nika\ camera and the implications for the
\nikad\ design.
The paper is organized as follows: Sect.~\ref{nika instrument} presents the
\nika\ instrument and the polarization
setup. 
Sect.~\ref{lab_characterization} discusses the laboratory
characterization of the polarization setup. Sect.~\ref{data_analysis}
presents the observational strategy and the dedicated polarization-data-reduction
pipeline.  Sect.~\ref{polcalibration} discusses
observations on quasars; Sect.~\ref{sec:extended} presents the polarization
maps of few extended sources, Orion OMC-1, M87 and Cygnus~A. 
We draw conclusions in  Sect.~\ref{conclusions}.
\begin{figure*}
  \begin{center}
    \includegraphics[width=7cm, keepaspectratio]{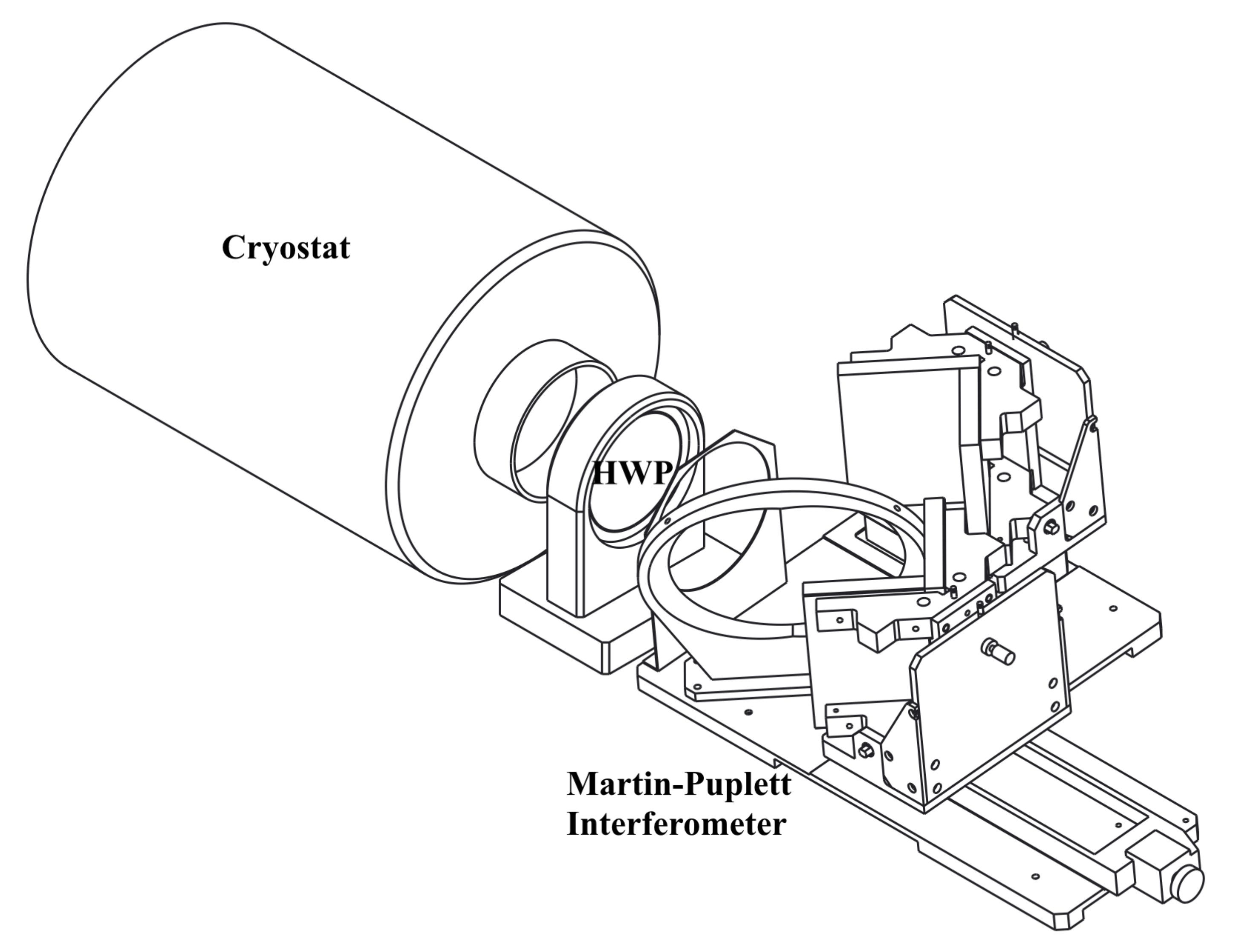}
    \includegraphics[width=7cm, keepaspectratio]{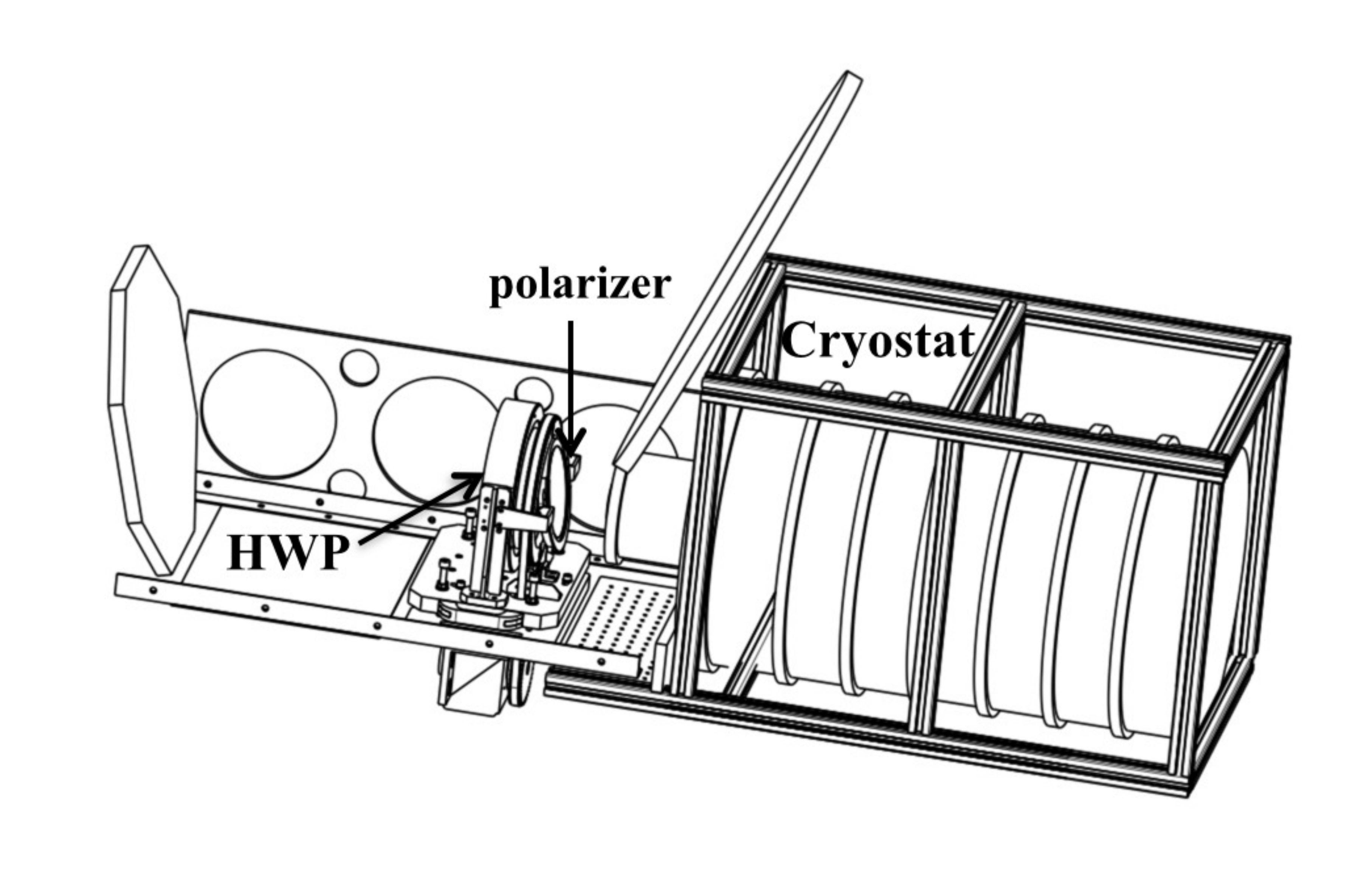}
    \caption{{\it Left} Laboratory instrumental setup: the \nika\ cryostat,
      the HWP in a fixed position, the polarizer mounted inside the cryostat and a Martin-Puplett interferometer. {\it Right}
      Instrumental setup for polarization measurements at the telescope with the last two mirrors of the optics chain and the polarization
        module with the HWP and the stepper motor mounted in front of the
        entrance window of the cryostat. The polarizer is tilted by approximately 10
        degrees with respect to the optical axis to avoid standing waves.}
         \label{polarsetup}
  \end{center}
\end{figure*}
\section{The \nika\ instrument}
\label{nika instrument}
Extensive descriptions of \nika\ and its intensity performances can be found
in \cite{adam2014,catalano2014, monfardini2010, monfardini2011, Calvo2013}.
\cite{adam2015,adam2016} also provides information about \nika\ performance, reporting the characterization of the processing transfer function and extra-noise induced by astrophysical contaminants. 
Here we give a brief summary of the instrument and focus on the extra module that was added to
provide \nika\ with polarimetric capabilities.
\subsection{\nika: a dual-band LEKID camera}
 \nika\ is a dual-band camera consisting of two arrays filled by LEKIDs with a Hilbert geometry
 \citep{roesch}. LEKIDs are superconducting resonators. When photons are
 absorbed, they break Cooper pairs in the superconducting resonant element. This
 changes the density of quasi-particles and modifies the kinetic inductance, and
 hence the resonant frequency of the LEKID \citep{Doyle2008}. The absorbed power
 can be directly related to the shift of the resonant frequency
 \citep{Calvo2013}. The two LEKID arrays were cooled down to their optimal
 temperature of approximately 100~mK using a 4K cryocooler combined to a closed-cycle
 $^3$He-$^4$He dilution. The optical coupling between the telescope and the
 detectors was achieved by warm aluminium mirrors and cold refractive optics
 \citep{catalano2014}. The camera observed the sky in two millimeter channels,
 at 1.15 mm and 2.05 mm, corresponding to bandwidths of 220-270 GHz and 137-172 GHz, respectively,
 with central frequencies of 260~GHz and 150~GHz, respectively. \nika\ had a
 Field of View (FoV) of approximately 1.8 arcminutes and angular resolutions (FWHM) of
 18.2 arcsec and 12 arcsec at 150~GHz and 260~GHz, respectively
 \citep{catalano2014, adam2014}.
\subsection{The polarimetric module}
\label{sec:pol_module}
The polarization setup of the \nika\ camera consisted of a continuously rotating
HWP and an analyzer, at room temperature, facing the cryostat window. The
Hilbert geometry \citep{roesch} of the \nika\ LEKIDS was specifically optimized
for intensity observations and, as a consequence, \nika\ LEKIDs are not sensitive
to polarization. Therefore, for polarization observations, the analyzer was placed
after the HWP at a distance of 6 cm from the cryostat window as shown in the two
configurations of Fig.~\ref{polarsetup}. The analyzer, consisting of a
lithographic kapton coper polarimeter, was tilted by 10 degrees with
respect to the optical axis to avoid standing waves with the cold optical
filters inside the cryostat. The HWP was placed before the analyzer and
  shared the same holder.  We used a hot-pressed metal mesh HWP designed and
built at Cardiff University \citep{pisano}. The HWP was designed to
allow an approximately constant phase shift of the transmitted radiation
over a broad spectral band including the two \nika\ bands. A two-layer broadband
anti-reflection coating was added to the HWP to maximize the transmission across
the band. The HWP was mounted into a mechanical modulator actioned by a stepper
motor synchronously controlled by the \nika\ acquisition system. The power of
the motor was chosen so that a typical stable rotation frequency of 5~Hz could
be achieved during observational campaigns of at least one week. 

The combined action of the continuously rotating HWP and the analyzer leads to a
modulation of the input linear polarization at four times the mechanical
rotation frequency, $\omega_{\rm P} = 2 \pi \nu_{P}$. Thus, setting aside
calibration and system imperfections, each LEKID in the focal plane measures the
following combination of the three Stokes parameters (commonly used to represent the time-averaged polarization state of electromagnetic radiation; for a review on polarization basics we refer to \cite{1992plfa.book.....C}) $I$, $Q$, and $U$:
\begin{equation}
m = I + \rho_{\rm pol} \lbrace Q \cos ( 2 \psi(t) + 4 \omega_{\rm P} \ t) + U \sin ( 2\psi(t) + 4 \omega_{\rm P} \ t) \rbrace,
\label{photoequ}
\end{equation}
where $\psi(t)$ is the angle between the analyzer and the reference axis of the polarization reference
frame, that is, the direction along which $U=0$ and $Q$ $\textgreater$ $0$, and 
$\rho_{\rm pol}$ is the polarization efficiency of the full system. 
\section{Laboratory characterization of the polarimetric module}
\label{lab_characterization}
We begin this section by introducing the main parameters used in the
characterization of the \nika\ HWP. Following \citet{savini2006}, the
Mueller matrix of a realistic HWP can be written as represented in Eq.~\ref{equ.HWP}.

\noindent $\alpha$ and $\beta$ represent the normalized
transmission coefficients on the ordinary and extraordinary axes of the HWP.
The HWP phase shift angle between the ordinary and extraordinary axes is noted
as $\phi$.  $\theta$ represents the angle of the HWP ordinary and extraordinary
axes with respect to the polarization reference frame.

 \begin{widetext}
 \begin{eqnarray} \label{equ.HWP}
  \begin{split}
     M_{HWP}=\frac{1}{2} \left(\begin{array}{lll} \\
       \alpha^2+\beta^2              & (\alpha^2-\beta^2)\cos2\theta & (\alpha^2-\beta^2)\sin2\theta \\
        (\alpha^2-\beta^2)\cos2\theta & (\alpha^2+\beta^2)\cos^22\theta +
       2\alpha\beta\sin^22\theta\cos\phi &
       (\alpha^2+\beta^2-2\alpha\beta\cos\phi)\cos2\theta\sin2\theta \\
        (\alpha^2-\beta^2)\sin2\theta &
        (\alpha^2+\beta^2-2\alpha\beta\cos\phi)\cos2\theta\sin2\theta &
        (\alpha^2+\beta^2)\sin^22\theta + 2\alpha\beta\cos^22\theta\cos\phi
      \end{array}\right)
   \end{split}
  \end{eqnarray}
\end{widetext}

\begin{figure}[t!]
  \begin{center}
   \includegraphics[width=9cm, keepaspectratio]{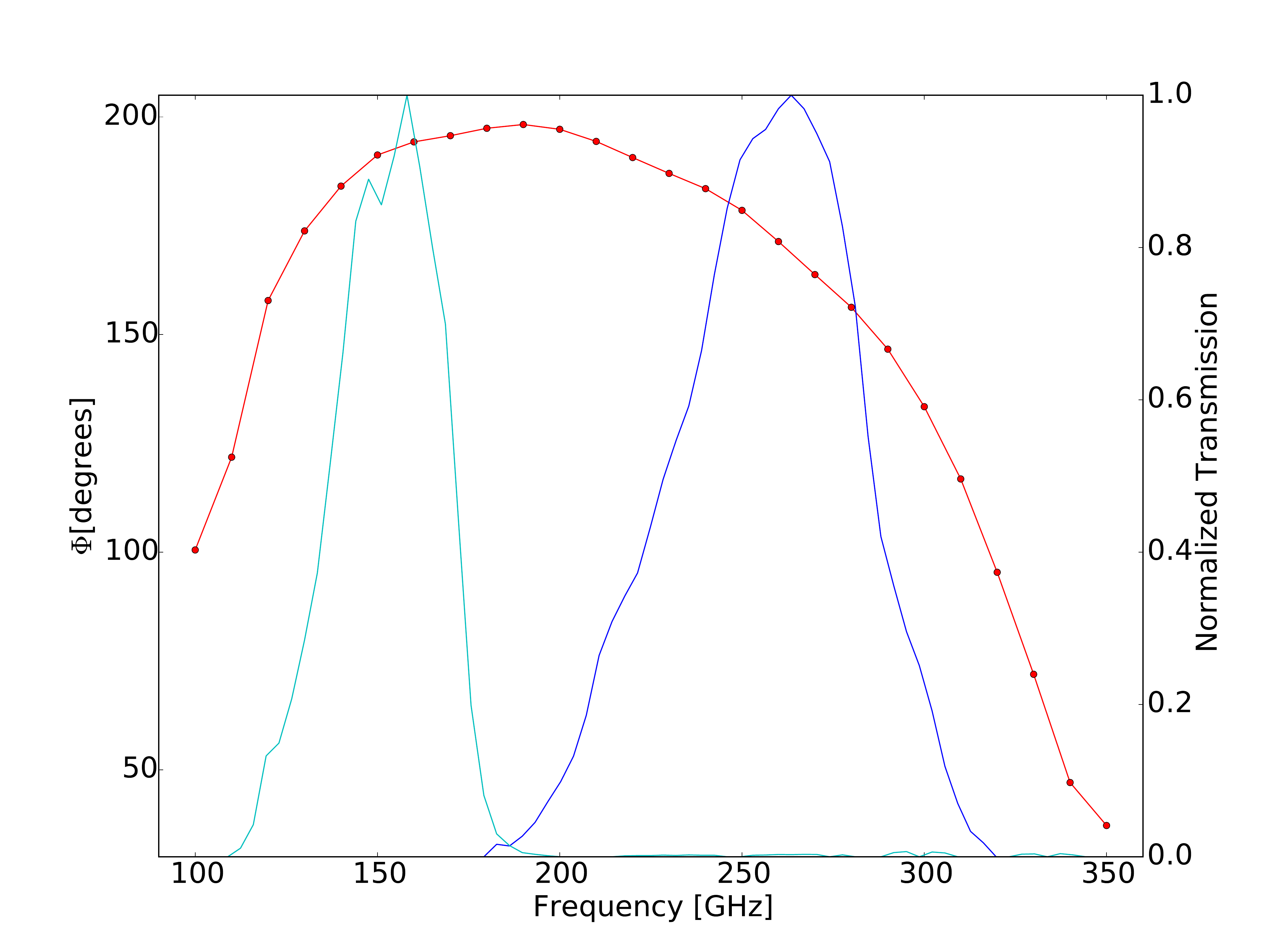}
   \includegraphics[width=9cm, keepaspectratio]{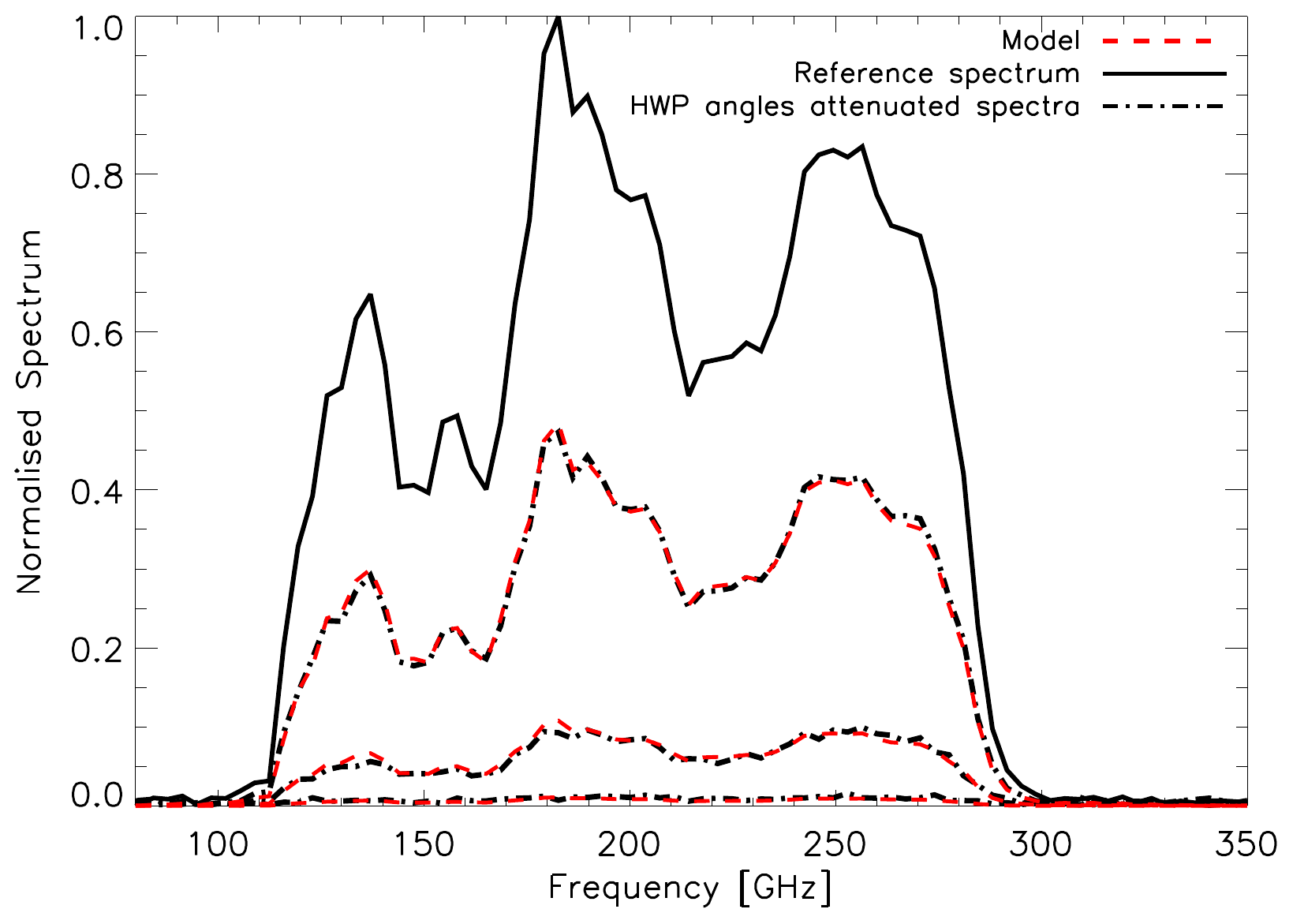}
    \caption{Top: Phase shift angle as a function of frequency for the \nika\ HWP. The transmission for the two \nika\ frequency bands
    at 1 (blue) and 2 (cyan) mm are also plotted for illustration. Bottom: 
    Spectral transmission of the \nika\ HWP for different angles. 
    The red curve corresponds to the best-fit model for the intermediate angle data. The maximum transmission
    corresponds to an angle of $46.8^{\circ}$ with respect to the HWP zero (optical axis). Attenuated spectra at $72^{\circ}$, $79^{\circ}$, and $86.4^{\circ}$
     (top to bottom curves) are also shown.}
        \label{fig:spectre}
  \end{center}
\end{figure}

A first characterization of the properties of the \nika\ HWP was carried out
after fabrication at Cardiff University. In particular, the HWP phase shift angle
$\phi$ was measured over the full bandwidth from 100 to 350~GHz, as shown on the
top panel of Fig.~\ref{fig:spectre}. In the bottom panel of the figure, we show the spectral transmission of the \NIKA\ HWP for 
different rotation angles; an attenuation of the signal is observed as expected. 
In order to estimate the performance of the whole \nika\ polarization
chain, we performed laboratory measurements of the system transmission.  We used
a polarizing Martin Puplett Fourier Transform Spectrometer (MPFTS) to
characterize the spectral transmission of the system.  The MPFTS produces the
difference between the power of two input polarized beams that come from two
black bodies at different temperatures (ambient ECCOSORB and warmed ECCOSORB)
modulated by a rotating wire-grid polarizer.  An array of LEKIDs was placed
facing the MPFTS inside a \nika-type dilution cryostat, which cools down the
optics, the analyzer and the LEKIDs to a temperature of approximately 100 mK.  The analyzer consists of a
wire-grid and it is considered as ideal in the following.  A schematic
view of the instrumental setup is shown in the left panel of
Fig.~\ref{polarsetup}.

We performed a total of eight independent measurements by varying the angle
of the \nika\ HWP axis with respect to the optical axis. During each transmission
measurement the HWP was kept fixed in a defined position. We achieved a spectral
resolution of approximately 3.5 GHz, which corresponds to an approximately 44 mm excursion of the
roof mirror of the MPFTS. We covered the bandwidth of interest by considering a
total of 80 steps for transmission measurement for a total of 9 minutes 
integration time.  As the MPFTS polarizer and the analyzer transmission axis
were set perpendicularly to each other, prior to any measurement, we rotated the
\nika\ HWP to find a zero-point initial position, which maximized the measured
LEKID signal.  By rotating the HWP for each measurement we rotated the
polarization of the MPFTS output signal.  As the analyzer was kept at a fixed
position, this induced an attenuation of the signal measured by the LEKIDs.
This can be observed in the bottom panel of Fig.~\ref{fig:spectre} where we
present the measured transmission as a function of frequency for four HWP
positions going from the maximum (black solid line) to the minimum of
transmission (dashed solid line) spectra. We find that the angle between
these two transmissions is $46.8\pm 1.8^\circ$, which agrees with the expected
45 degrees from Eq.~(\ref{equ.HWP}).
A $1.8^{\circ}$ uncertainty is taken considering that the motor completes 100 steps per tour of the HWP.
Therefore, it is the precision associated to the determination of the HWP zero, corresponding to its optical axis
in the cabin reference frame.

We used previous transmission spectrum measurements to determine
the $\alpha$ and $\beta$ transmission coefficients describing the HWP.
Taking the maximum transmission spectrum described above as a reference, we fitted for the other transmission spectra taken at different angles of the HWP
with respect to the reference polarization axes. We assumed the 
HWP model in Eq.~\ref{equ.HWP} and accounted for the analyzer facing
the LEKID array. The phase shift angle, $\phi$, was set using the values per frequency measured
at Cardiff University and presented above.
Note that as we are using the maximum transmission
spectrum as a template, we are not sensitive to an absolute attenuation amplitude.
Therefore, to break the degeneracy between  $\alpha$ and $\beta$, we fixed $\alpha$ to unity and
only fitted for $\beta$. Using a least square minimization, we find $\beta = 0.999\pm0.005$ at 1.15
mm and $\beta = 0.924\pm0.005$ at 2.05 mm. The best fit models obtained for each position
of the HWP are plotted in red in Fig.~\ref{fig:spectre}.

We define  the polarization efficiency from Eq.~\ref{equ.HWP} as $\rho_{pol} = (1-2\gamma)/2$, where
$\gamma = \frac{\alpha \beta \cos(\phi)}{\alpha^2 + \beta^2}$. Using the effective HWP phase-shift in the \nika\ bands, we find $\rho_{\rm
  pol} = 0.9956 \pm 0.0002$ at 1.15 mm and $0.9941 \pm 0.0002$ at 2.05 mm. These values
are accounted for in the final absolute calibration of our polarization
maps. Finally, as a consequence of the small difference observed between the
$\alpha$ and $\beta$ transmission coefficients, we expect the modulation of the
incoming intensity and polarization around the second harmonic of the HWP
rotation (see Fig.~\ref{spectre_iqu}). However, as presented in Sect.~\ref{se:demod_mapmaking}, we are not
sensitive to this effect in the final maps as it is accounted for in the data
processing.

\begin{figure}
\includegraphics[width=1.\linewidth,keepaspectratio]{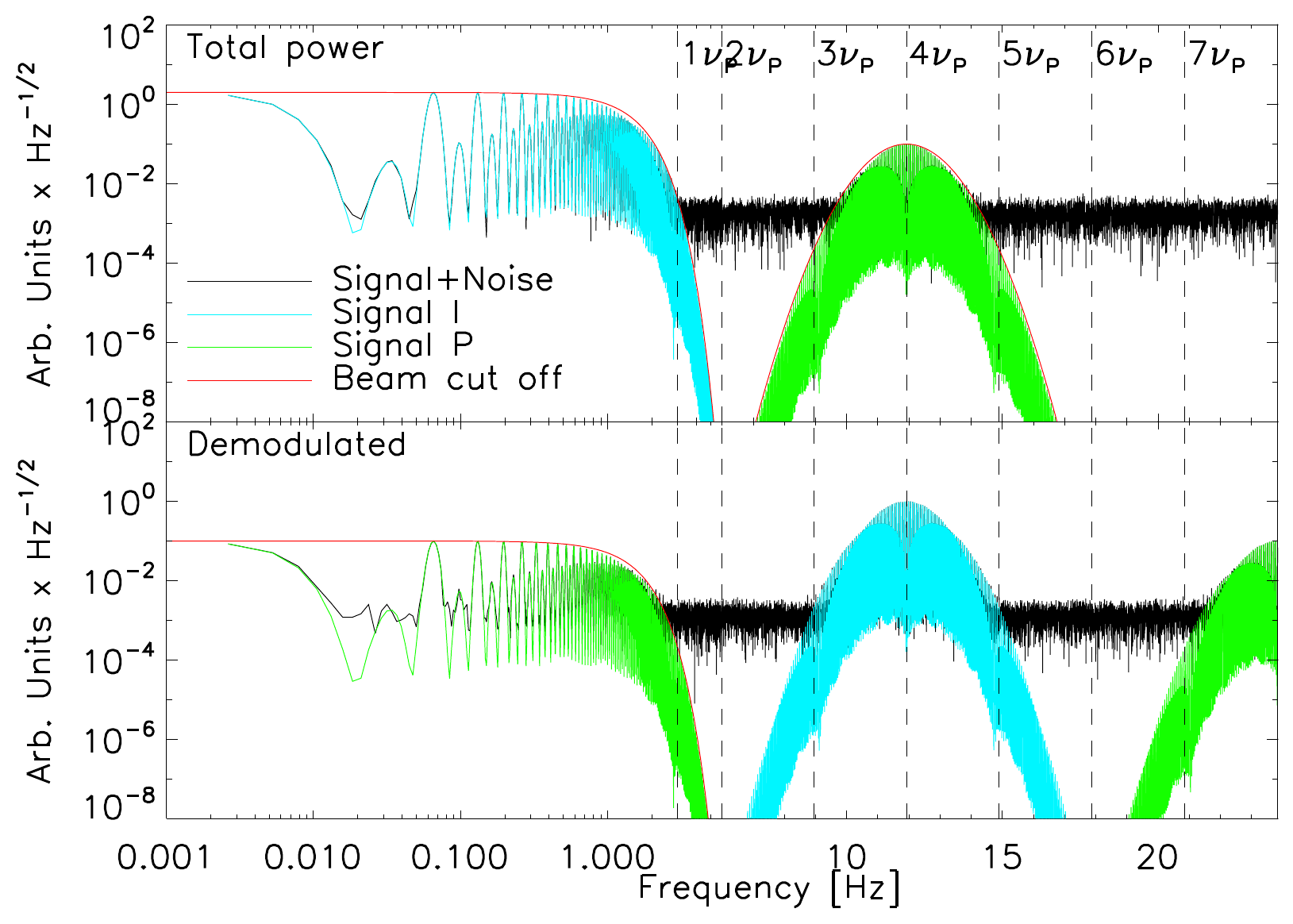}
\caption{The top panel shows the power spectrum of a simulated 
  TOI (Time Ordered Informations) for a polarized point source observed under a
  raster scan with a continuously rotating HWP facing
  a polarization analyzer. The raw signal plus noise TOI (black) has its total
  intensity content highlighted in cyan and the polarized content in green. The
  polarization signal band is centered on the fourth HWP harmonics while the
  intensity signal band lays at lower frequencies. On the bottom plot, we
  present the TOI after demodulation (see. Sect.~\ref{se:lockin}). We observe
  that half of its Stokes $Q$ content has been put at low frequency while the
  Stokes $I$ content and the remaining polarized contents are shifted at
  frequencies higher than the signal band.}
   \label{fig:toi_simu}
\end{figure}  

\begin{figure*}
  \begin{center}
  \includegraphics[%
  width=8.2cm, keepaspectratio]{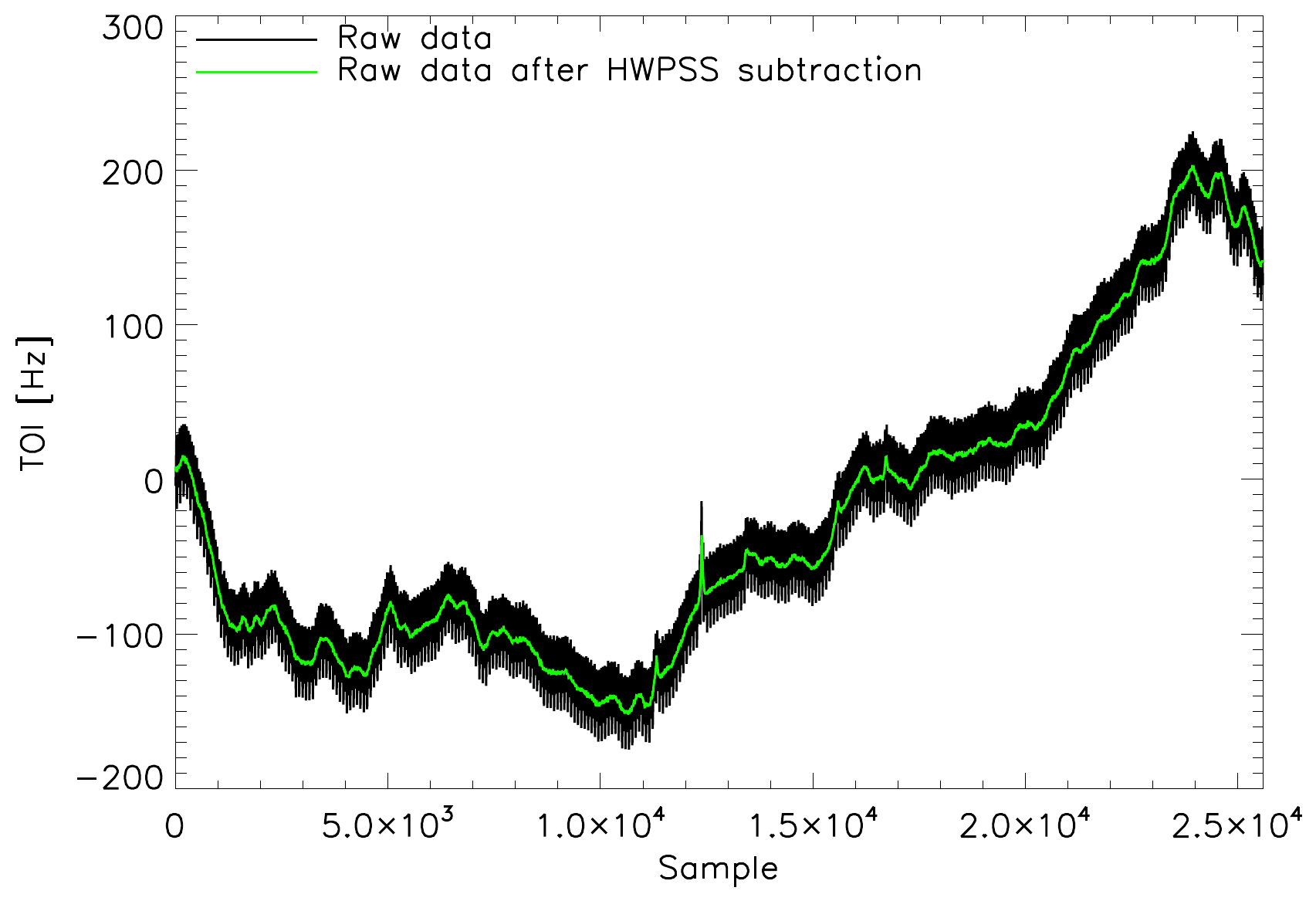}
  \includegraphics[%
  width=7.85cm, keepaspectratio]{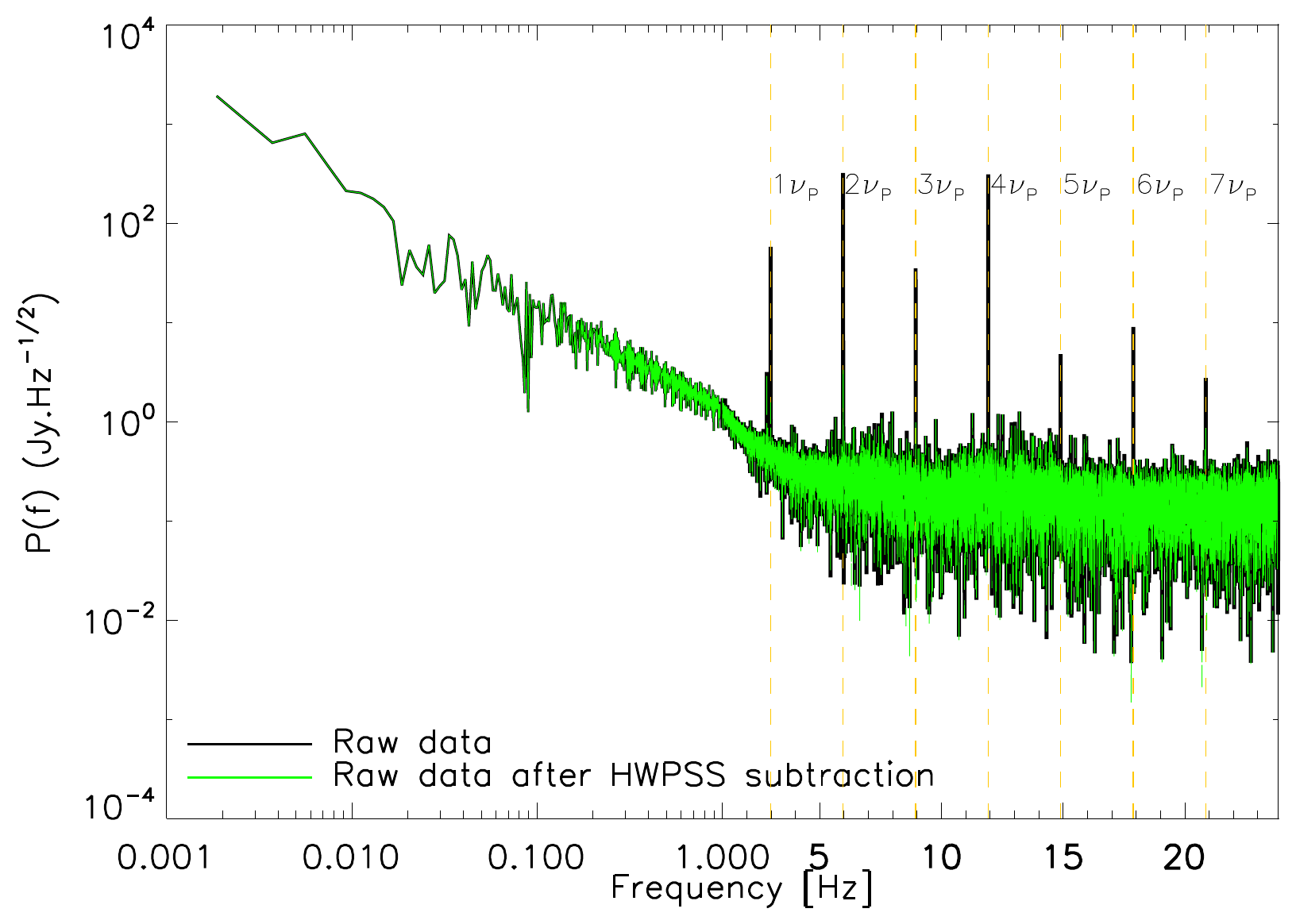}
\caption{TOI (left) and power spectrum (right) of an observation of Orion OMC-1
  for a single KID. Raw data are presented in black and the Half Wave Plate systematic signal (HWPSS) subtracted
  data in green.}
  \label{toi_i}
   \end{center}
   \end{figure*}  
    \begin{figure}
  \begin{center}
    \includegraphics[%
      width=10cm,keepaspectratio]{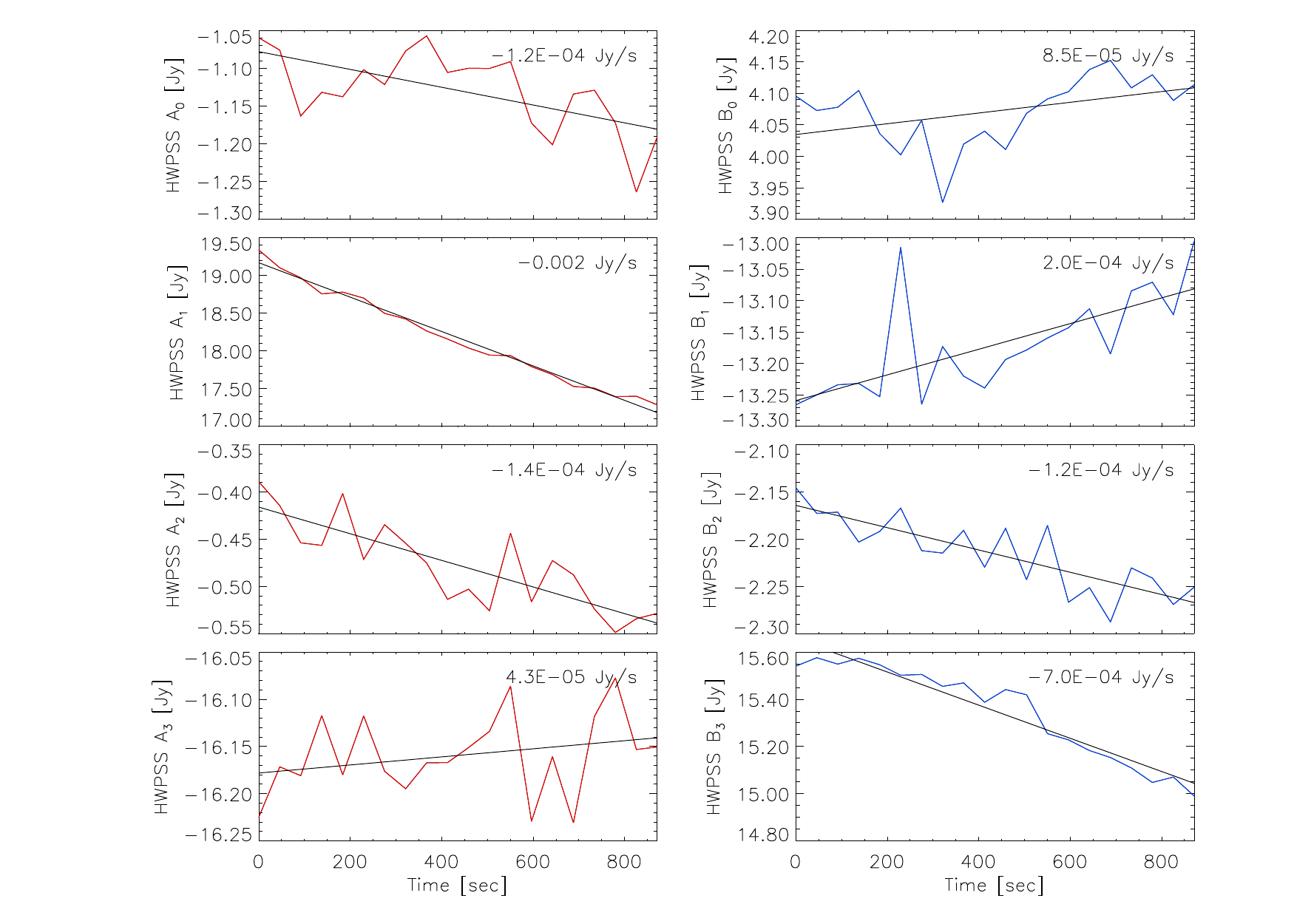}
    \caption{Absolute amplitudes of the four main harmonics of the HWPSS as a function of time. Each point is a measurement of these amplitudes on a chunk of approximately 30 seconds. The amplitudes show a slow and linear drift in time.}
    \label{time_drift}
  \end{center}
\end{figure}
 \begin{figure*}
   \begin{center}
     \includegraphics[%
       width=6cm,keepaspectratio]{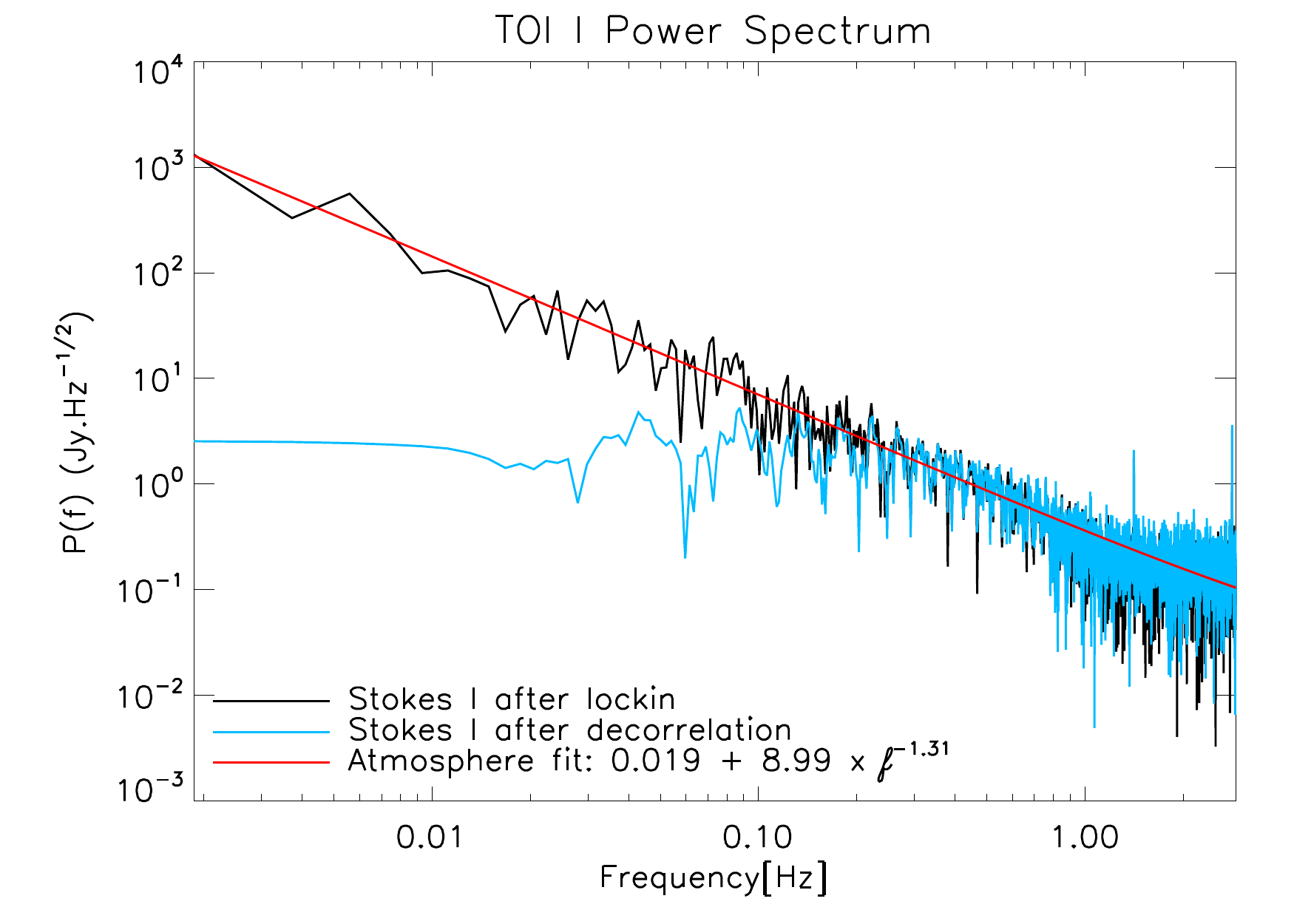}
     \includegraphics[%
       width=6cm,keepaspectratio]{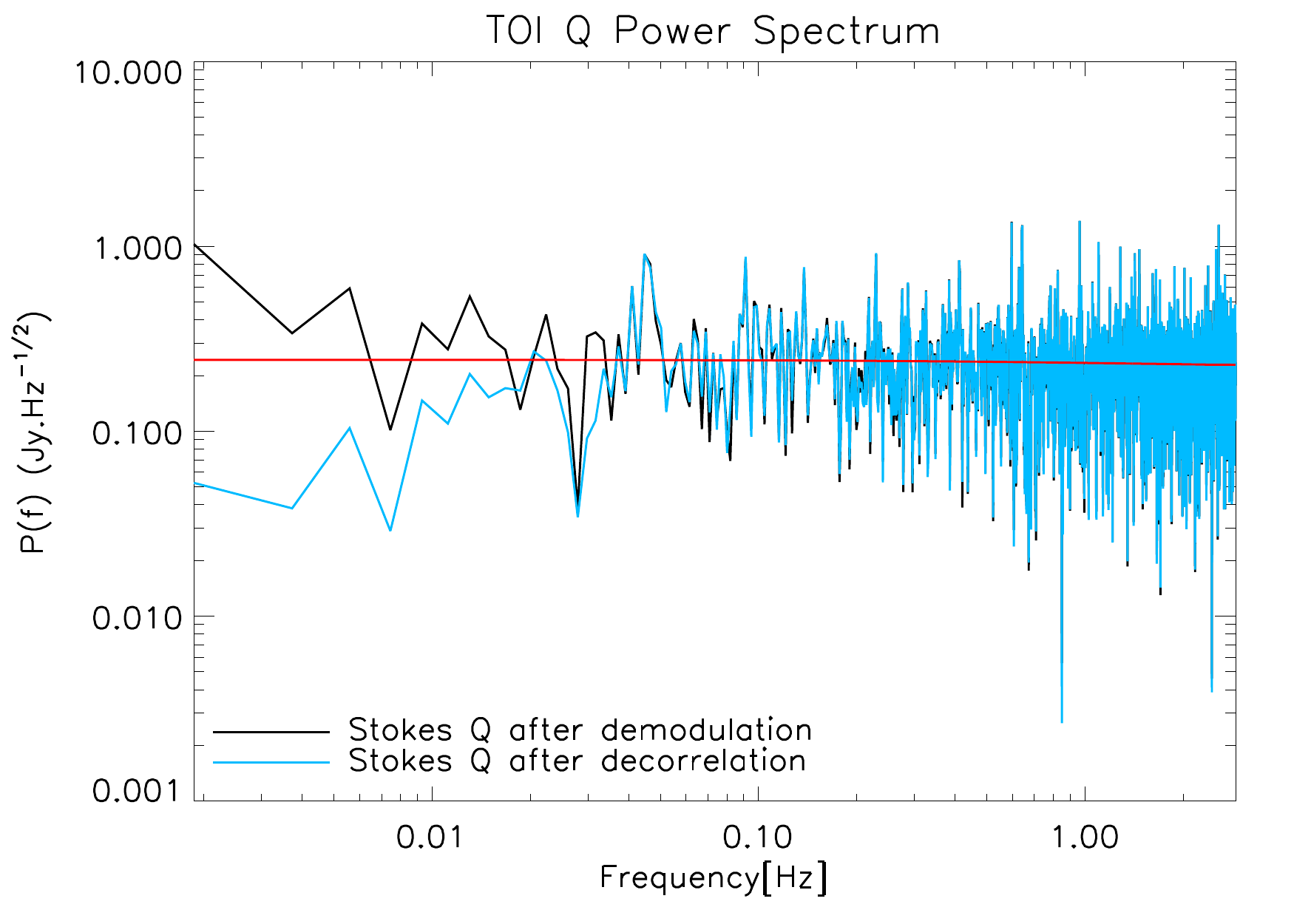}
     \includegraphics[%
       width=6cm,keepaspectratio]{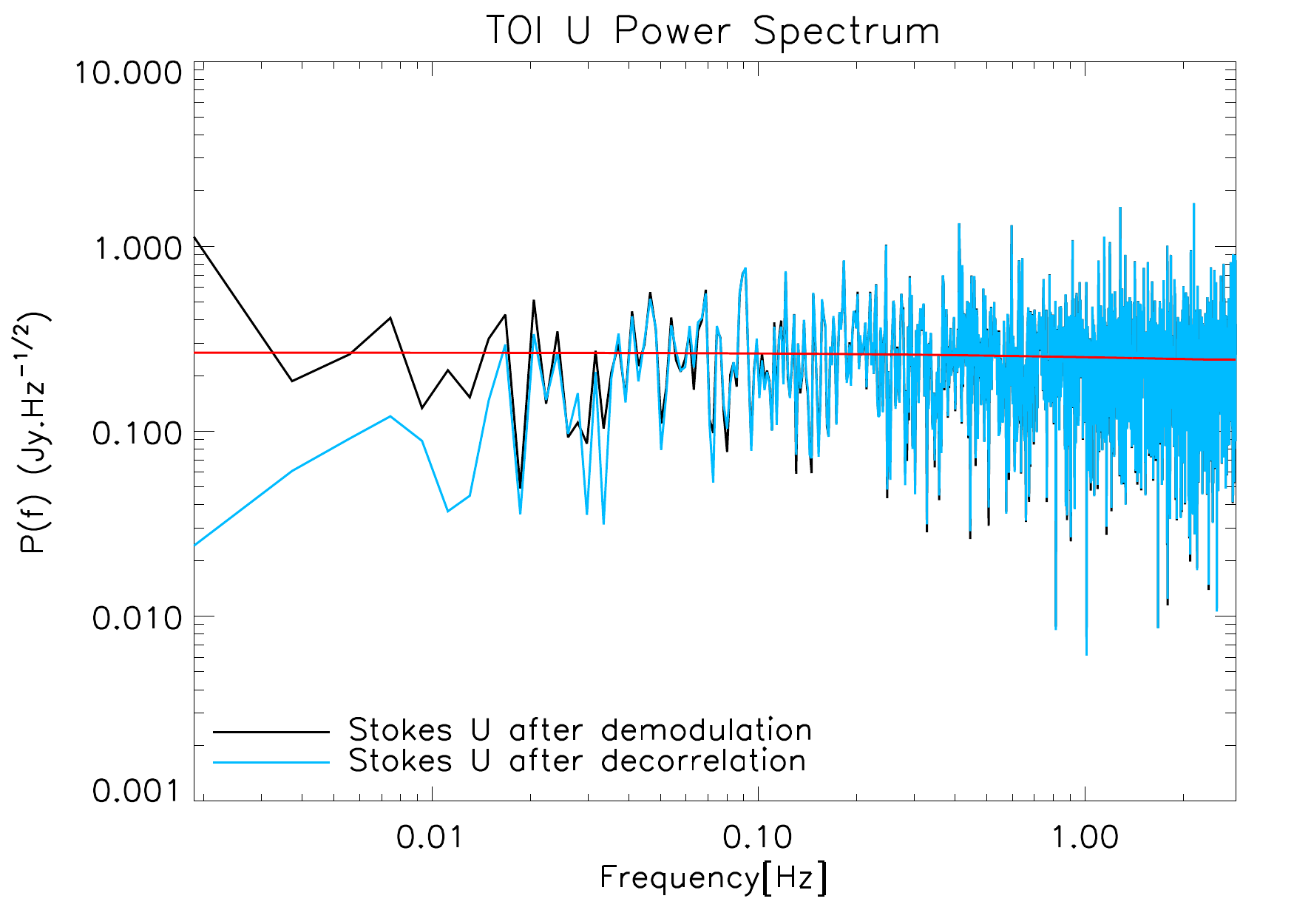}
     \caption{From left to right: power spectra of the Stokes $I$, $Q$, and $U$ pure
       TOIs (black) after applying the lock-in procedure to the raw data on
       Fig.~\ref{toi_i}.  A bandpass filter ($[0.01,2.9]$~Hz) was applied to
       reject any high-frequency noise and half wave plate systematic signal (HWPSS) residual at the fourth
       harmonics. In blue we also show the power spectra after applying the
       decorrelation procedure to the demodulated data.  See
       Sect.~\ref{se:demod_mapmaking} for details.}
     \label{spectre_iqu}
   \end{center}
 \end{figure*}

\section{Reconstruction of the \nika\ Stokes parameter maps}
\label{data_analysis}

We here describe the specificities of \nika's polarization modulation strategy
and data analysis. We start by giving more details on the fast and continuous
rotation of the HWP and how it impacts the signal. We then present specific
systematics associated to the rotating HWP and the optics, and how we handle
them. 
 
\subsection{Polarization modulation with a continuously rotating HWP}
\label{se:lockin}
The key aspect for reconstructing polarization with \nika\ is the rotating HWP,
which modulates the input polarization signal.  Thus, we recall here the main
lines guiding polarization modulation by a continously rotating HWP and its
subsequent data analysis. For the sake of clarity, we consider the case of a
polarized point source, which is observed under a classical raster scan strategy
with constant declination while scanning along right ascension $\alpha$ at speed
$\dot{\alpha}$. This type of scan being pseudo-periodical, the Fourier transform
of the detector-observed time-ordered data (TOI, equivalent to Time Ordered Information) shows peaks at harmonics of the
scanning frequency (see Fig.~\ref{fig:toi_simu}), each peak containing the sum of
the unpolarized and polarized fluxes of the source (see
Eq.~\ref{photoequ}). These peaks are damped by the instrument resolution at
high frequency. The cut off at
high-spatial-resolution turns into a high-temporal-frequencies cutoff with
typical Gaussian width $FWHM_\nu = \dot{\alpha}/2\pi FWHM$. This beam cutoff
defines the signal band. According to Eq.~(\ref{photoequ}), when we rotate a HWP
in front of an analyzer, the polarized fraction of the signal is modulated at
four times the mechanical rotation frequency of the HWP. Therefore, this shifts the
polarized content of the signal at higher frequencies and recenters the signal
band around the fourth harmonic of the mechanical rotation of the HWP (see
Fig.~\ref{fig:toi_simu}-top). It is therefore clear that a low pass filter applied to
the data above the beam cutoff and below the polarization signal band will
preserve the unpolarized signal band while rejecting the polarized content and high-frequency noise. We refer to this type of TOI as ``pure-$I$''
TOI in the following. To recover the polarization Stokes parameters $Q$
and $U$, we use a classic demodulation procedure discussed, for example, in \citet{haykin2008communication} and also
adopted by \citet{johnson2007}. We
build two reference signals:
\begin{eqnarray}
ref_Q = \cos(2\psi(t) + 4\omega_Pt), \nonumber \\
ref_U = \sin(2\psi(t) + 4\omega_Pt).
\end{eqnarray}
We multiply our TOIs by these reference signals and according
to Eq.~(\ref{photoequ}), obtain, for example, {\it e.g.} $Q/2 + I\cos[2\psi(t)+4\omega_Pt] +...$ $Q$
and $U$ terms around the 8th harmonics of the HWP rotation. We have thus
``demodulated'' the $Q$ content of the TOI and brought it back to low
frequencies, while rejecting the $U$ and $I$ content at high frequencies. A low
pass filter above the beam cutoff and below the tail of the modulated intensity
therefore provides a ``pure-$Q$''  TOI (see the bottom panel of
Fig.~\ref{fig:toi_simu}). Equivalently multiplying by
$\sin\left[2\psi(t)+4\omega_Pt\right]$ and filtering we obtain a ``pure-$U$'' TOI.

\subsection{Polarization measurements at the telescope}
\label{polsetup}
The \nika\ polarization setup at the telescope was similar to the one in the
laboratory as shown on the right panel of Fig.~\ref{polarsetup}.  The HWP and
analyzer were placed in the same mount facing the cryostat window in the optical
axis of the Nasmyth cabin. Thus, the polarization reference frame was defined
perpendicularly to the optical axis in Nasmyth coordinates. 

During polarization observations the HWP was continuously rotated to modulate
the input polarization signal as discussed in Sect.~\ref{se:lockin}. The optimal
HWP rotation speed is constrained by several factors including scanning speed
and the scale of atmospheric variations. Usual scanning speeds for \nika\ were approximately a few tens of arcsecs/s. Atmospheric turbulence and variations
across the scan transform into $1/f$-like noise with typical knee frequency
below 1-2~Hz for reasonable atmospheric conditions, as illustrated on
Fig.~\ref{toi_i} that shows measured raw TOI (black curve) for a single detector as a
function of sample number (left) and its corresponding power spectrum
(right). In addition to the atmospheric noise, a subdominant detector
correlated-noise component had been found in the \nika\ data and identified as
electronic-based noise. 

Rotating the HWP such that four times its rotation speed places the polarization
signal well above 2~Hz  permits a natural rejection of these two major
noise components. If the rotation is also fast compared to the scanning speed
and the angular resolution, the three Stokes parameters can be derived
quasi-simultaneously, thus rejecting further residual low-frequency
drifts. Finally, a fast rotation places possible parasitic signals at harmonics
of the rotation frequency outside the signal band (see Sect.~\ref{se:hwpss} for
more details). Both the polarization modulation and parasitic signals can be
clearly observed on Fig.~\ref{toi_i} as high-frequency peaks in the TOI
spectrum at harmonics of the HWP rotation frequency. A fast rotation sets tight mechanical constraints on the stepper motor 
and imposes a faster data acquisition with respect to unpolarized observations. That
is why we chose, as discussed above, to acquire data at 47.7~Hz, rotate
the HWP at 2.98~Hz, and to limit our scanning speed to 26~arcsecs/s. This
provides five measurements of $I$, $Q$, and $U$ per FWHM, well within the Nyquist limit,
even at a LEKID timeline level. Each of these five measurements results from the
  combination of four data samples taken at four different HWP angles. To
  conclude, we are then able to reconstruct the Stokes parameters per
  detector, with high spatial and temporal redundancy and to reject atmospheric
  noise from the polarization signal band.
\subsection{Systematic HWP synchronous signal correction}
\label{se:hwpss}
Imperfections of the HWP modulate the background and lead to a strong
additional parasitic signal, highly peaked at harmonics of the HWP rotation
frequency $\nu_P$ as shown in the right panel of Fig.~\ref{toi_i}. Such a HWP
synchronous signal (HWPSS) was previously observed by Maxima \citep{johnson2007}
and EBEX \citep{ebex}, although the HWP driving mechanisms were
  different. Like them, we find that the signal is well fitted by a sum of
harmonics of the HWP rotation frequency, $\nu_P$ with amplitudes slowly and
linearly drifting. This is illustrated in Fig.~\ref{time_drift} where we plot
the time variation of the cosine $A_n$ (red) and sine $B_n$ (blue) component
amplitudes of the first four harmonics of the HWP rotation frequency along a scan of Uranus. To derive these amplitudes, we use a simplified model of Eq.~(\ref{eq:hwpss}) to fit only the constant part of amplitudes $A_n$ and $B_n$ over chunks of 30 seconds. We observe that the relative
variation is at most 2 mJy/s, mainly dependent on the background (Stokes $I$). 
We therefore model this additional parasitic signal as a Fourier series of 
the harmonics of the HWP rotation frequency
\begin{eqnarray}
HWPSS(t) &= \sum_{n=1}^{8}& (A_n^0 + \epsilon_{A_n}t)\cos n\omega_Pt\nonumber\\
&& + (B_n^0 + \epsilon_{B_n}t)\sin n\omega_Pt.
\label{eq:hwpss}
\end{eqnarray}
We consider up to eight harmonics of $\nu_{P}$ and explicitly include a linear variation of the
amplitude coefficients both for the sine and cosine components when we fit this model on data.
We perform a simple linear fit of the previous model to the full calibrated and
opacity corrected (see Sect.~\ref{se:calib} for details) raw TOI to derive the
best-fit amplitude coefficients. With these coefficients in hand, we
construct a template of the HWPSS, which is then subtracted from the raw
calibrated TOI. An illustration of this procedure is shown in the left panel of
Fig.~\ref{toi_i} where the raw TOI is shown before (black) and after (green)
subtraction of the HWPSS template. Equivalently, the right panel of the figure
presents the power spectrum of the raw TOI before (black) and after (green) the
subtraction of the HWPSS template. Thanks to this procedure, the
HWPSS residuals are reduced to the noise level.
\subsection{\nika\ demodulation procedure and map making}
\label{se:demod_mapmaking}
After subtraction of the HWPSS, we use the lock-in procedure described in
Sect.~\ref{se:lockin} to separate the raw TOI per KID into pure Stokes $I$,
$Q$, and $U$ TOIs. These TOIs are decorrelated from a common mode to remove the
residual atmospheric and electronic noise, and then projected into Stokes $I$, $Q$, and
$U$ maps.  We use the same decorrelation procedures developed for the
intensity-only \nika\ observations \citep[see][for
 details]{adam2014,catalano2014}. In particular we use a common mode decorrelation, masking the source
 and using only the pixels outside the source to decorrelate.
 A low pass filter is also applied to the pure
Stokes $I$, $Q$, and $U$ TOIs to reject high-frequency noise. The frequency
cutoff is set slightly below the HWP rotation frequency.  For the projection we
use an inverse noise-weighting procedure and account for instrumental flags
indicating unreliable data samples or detectors \citep[see][for details]{adam2014,catalano2014}.
In the case of the $Q$ and $U$ maps, this map-making procedure is almost optimal map making because the noise is
expected to be nearly white in the pure Stokes $Q$ and $U$ TOIs. This is
illustrated in Fig.~\ref{spectre_iqu} where we present, in black, the power
spectra of the pure Stokes $I$, $Q$, and $U$ TOIs obtained by applying the lock-in
procedure to the raw TOI of Fig.~\ref{toi_i}. We observe that the power
spectra of the pure Stokes $Q$ and $U$ TOIs are almost flat indicating, as
expected, a significant, although not complete, reduction of the contribution from
atmospheric fluctuations that shows a $1/f$-like component on a pure $I$
TOI.
The best-fit power spectrum models are presented in red in
Fig.~\ref{spectre_iqu}. Using simulations, we have proven that the residual $1/f$-like component in $Q$ and $U$ is consistent with residual atmospheric emission
induced by intensity to polarization leakage as discussed in
Sect.~\ref{sec:polleak}. We also show the power spectra, after applying
the decorrelation procedure, in cyan. This leads to a significant reduction of the $1/f$-like noise in the pure Stokes $I$ TOI power spectrum and of the residual low-frequency tail on the pure Stokes $Q$ and $U$ TOIs.
 
\subsection{Absolute calibration and inter-calibration}
\label{se:calib}

Absolute calibration is performed in the same way as for the intensity-only
\nika\ observations \citep{adam2014,catalano2014}. We use Uranus as our main
absolute flux calibrator and compute calibration factors per KID by fitting a 2D
Gaussian to the data. We take the measured FWHMs of 12~arcsec at 1.15~mm and
18.2~arcsec at 2.05~mm. For the data presented in this paper, we have 14\% uncertainty on
absolute calibration at 1.15~mm and 5\% at 2.05~mm.  
The standard deviation of the flux distribution on Uranus directly gives the calibration 
error associated to the estimation of point source flux values.
After calibration, the data
are given in units of Jy/beam.  We also correct the raw data from atmospheric
absorption using the \nika\ instrument as a taumeter following the procedure
described in~\cite{catalano2014}. 

\subsection{Instrumental polarization and leakage}
\label{sec:polleak}

\begin{figure*}
  \begin{center}
\setlength{\unitlength}{\columnwidth} 
\begin{picture}(2,2.5)
    \put(0.05,2.48){(a) 1.15 mm raw (top row) and leakage corrected (bottom row) Stokes $I$,$Q$ and $U$ maps.}
     \put(0,1.85){ \includegraphics[width=0.33\linewidth,keepaspectratio]{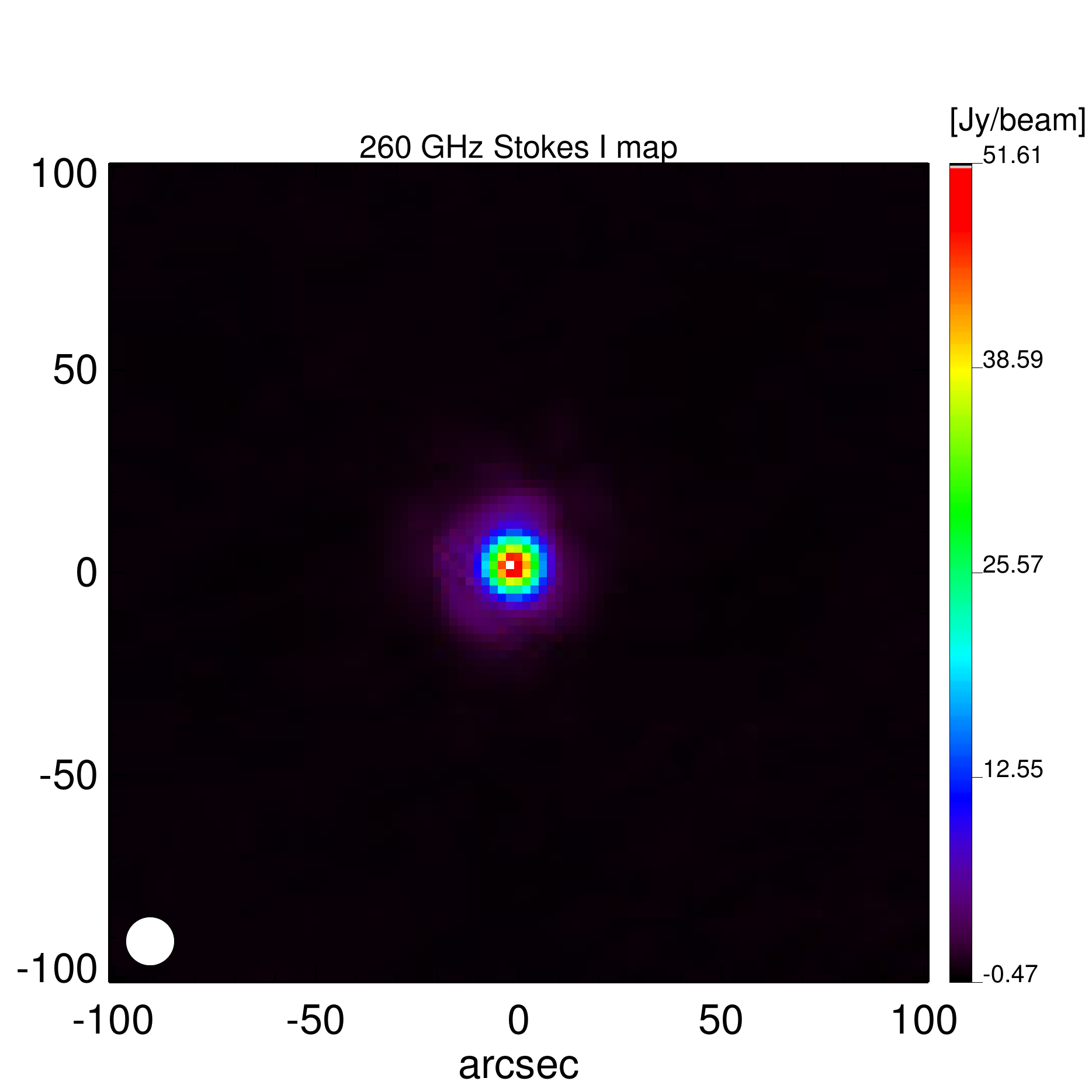}}
     \put(0.7,1.85){\includegraphics[width=0.33\linewidth,keepaspectratio]{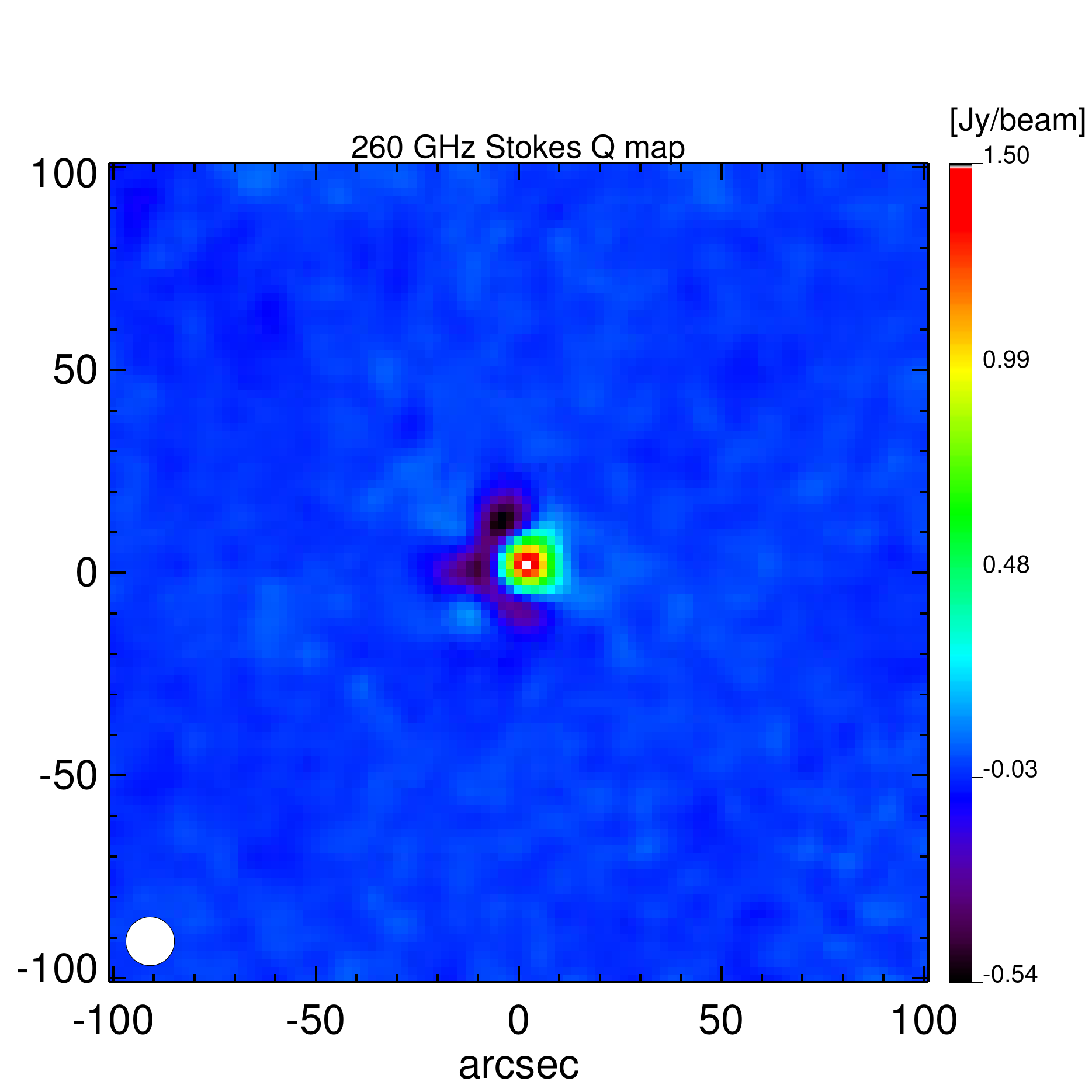}}
       \put(1.4,1.85){\includegraphics[width=0.33\linewidth,keepaspectratio]{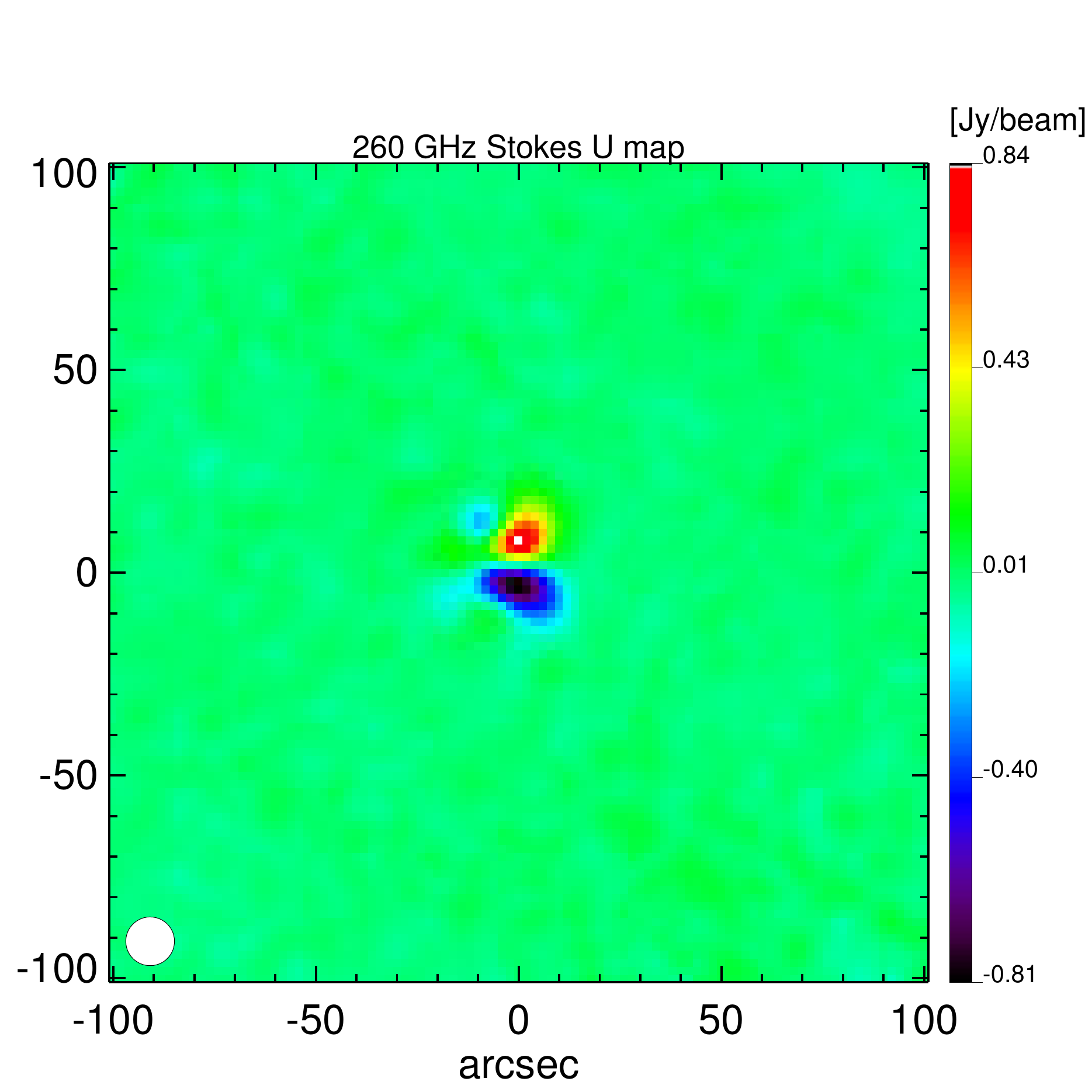}}
 
     
      \put(0,1.25){\includegraphics[width=0.33\linewidth,keepaspectratio]{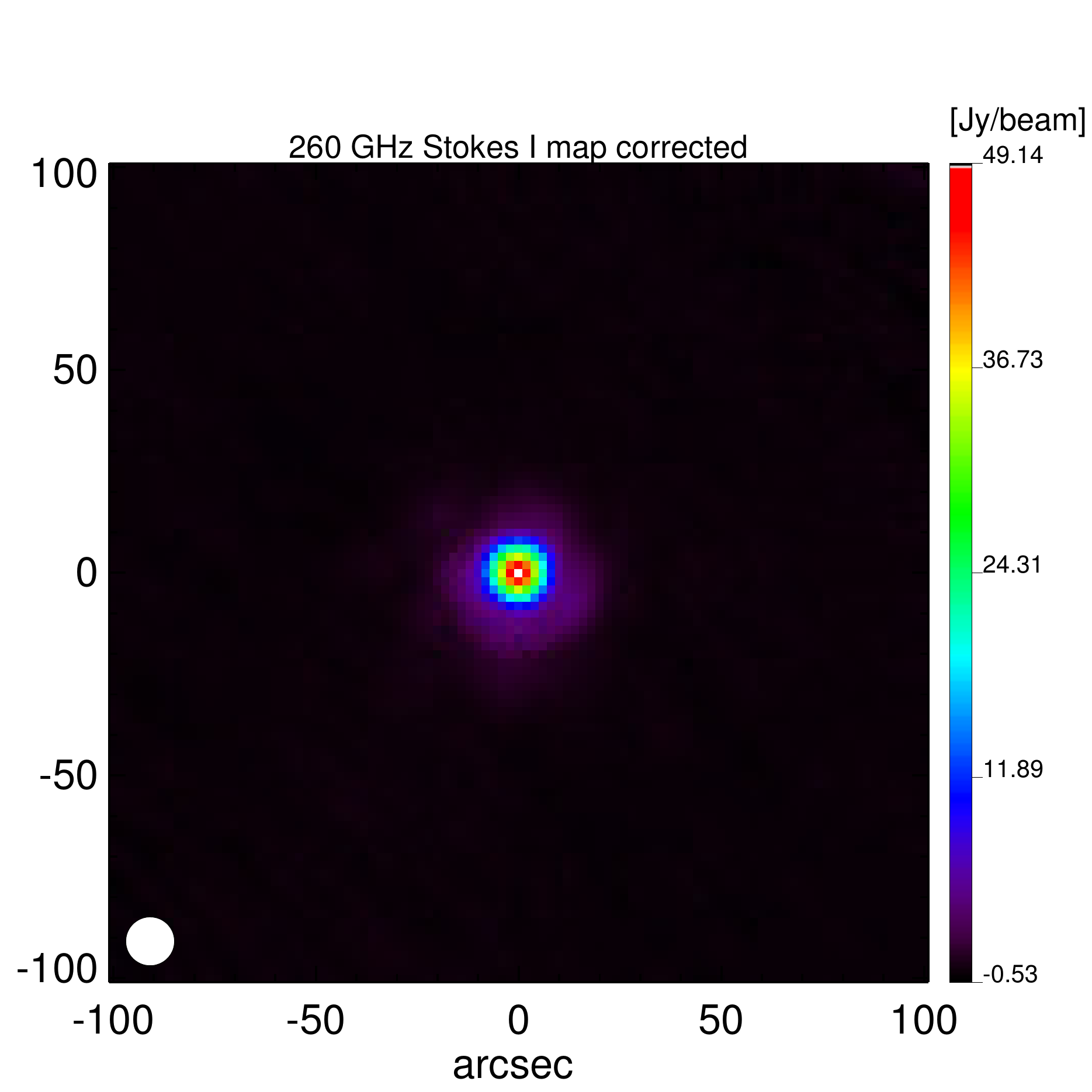}}
     \put(0.7,1.25){\includegraphics[width=0.33\linewidth,keepaspectratio]{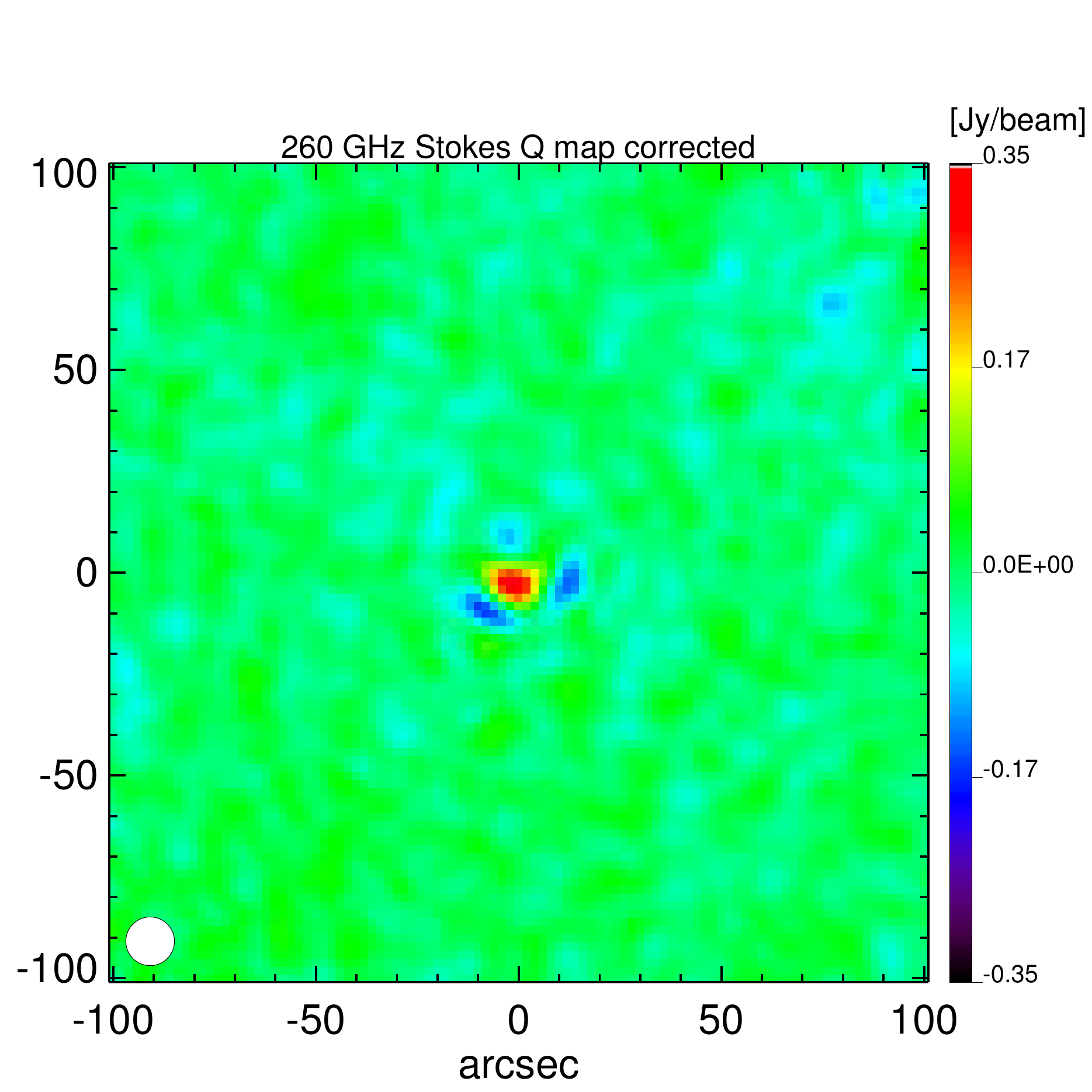}}
     \put(1.4,1.25){ \includegraphics[width=0.33\linewidth,keepaspectratio]{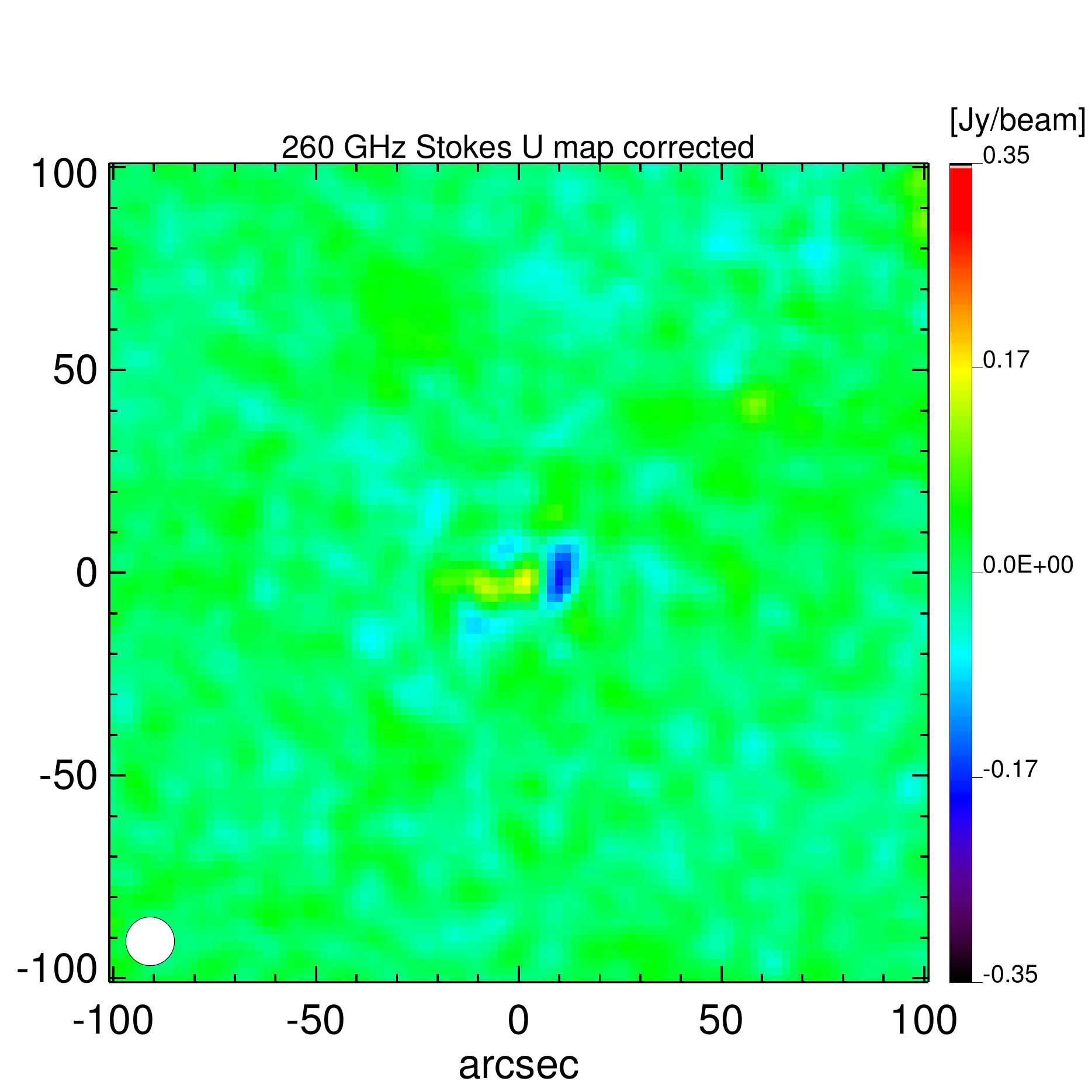}}

  
    \put(0.05,1.22){(b) 2.05 mm raw (top row) and leakage corrected (bottom row) Stokes $I$, $Q$ and $U$ maps.}

      \put(0,0.6){\includegraphics[width=0.33\linewidth,keepaspectratio]{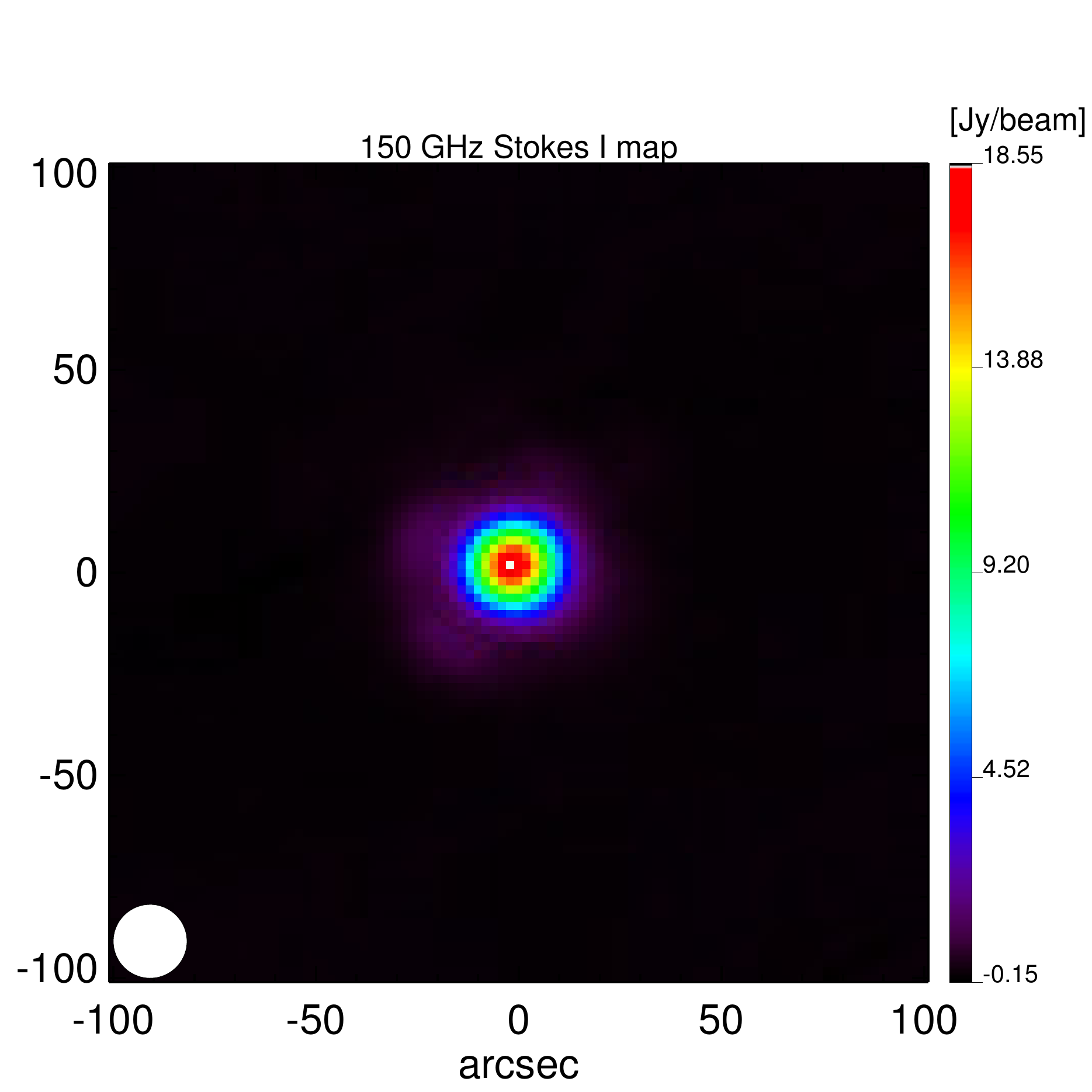}}
      \put(0.7,0.6){\includegraphics[width=0.33\linewidth,keepaspectratio]{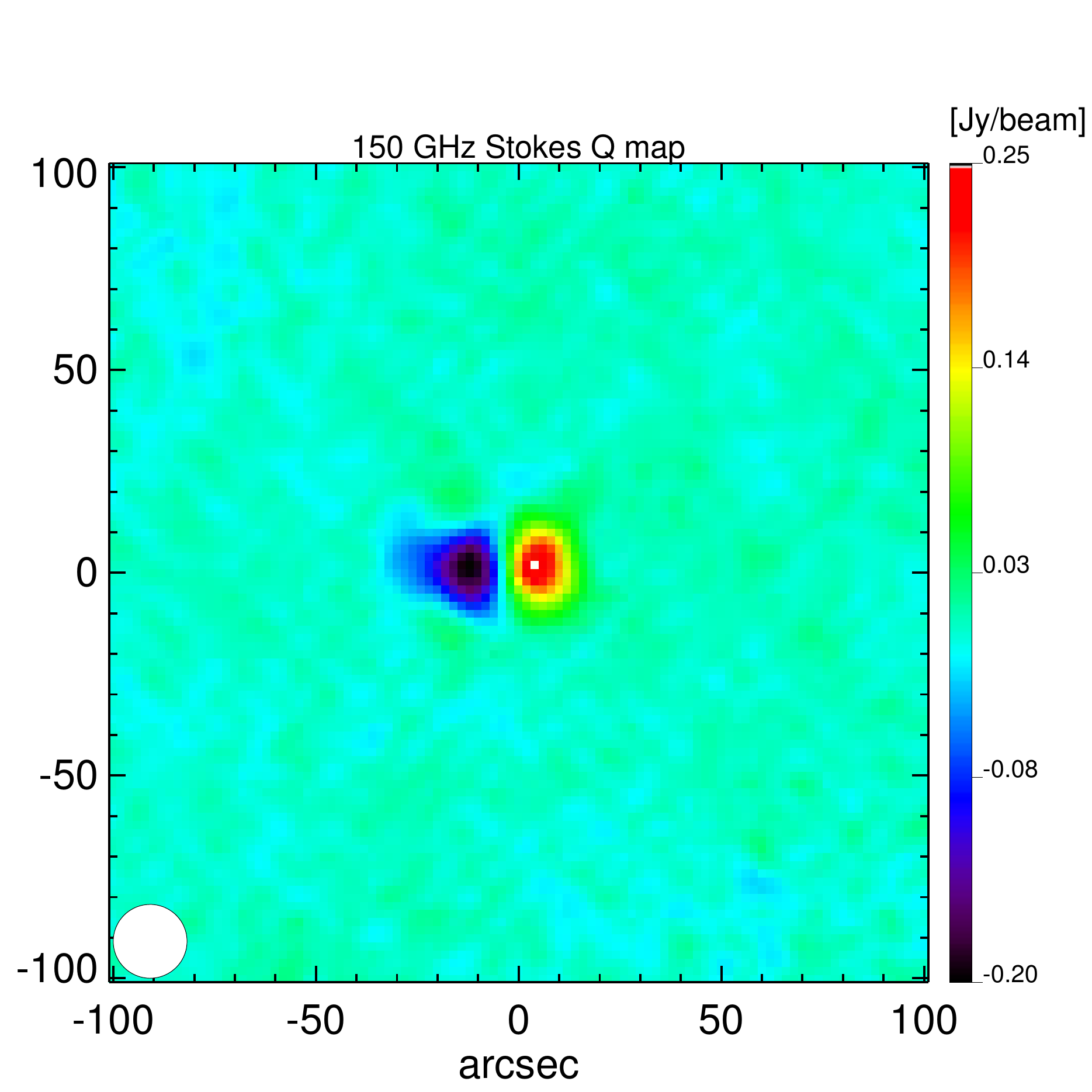}}
     \put(1.4,0.6){\includegraphics[width=0.33\linewidth,keepaspectratio]{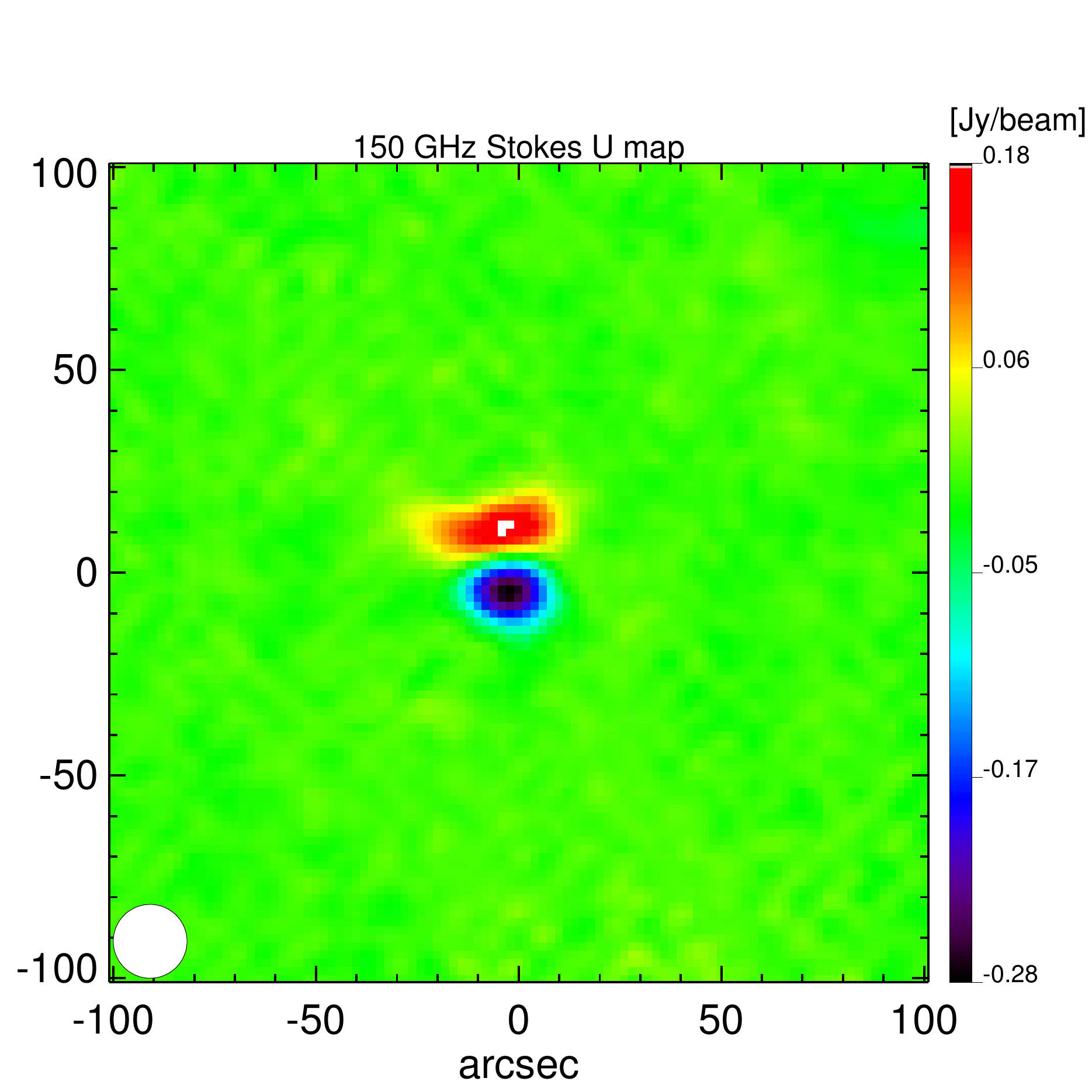}}
      \put(0,0){\includegraphics[width=0.33\linewidth,keepaspectratio]{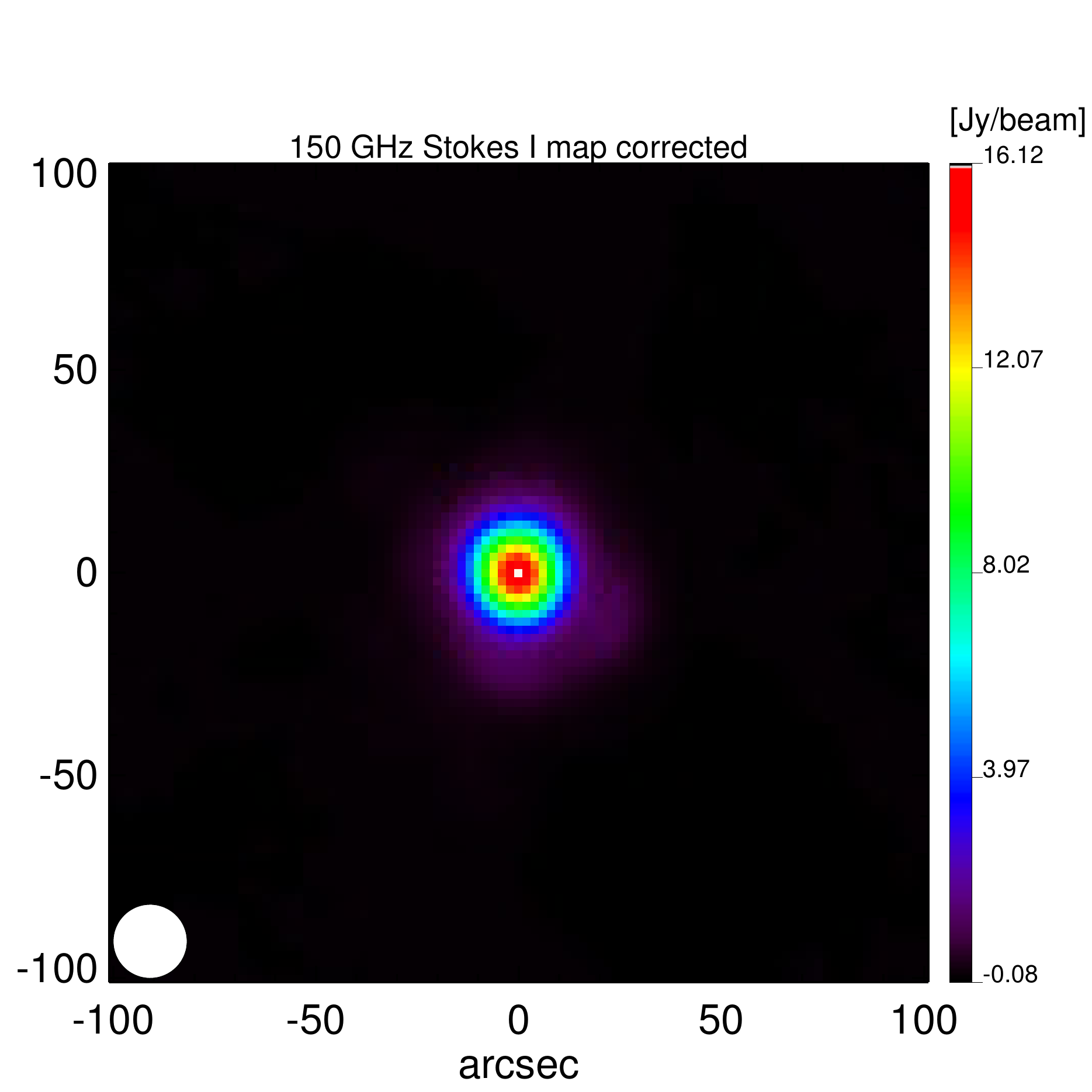}}
     \put(0.7,0){\includegraphics[width=0.33\linewidth,keepaspectratio]{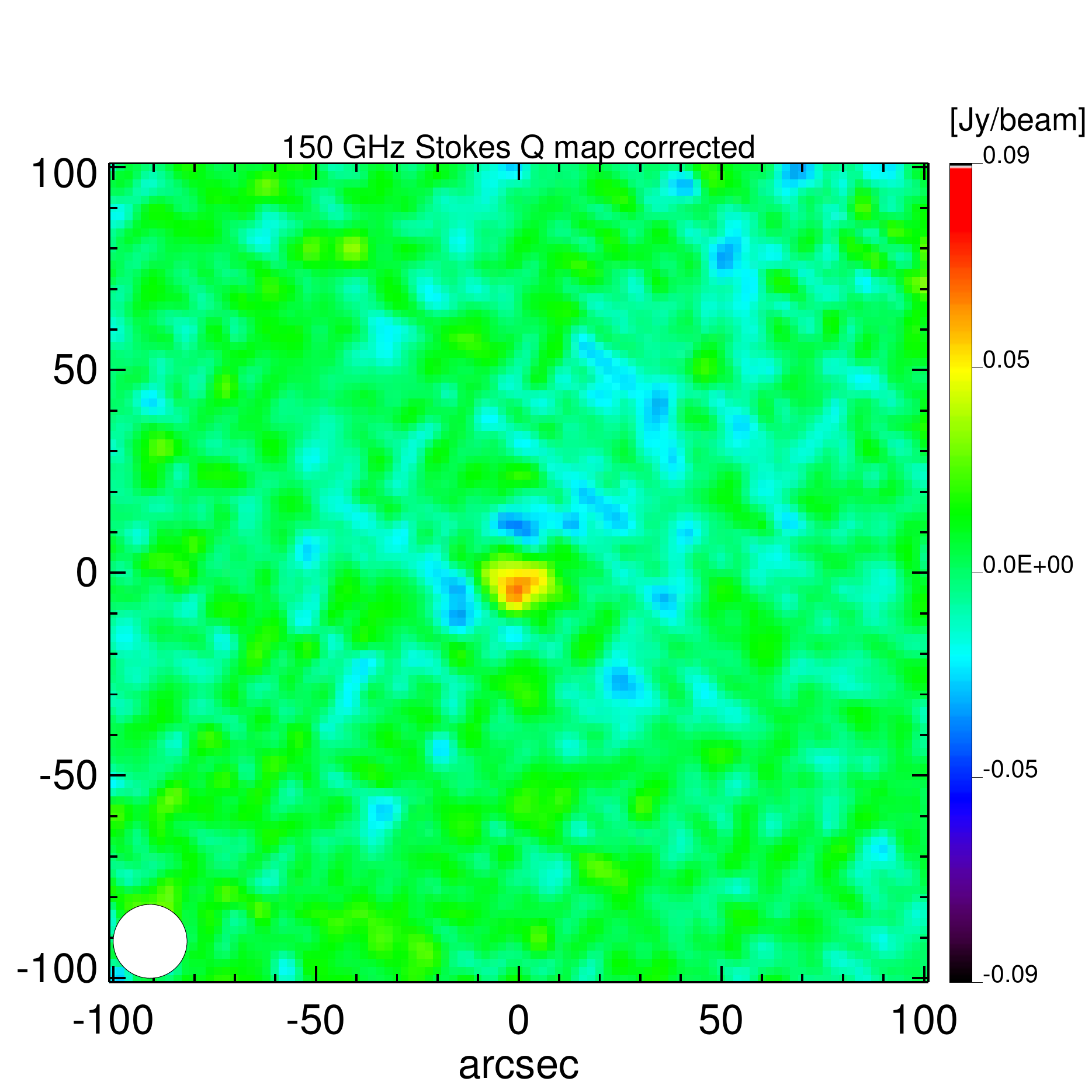}}
     \put(1.4,0){\includegraphics[width=0.33\linewidth,keepaspectratio]{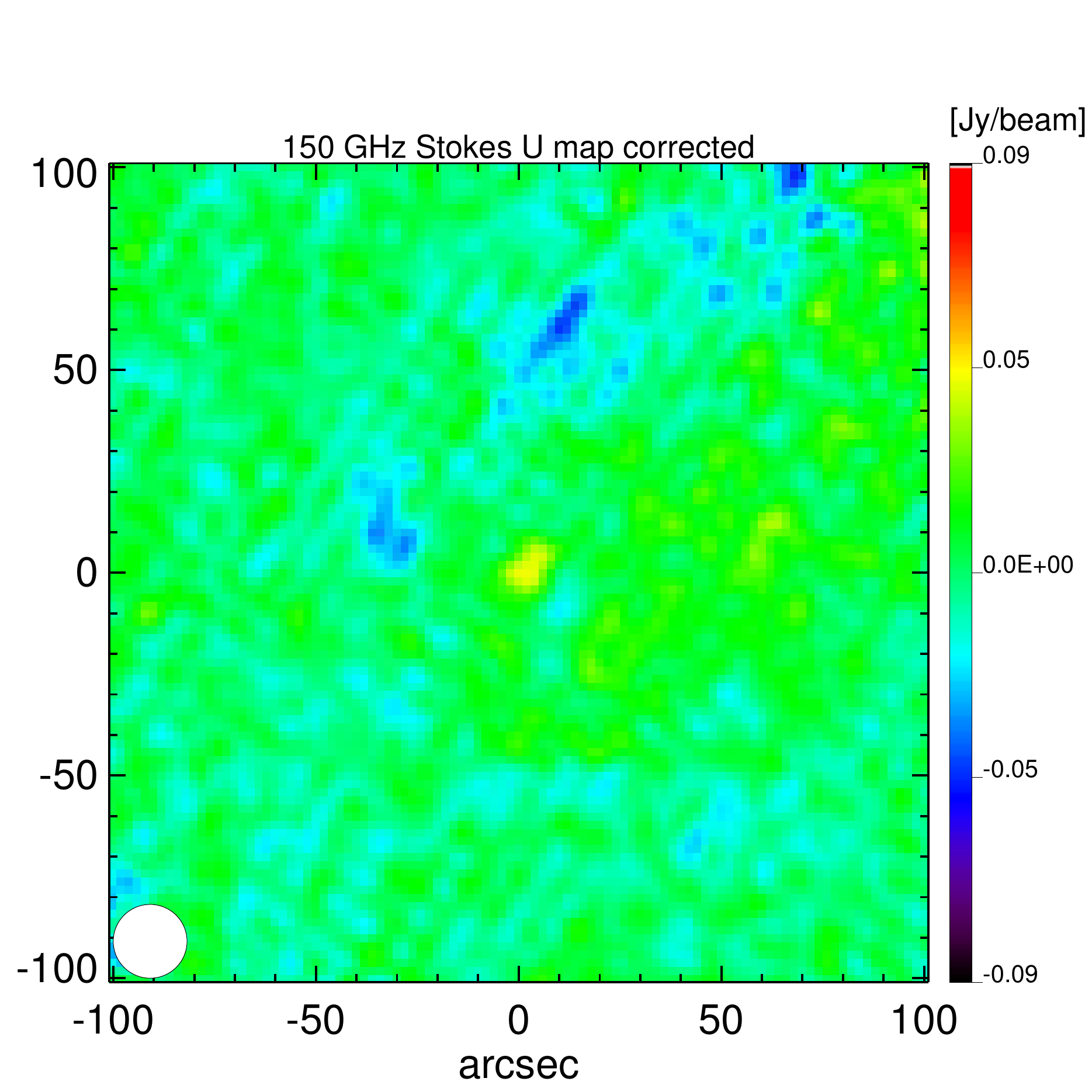}}

\end{picture}

  \caption{ Uranus Stokes $I$, $Q$ and $U$ maps in Nasmyth coordinates at 260
    GHz (a) and 150 GHz (b) before and after leakage correction. After the
    leakage correction, we are left with a residual instrumental polarization below 1\% (0.7 \% at 1.15 mm and 0.6 \% at 2.05 mm).}
   \label{fig:uranus_lkg}
  \end{center}
\end{figure*}

Like most authors, we define instrumental polarization as the ability of the
instrument to convert incident unpolarized total power into a polarized
signal. This conversion can then take various forms, as we describe in this
section. To estimate the level of instrumental polarization at the telescope we
repeatedly observed Uranus, which is assumed to be unpolarized, (\citealt{polka_apex} find a
polarization of $\sim 0.1 \%$), and bright (47.2~Jy at 1.15 mm, 16.4~Jy at
2.05 mm). It has an apparent diameter at the time of observations of 3~arcsec and
can therefore be approximated as a point source compared to our
beams. Fig.~\ref{fig:uranus_lkg} shows Stokes $I$, $Q$, and $U$ maps of Uranus in
Nasmyth ({\it i.e.},~cabin) coordinates at 1 and 2.05 mm in panels (a) and (b),
respectively.  For each panel, the top row shows the raw \nika\ maps after
projection of the decorrelated pure Stokes $I$, $Q$, and $U$ TOIs. We observe a
significant signal in $Q$ and $U$ that indicates a non-zero level of
instrumental polarization. We mainly identify a bipolar pattern, partially
consistent between the two bands, with a peak to peak amplitude at the level of
3\% of the total intensity peak. Such an effect has already been observed in
other experiments ({\it e.g.}~\citealt{thum2008} and
\citealt[][]{2015ApJ...806..206B}).  We performed a large number of
observations of Uranus at different elevation angles. From these observations we concluded that the observed leakage effect is fixed in Nasmyth
coordinates. Although we still lack a convincing physical interpretation of the
observed signal, we can model it as leakage from total intensity $I$ into $Q$
and $U$, and write the observed Stokes parameters in Nasmyth coordinates as
 \begin{eqnarray}
 \hat{I}_{N}  & = & B_{I} * I_{N} + N_{I}, \label{eq:i_leak}\\
 \hat{Q}_{N}  & = & B_{I} * Q_{N} + {\cal{L}}^{IQ}_{N} * I_{N} + N_{Q}, \label{eq:q_leak} \\
 \hat{U}_{N}  & = & B_{I} * U_{N} + {\cal{L}}^{IU}_{N} * I_{N} + N_{U}; \label{eq:u_leak}
 \end{eqnarray}
 where $I_{N}$, $Q_{N}$, and $U_{N}$ are the original sky Stokes parameters in
 Nasmyth coordinates. $B_{I}$ represents the \nika\ response pattern and $*$ denotes
 spatial convolution. The different noise contributions discussed above are
 accounted for in $N_{I,Q,U}$. Finally, we model the leakage term as the
 convolution of the original intensity map with response-pattern-like kernels
 ${\cal{L}}^{IQ}_{N}$ and ${\cal{L}}^{IU}_{N}$ for $Q$ and $U$,
 respectively. These two kernels are directly estimated from the $Q_{N}$ and
 $U_{N}$ maps of Uranus presented in Fig.~\ref{fig:uranus_lkg}, which, as
 discussed above, can be considered as a point source. Note that here we assume no modification of the intensity signal and we account for any loss of power at
 the calibration stage.

One way to correct for this instrumental polarization is to convolve the
observed polarization maps (Eqs.~\ref{eq:q_leak} and \ref{eq:u_leak}) by the main instrumental gaussian beam
$B_I$ and subtract the convolution of the observed intensity map $\hat{I}_N$
(Eq.~\ref{eq:i_leak}) by the leakage kernels. The results are polarization maps
that are free from instrumental polarization leakage, but with a degraded resolution at
17 and 25.5~arcsec at 1.15 and 2.05~mm respectively, and with an extra fraction
of the total intensity noise convolved by the leakage kernels. To avoid these
two artefacts, we have devised a dedicated algorithm:
 \begin{enumerate}
 \item With the demodulation and projection techniques presented in
   Sect.~\ref{data_analysis}, we build maps of Stokes $I$, $Q$, and $U$ of the
   observed signal in equatorial coordinates. These maps can be the result of
   multiple observation scans to obtain the best possible signal to
   noise. We only need the $I$ map in the following to derive the leakage signal that we want to subtract.
 \item We rotate the $I$ map into Nasmyth coordinates to obtain $\hat{I}_N$ for
   a given scan. The needed rotation angle, which is the combination of the
   elevation and the parallactic angles, varies along the scan. However, we find
   that not accounting for this variation during a given scan leads to negligible differences.
 \item Build Fourier space convolution/deconvolution kernels of the form
   ${\cal{L}}_{IQ}/B_I$ and ${\cal{L}}_{IU}/B_I$ from observations of
   Uranus.
 \item Multiply the Fourier transform of $I^N$ by the above kernels 
   and transform the result back into real space to build maps of leakage from $I$ into $Q$ and $U$.
 \item Deproject the obtained maps with the actual scanning strategy to produce $Q$ and $U$ TOIs that
  are then subtracted from the decorrelated pure Stokes $Q$ and $U$ TOIs presented in Sect.~\ref{se:demod_mapmaking}.
 \item Project these corrected TOIs onto final maps following the same map-making
   procedure as in Sect.~\ref{se:demod_mapmaking}.
 \end{enumerate}
 A similar technique developed to reduce the leakage effect observed with XPOL instrument from the IRAM 30m telescope is presented in \citet{2013A&A...558A..45H}.
 In Fig.~\ref{fig:uranus_lkg}, the bottom rows of panels (a) and (b) show the
 final Nasmyth-coordinate-Uranus-Stokes $I$, $Q$, and $U$ maps after leakage
 correction using the above algorithm. Note that to compute the leakage kernels
 we use a set of independent Uranus observations to cross check the efficiency of
 the procedure.  We observe that after leakage correction the residual leakage in
 the $Q$ and $U$ maps of Uranus drops below 1 \%. We now see a residual
 signal that we interpret as ``straightforward'' instrumental polarization, that is, an induced polarization directly proportional to $I$. This instrumental
 polarization is below 1\% for both $Q$ and $U$, and is removed
 by subtracting the relative fraction of the total intensity map from
 our polarization maps.

\begin{figure*}
   \begin{center}
   \includegraphics[%
       width=0.324\linewidth,keepaspectratio]{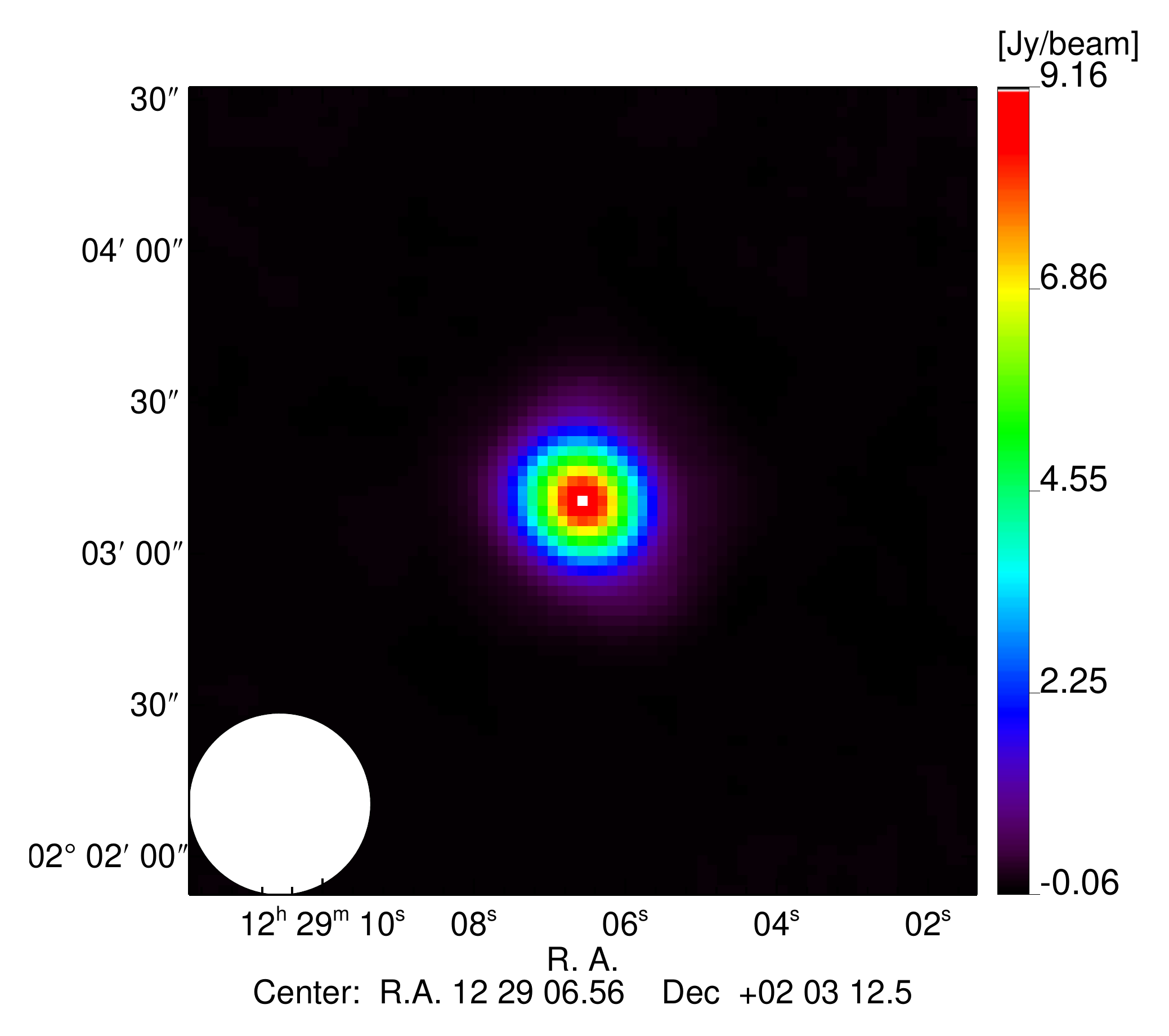}
     \includegraphics[%
       width=0.33\linewidth,keepaspectratio]{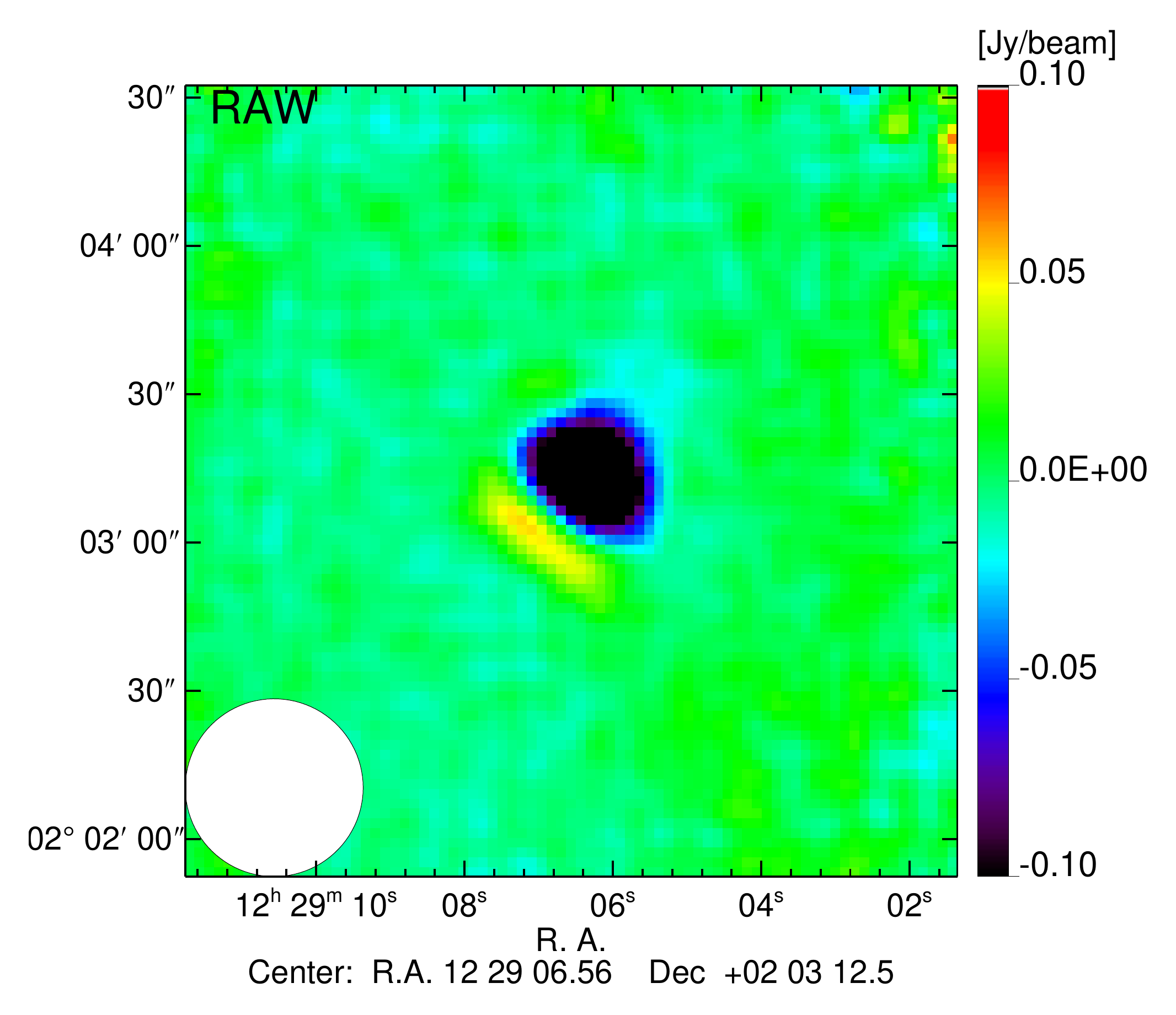}
         \includegraphics[%
       width=0.324\linewidth,keepaspectratio]{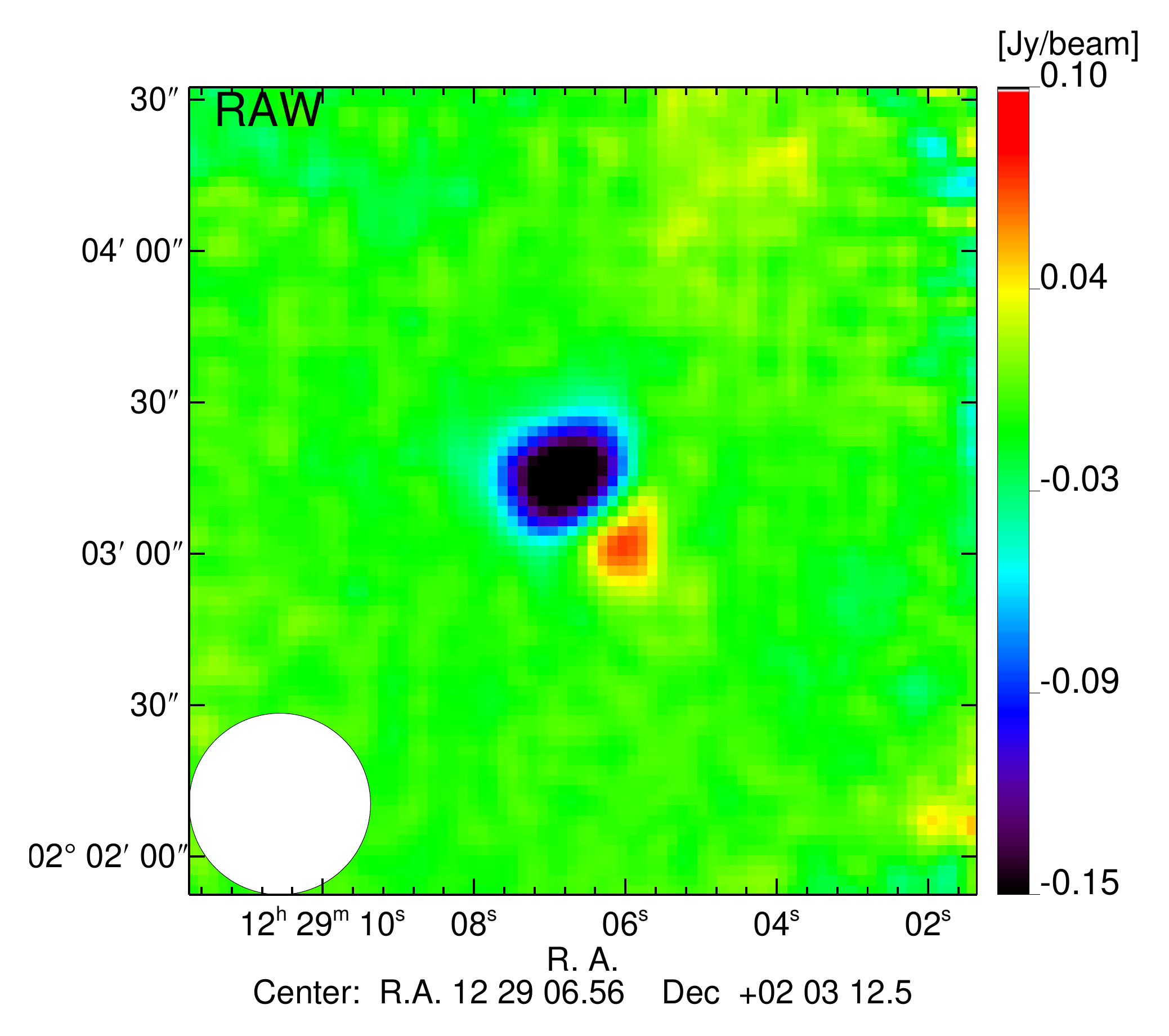}
        \includegraphics[%
       width=0.324\linewidth,keepaspectratio]{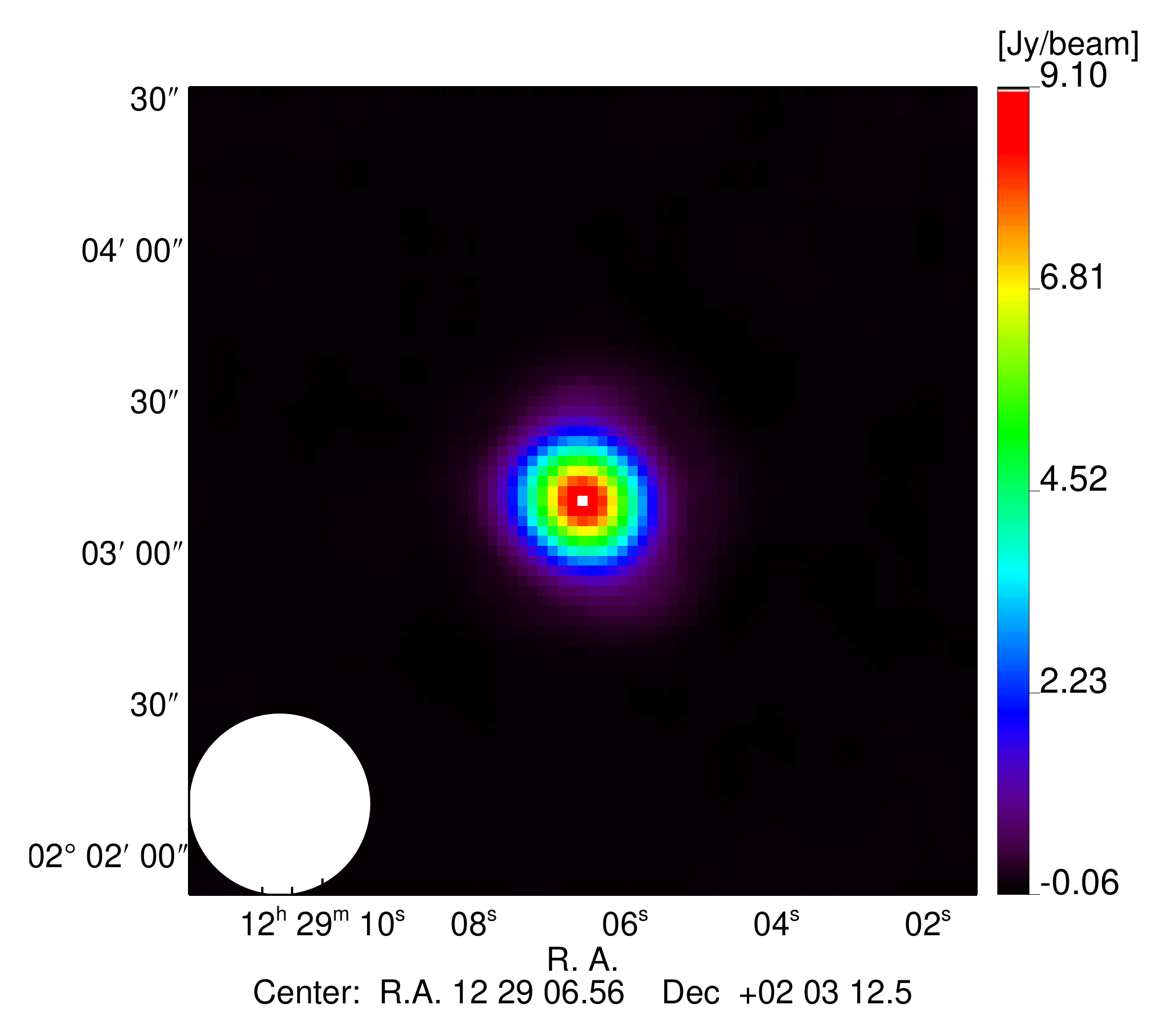}  
    \includegraphics[%
       width=0.33\linewidth,keepaspectratio]{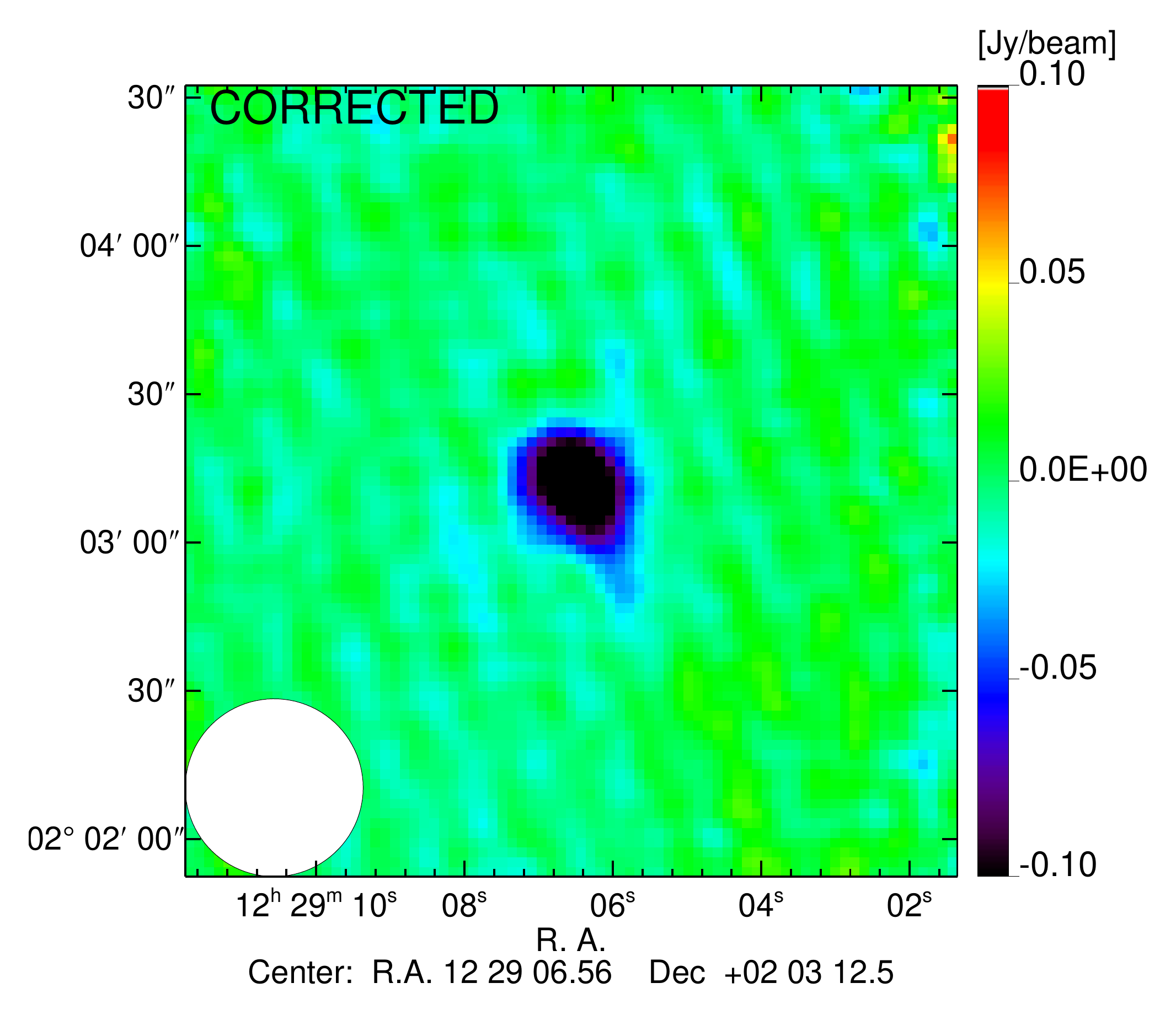}
     \includegraphics[%
       width=0.324\linewidth,keepaspectratio]{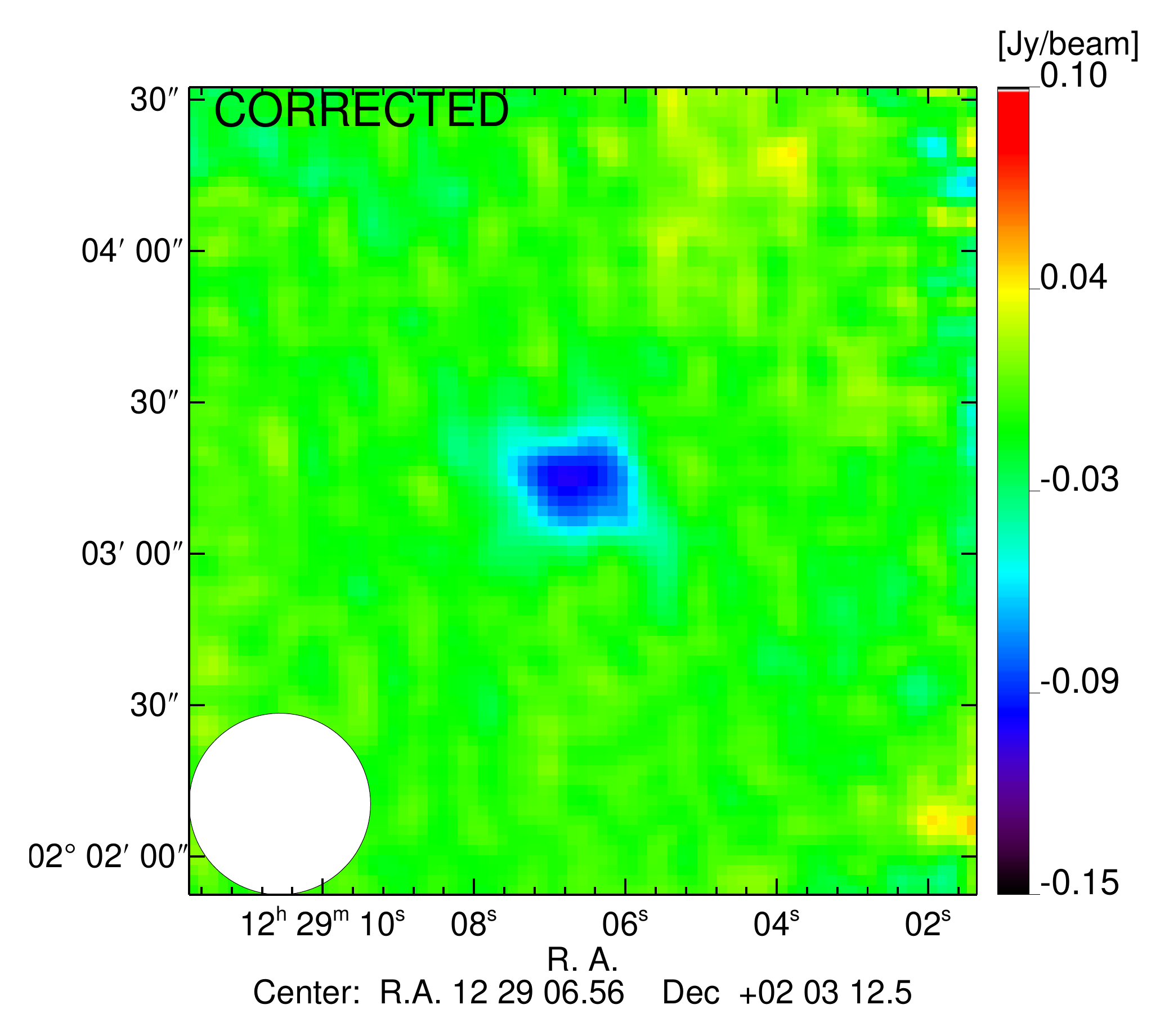}
       \caption{$I$, $Q$, and $U$ maps of the quasar 3C 273 observed at 150 GHz
       before (top) and after (bottom) correction for the leakage effect. The polarization angle and degree are reported in Tab. \ref{tab:table_quasar}.}
     \label{3c273_ex}
   \end{center}
 \end{figure*}

Applying this data-reduction algorithm to observations of sources with polarized
emission allows us to correct for leakage effects in this context as well. The top
row of Fig.~\ref{3c273_ex} shows the \nika\ Stokes $I$, $Q$, and $U$ maps at 2.05
mm of the quasar 3C273 before leakage correction. We clearly see in the $Q$ and
$U$ maps a bipolar structure similar to the one observed on the Uranus maps. The
bottom row of the figure presents the leakage-corrected maps, which show no
residual bipolar structure but slightly increased noise contribution. Indeed,
the division by $\tilde{B_I}$ boosts the signal on small angular scales and
therefore the noise, but the damping of ${\cal{L}}_{IQ}$ compensates and ensures
regularization. In the case of point-source observations, we can make this noise
increase even smaller by simulating a point source with the measured flux and
convolve it directly by ${\cal{L}}_{IQ}$ and ${\cal{L}}_{IU}$ to derive the
leakage corrections. The noise in total intensity is then not involved in the
process.

 \begin{figure} 
  \begin{center}
     \includegraphics[%
      width=1.3\linewidth,keepaspectratio]{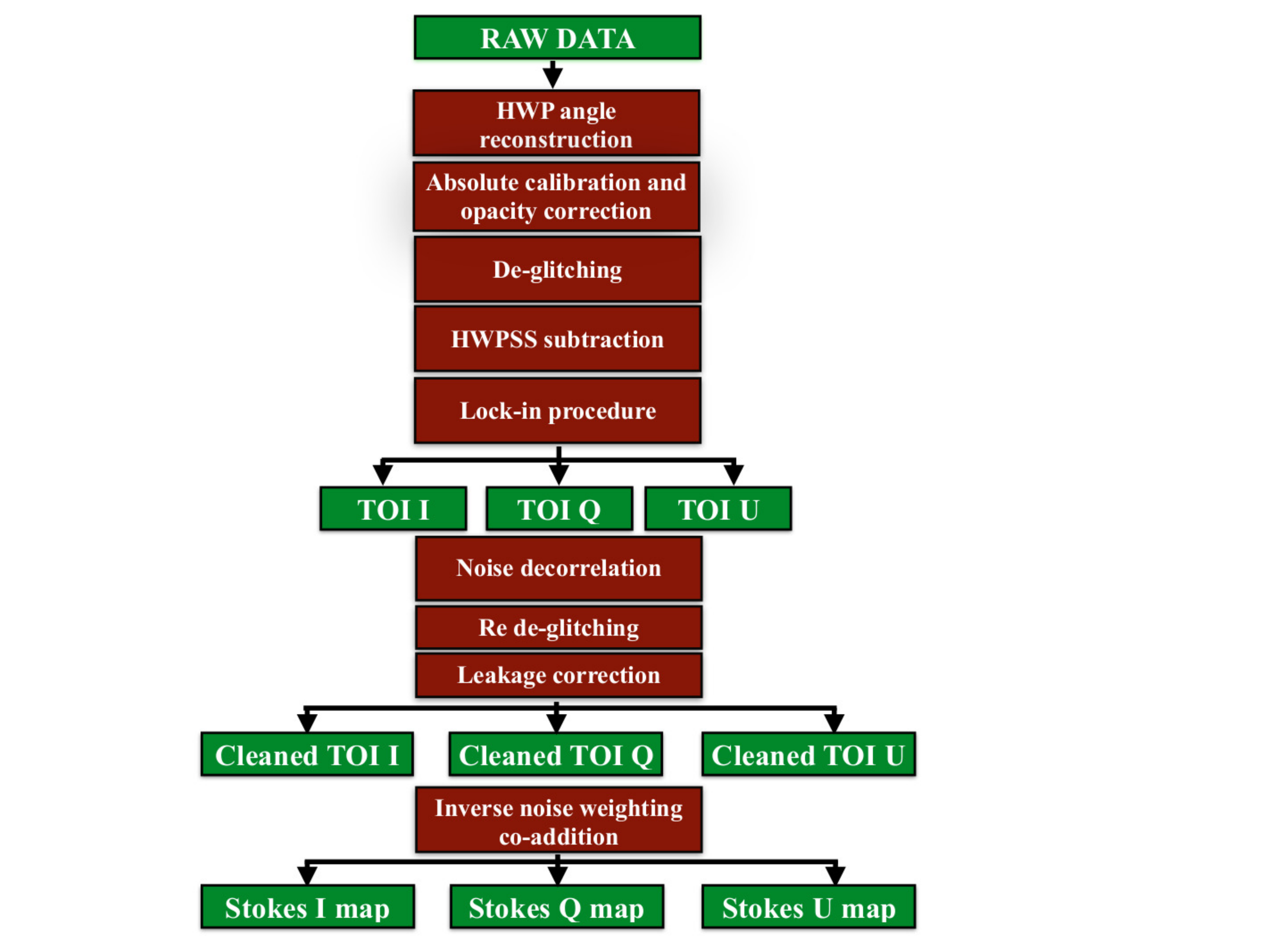}
\caption{Schematic view of the main procedures of the data processing pipeline from raw data to sky maps.}
\end{center}
\label{fig:scheme_pipe}

\end{figure}
 \subsection{Summary of data processing pipeline}\label{se:sumpipe}
 Fig.~\ref{fig:scheme_pipe} presents a schematic view of the main procedures used
 to convert the raw \nika\ data into leakage corrected Stokes $I$, $Q$, and $U$
 maps. The main steps are
 \begin{enumerate}
 \item Read raw data.
 \item Reconstruct the position of the HWP and the corresponding angle with respect to the zero reference.
 \item  Compute absolute calibration and correct for atmospheric absorption.
 \item  Apply a basic de-glitching algorithm to remove spikes on the raw \nika\ TOIs.
 \item  Reconstruct and subtract the HWPSS.
 \item  Apply lock-in procedure to the raw \nika\ TOIs to build pure Stokes $I$, $Q$, and $U$ TOIs.
 \item  Apply the decorrelation procedure to the pure Stokes $I$, $Q$, and $U$ TOIs.
 \item  Apply a basic de-glitching algorithm to remove spikes on the pure Stokes $I$, $Q$, and $U$ TOIs.
 \item  Apply the leakage-correction algorithm to the decorrelated pure Stokes $I$, $Q$, and $U$ TOIs.
 \item  Apply the map-making procedure to the decorrelated and leakage-corrected pure Stokes $I$, $Q$ and $U$ TOIs.
 \item Project the cleaned TOIs into maps.
 \end{enumerate}
 
 \section{Validation of the quality of the \nika\ polarization reconstruction on quasars} \label{polcalibration}
  \subsection{Quasar selection and previous observations}
 During the \nika\ February 2015 technical campaign we  observed a selection
 of quasars to validate the quality of the reconstruction of the polarization
 signal with the \nika\ camera. We have considered both bright and highly
 polarized quasars. As shown in Tab.~\ref{tab:table_quasar}, the selected
 targets were 3C279, 3C273, 3C286, and 0923+392. We list in the following the
 main physical characteristics and polarization properties of the most commonly
 observed quasars at mm wavelength: 3C286, 3C279 and 3C273. 

 \subsubsection*{3C 286}
 3C 286 is a compact steep-spectrum quasar at redshift z = 0.846.  The stability
 of this quasar in intensity and polarization in a large frequency range and the
 slow wavelength dependence make it a primary calibrator for polarization
 measurements. At centimeter wavelengths, where 3C 286 is commonly used as a
 primary polarization calibrator, observations show that the 3C286 polarization
 angle (PA) has been stable for decades \citep{perley&butler}. At millimeter
 wavelengths the XPOL polarimeter \citep{thum2008} has monitored 3C286 from
 2006 to 2012 \citep{xpol}. As presented in Tab.~\ref{tab:tab_quasar}, XPOL
 observations show that 3C 286 is highly polarized, up to 14\%, with PA
 increasing slowly with frequency. These results have been confirmed by
 observations at 1.3 mm with CARMA \citep{carma} in May 2015 (see
 Tab.~\ref{tab:tab_quasar}).
 
 \subsubsection*{3C 279}
 The blazar 3C 279 is one of the brightest and best monitored
 flat-spectrum quasars. It was the first object to exhibit apparent superluminal
 motion. The source of its strong radio to $\gamma$-ray emission is a
 relativistic jet of material ejected from the black hole in its centre
 \citep{apex3c279}.  3C279 is a variable source but strongly polarized up to 11
 \%.  In Tab.~\ref{tab:tab_quasar}, we present results in terms of degree of
 polarization and PA from recent observations of 3C279 by the SHARP polarimeter
 \citep{sharp3c279}. These observations were performed in March 2014 at 350
 $\mu$m. Simultaneous observations have been performed with the very long baseline
 interferometer (VLBI) \citep{2014AJ....147...77L} at 3.5, 7, and 13 mm.
 \subsubsection*{3C273}
 3C 273, the first quasar ever to be identified, is located in the constellation
 of Virgo at a redshift z = 0.158 \citep{3c273madsen}. 3C 273 is the brightest
 and hence one of the best monitored active galactic nuclei (AGN). From radio to
 millimeter wavelengths, flares from the relativistic jet dominate the variability
 of 3C 273 \citep{Abdo2010}. 3C 273 shows relatively low polarization, approximately 3-4
 \% (see Tab.~\ref{tab:tab_quasar}), at mm wavelengths.
  \subsection{Polarization reconstruction accuracy}
  Tab.~\ref{tab:table_quasar} presents the Stokes $I$, $Q$, and $U$
 fluxes measured by \nika\ at 1.15 and 2.05 mm for the selected quasars. The
 fluxes have been measured using a simple aperture photometry procedure. The
 reported uncertainties account for inhomogeneities as well as for correlated
 noise in the maps \citep[see][for details]{adam2016}. 
  
In observations of linear polarization, it is common to represent the polarized signal in terms of polarization
degree, $p$, and angle, $\psi$:
\begin{eqnarray}
 p    &=& \frac{\sqrt{Q^2 + U^2}}{I},\label{p_degree}\\
 \psi &=& \frac{1}{2}\arctan\frac{U}{Q}.\label{angle_polar}
 \end{eqnarray}
 These definitions are not linear in $I$, $Q$, and $U$ and their naive estimation
 is biased by the noise. A non-Gaussian behavior is expected for $p$ and $\psi$
 leading to both biases and wrong estimates of the uncertainties of $p$ and
 $\psi$.  In particular regions of low polarized signal-to-noise ratio, when $Q
 \simeq U \simeq 0$, the noise measured on $Q$ and $U$ maps will yield a non-zero
 degree of polarization estimation.  Ways to correct the bias in the estimation
 of $p$ have been proposed by \cite{1980A&A....91...97S} and more recently by
 \cite{1985A&A...142..100S} and \citep{montier} to whom we refer
 here.
Although we have implemented the full
   likelihood based estimators of $p$ and $\psi$ according to the ``1D-marginal
   distributions'' of \citet{montier}, in this first paper on test data, we
   focus on high-S/N results (especially on $I$) and are therefore in the limit
   where the estimator of the degree of polarization approaches:
   \begin{equation}
   \hat{p} \simeq \sqrt{Q^2 + U^2 - \sigma_q^2 - \sigma_u^2}/I.
   \end{equation}
   \\
and
   \begin{equation}\label{p_c_uncertainty}
   \sigma_{\hat{p}} = \frac{\sqrt{Q^2\sigma_Q^2 + U^2\sigma_U^2 + \hat{p}^4I^2\sigma_I^2}}{\hat{p}I^2}.
   \end{equation} 
As far an the angle estimator is concerned, in the case of high S/N, we
  also reach the limit where the classical estimator $\psi=1/2\arctan(U/Q)$ is
  valid, with its associated uncertainty given by
  \begin{equation}\label{angle_uncertainty}
    \sigma_{\psi} = \frac{\sqrt{Q^2\sigma_q^2 + U^2\sigma_u^2}}{2(\hat{p}I)^2}.
    \end{equation}
To express polarization angles, we use the IAU convention, which counts East from North in the equatorial coordinate system. The results obtained on the quasars observed are reported in Tab. \ref{tab:table_quasar}. Notice that the uncertainties in $Q$ and $U$ are generally comparable, as expected. 
Differences between the uncertainty values in $Q$ and $U$ of the quasar 3C279 can be explained by residual correlated pixel-to-pixel noise. In addition, the absolute calibration uncertainty estimated at a level of 14\% at 1.15 mm and 5\% at 2.05 mm has to be added.
These values come from the intensity flux dispersion measured on all scans of Uranus during an observational campaign. 
An additional uncertainty is linked to the HWP zero-position determination as discussed in Sect.~\ref{lab_characterization}. 
 \begin{figure*}[h!]
  \begin{center}
   \includegraphics[%
   width=0.48\linewidth,keepaspectratio]{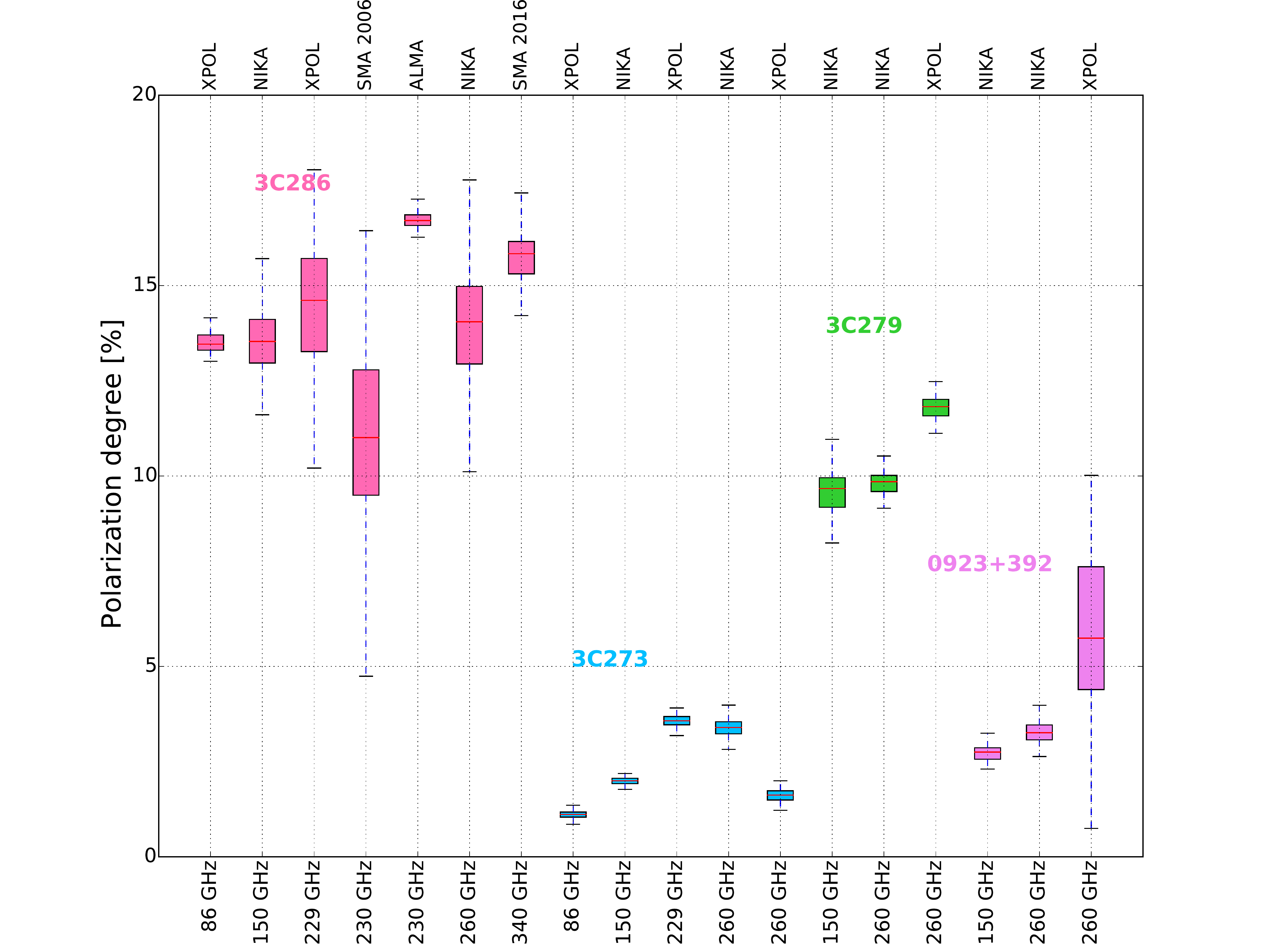}
   \includegraphics[%
   width=0.48\linewidth,keepaspectratio]{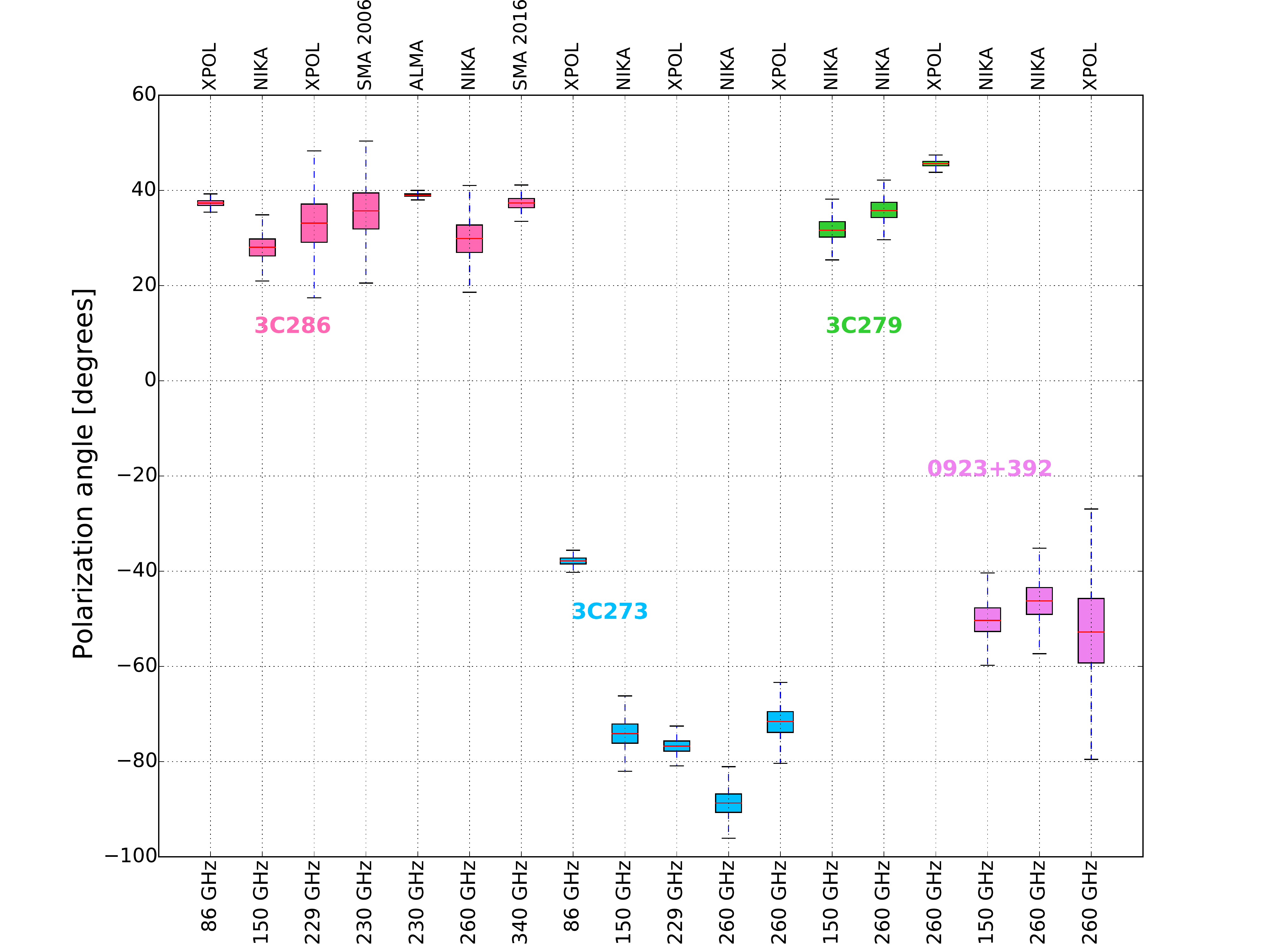}
   \caption{Results on polarization degree (left) and angle (right) obtained by \nika\ in comparison with other experiments. The values represented correspond
   to those reported in Tab.~\ref{tab:table_quasar} and \ref{tab:tab_quasar}. Dashed lines represent 2$\sigma$ error bars and the colored line in the box represents 1$\sigma$ error bars.}
   \label{pol_ang_deg}
  \end{center}
   \end{figure*}
   
 We begin by comparing the \nika\ results obtained in terms of degree and angle of
 polarization for the quasar 3C286, which is considered a polarization calibrator at millimeter
 wavelengths, to those of XPOL \citep{xpol}, SMA \citep{2006PhDT........32M}, CARMA \citep{carma}, SMA \citep{2016arXiv160806283H}, and ALMA \citep{2016ApJ...824..132N}
 experiments presented in Tab.~\ref{tab:tab_quasar} and Fig.~\ref{pol_ang_deg}. 
 The results found by \NIKA\ are consistent with those of XPOL within 1$\sigma$ error bar, whereas they are consistent within 2$\sigma$ error bars with the other experiments. Forthcoming observations with the \nikad\ instrument could improve the understanding of this discrepancy. The results obtained confirm that the quasar 3C286 is highly polarized.
 
  Similar results are found by comparing the \nika\ data for 3C279 with
 those of SHARP \citep{sharp3c279} at  (857, 85.7, 42.8, 23) GHz. Lastly, we
 had the opportunity to observe 0923+392 with XPOL in January~2016 during a test
 run of \nikad\ and found consistent results with our \nika\ observations of
 this quasar taken in February~2015. Our measurements of polarization degrees and
 orientation therefore agree with other experiments on significantly polarized
 sources down to $\sim 60$~mJy (polarized flux) and polarization degree as low
 as $\sim 3\%$. 
 In order to ease the comparison of the results obtained by \nika\ and other experiments we show them in Fig.~\ref{pol_ang_deg}.
    \begin{table*}
  \begin{center}
    \caption{\nika\ measured intensity and polarization fluxes, polarization
      degree and angle at 260 GHz and at 150 GHz for the quasars observed during
      the February 2015 campaign.}
    \begin{tabular}{ccccccccc}
      \hline
      \hline
      Source & Frequency & I flux & Q flux &  U flux  & p & $\psi$  \\ 
      & [GHz]         & [Jy] & [Jy] & [Jy] & [\%] & [$^\circ$] \\ 
      \hline
      \hline
      3C279 & 260 &  8.52$\pm$0.28 &  0.26$\pm$0.01 & 0.79$\pm$0.03 & 9.8 $\pm$0.4 & 35.9$\pm$0.5(stat)$\pm$1.8(syst)  \\
      & 150 &  12.21$\pm$0.58 &  0.51$\pm$0.02 &  1.04$\pm$0.05 &  9.5$\pm$0.6 & 31.9$\pm$0.7(stat)$\pm$1.8(syst) \\
      \hline
      3C273 & 260 &  6.35$\pm$0.22 & -0.22$\pm$0.01 & -0.01$\pm$0.01 &  3.4$\pm$0.3 & -88.7$\pm$0.9(stat)$\pm$1.8(syst) \\
      & 150 & 9.95$\pm$0.48 & -0.17$\pm$0.01 & -0.11$\pm$0.01 &  2.0$\pm$0.1 & -74.1$\pm$1.1(stat)$\pm$ 1.8(syst) \\
      \hline
      3C286 & 260 &  0.27$\pm$0.01 & 0.021$\pm$0.003 & 0.033$\pm$0.004 &  14.3$\pm$1.7 & 30.0$\pm$2.5(stat)$\pm$1.8(syst) \\
      & 150 & 0.51$\pm$0.03 & 0.039$\pm$0.002 & 0.056$\pm$0.002 & 13.6$\pm$0.8 &  28.0$\pm$0.9(stat)$\pm$1.8(syst) \\
      \hline
      0923+392 & 260 & 2.04$\pm$0.06 & -0.002 $\pm$0.005 & -0.066$\pm$0.005 & 3.2$\pm$0.3 & -46.1$\pm$2.4(stat)$\pm$1.8(syst) \\
      & 150 & 3.24$\pm$0.14 & -0.016$\pm$0.005 &  -0.087$\pm$0.006 & 2.7$\pm$0.2 & -50.4$\pm$1.7(stat)$\pm$1.8(syst)  \\
      \hline
      \hline
    \end{tabular}
    \label{tab:table_quasar}
  \end{center}
\end{table*}

\begin{table*}
  \begin{center}
    \caption{Degree and angle of polarization of the \nika\ observed quasars as measured by other experiments.}
    \begin{tabular}{ccccccc}
      \hline
      \hline
      Source & Experiment & Frequency & p       & $\psi$     & Observation date & Comments\\
             &            &            [GHz] & $[ \%]$ & [$^\circ$] &
      & \\
      \hline
      \hline
      3C279 &  SHARP and VLBI &  (857, 85.7, 42.8, 23) &  10 $\%$-12 $\%$ & 32-41 &  2014, March & \citep{2015ApJ...808L..26L} \\
      &  XPOL & 260 & 11.79  $\pm$ 0.29  & 45.6 $\pm$ 0.7 & 2016, January & \nika\ - XPOL joint session \\	
      \hline
           3C273 &   XPOL &    86    &   1.1$\pm$0.0 & -37.8$\pm$0.9 & 2015, February & \nika\ - XPOL joint session\\
           & XPOL &   229 &  3.6 $\pm$0.2 & -76.8$\pm$1.6 & 2015, February & \nika\ - XPOL joint session \\
            & XPOL &  260 & 1.59  $\pm$ 0.18  & -71.8 $\pm$ 3.1 & 2016, January & \nika\ - XPOL joint session \\
             \hline
      3C286 & XPOL & 86 & 13.5 $\pm$0.3 & 37.3 $\pm$0.8 & 2006-2012 &\citep{xpol}\\
      	& XPOL & 229 & 14.4 $\pm$ 1.8 & 33.1 $\pm$5.7 & 2006-2012 & \citep{xpol}\\
            & CARMA & 230 & & 39.1$\pm$ 1 & 2015, May & \citep{carma}\\
            & SMA & 230 & 11.5 $\pm$ 2.5 & 35.6 $\pm$ 5.9 & 2006  & \citep{2006PhDT........32M}\\
            & SMA &  340 &  15.7 $\pm$ 0.8 & 37.4 $\pm$ 1.5 & 2016 &\citep{2016arXiv160806283H}\\
            & ALMA & 230 & 16.7 $\pm$ 0.2 & 39 $\pm$ 0.4 & 2015, July & \citep{2016ApJ...824..132N}\\
      \hline
    0923+392 &  XPOL & 260 & 6.1  $\pm$ 2.3  & -52.59 $\pm$ 10.97 & 2016, January & \nikad\ - XPOL joint session \\
      \hline
      \hline
    \end{tabular}
    \label{tab:tab_quasar}
  \end{center}
\end{table*}

\subsection{Noise equivalent flux density (NEFD) in polarization observations}
The NEFD gives an estimation of the sensitivity of the instrument per frequency
band. It represents the uncertainty on the measure of the flux of point source
in one second of integration. We must estimate the \nika\ NEFD in both total
intensity and polarization, and make sure that it is consistent with \nika's
sensitivity when used in total power mode only, as reported in
\citet{catalano2014}, up to a factor two due to the presence of the analyzer after
the HWP that rejects half the incident photons. During our observation run, the
best observations we could use for this noise monitoring were 1h40m of
integration on 3C286. However, this is still a limited amount of time that
leaves atmospheric residuals on our total intensity maps that prevents us from
using them to derive NEFDs directly. However, with our HWP modulation,
polarization is not affected by low-frequency atmospheric or electronic
noise. The measured NEFDs in $Q$ and $U$ (equal to each other) are therefore
more reliable. We find 120 mJy.s$^{1/2}$ at 260 GHz and 50 mJy.s$^{1/2}$ at 150
GHz. Trusting these values, we can derive the expected NEFDs in $I$ that must be
a factor $\sqrt{2}$ lower, that is to say 85 mJy.s$^{1/2}$ at 260 GHz and 35.4
mJy.s$^{1/2}$ at 150 GHz. Now accounting for the expected factor two on absolute
calibration (the \nika\ primary calibrator is Uranus, which is
  unpolarized), due to the analyzer as mentioned before, we end up with
85/2=42.5 and 35.4/2=17.7 mJy.s$^{1/2}$ at 260 and 150 GHz, respectively, in
very good agreement with the measured values of 48 and 23 on \nika\ in total
power as reported in \cite{catalano2014}.
\noindent
Tab.~\ref{tab:nika_performance} reports the summary of the \nika\ polarimeter performance. 
\begin{table}[t!] 
  \begin{center}\footnotesize
    \caption{Performance of the \nika\ polarimeter.}
    \begin{tabular}{lcccccccc}
      \hline
      \hline
      Array & 1.15 mm & 2.05 mm \\
      \hline
      \hline
      Valid pixels &132 & 224 \\
      Field of View (arcmin) & 1.8 & 1.8 \\
      Band-pass (GHz) & 190 - 310 & 110 -180\\
      FWHM (arcsec) & 12 & 18.2 \\
      Polarization capability & yes & yes \\
      Sensit. on polarization ($Q$)    (mJy.s$^{1/2}$) & 120 & 50 \\
      Sensit. on $I$ in pol. mode (mJy.s$^{1/2}$) & 85 & 35 \\
      Sensit. on $I$ in tot. power mode (mJy.s$^{1/2}$) & 42.5 & 17.7 \\
      Instrumental polarization residual & 0.7 \% &  0.6 \% \\
      Syst. uncertainty on pol. angle & 1.8$^{\circ}$ & 1.8$^{\circ}$ \\
 \hline
      \hline
    \end{tabular}
    \label{tab:nika_performance}
  \end{center}
\end{table}
 \subsection{Photometric accuracy}
\begin{figure*}[h!]
  \begin{center}
   \includegraphics[%
   width=0.45\linewidth,keepaspectratio]{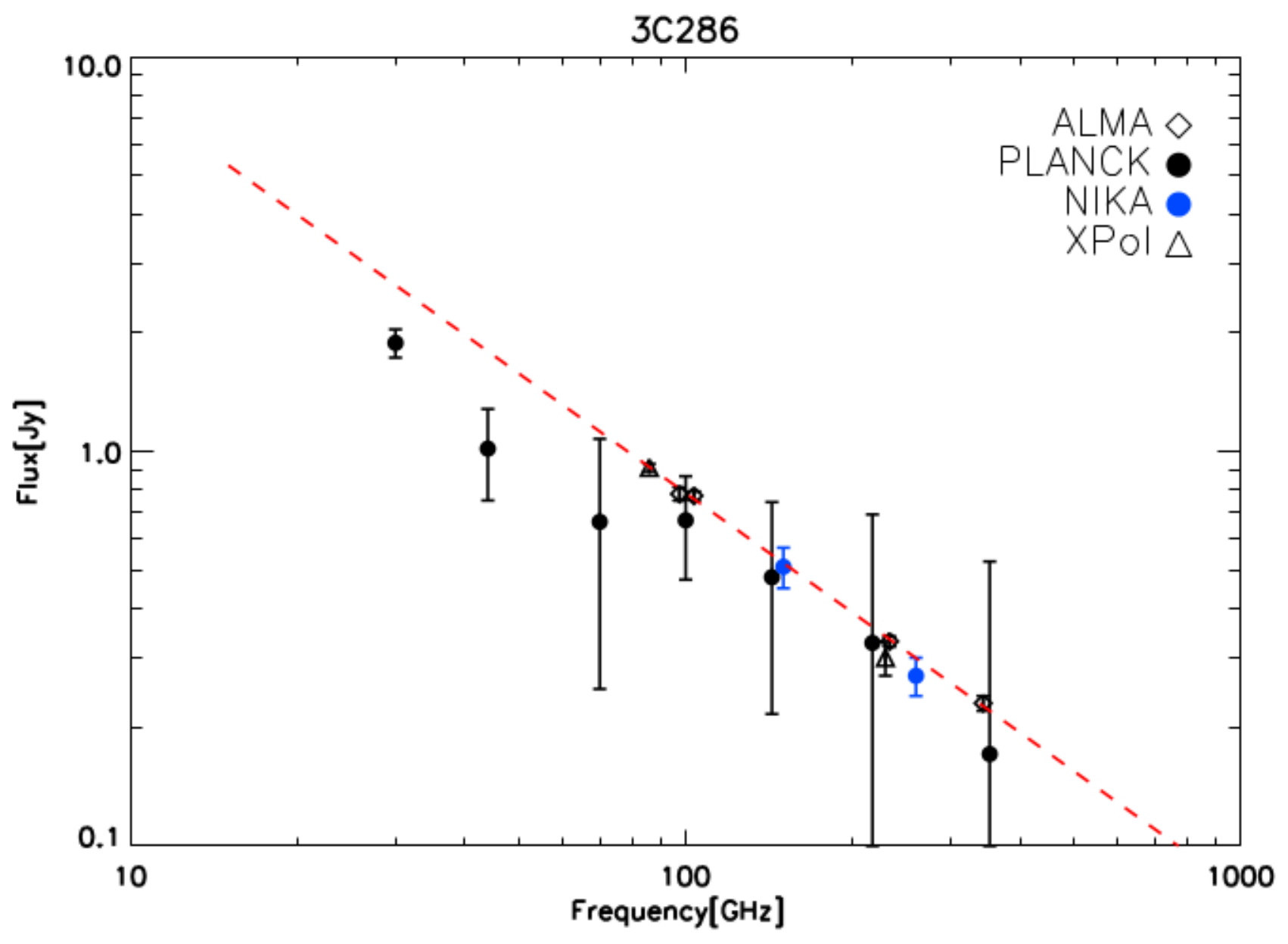}
   \includegraphics[%
   width=0.45\linewidth,keepaspectratio]{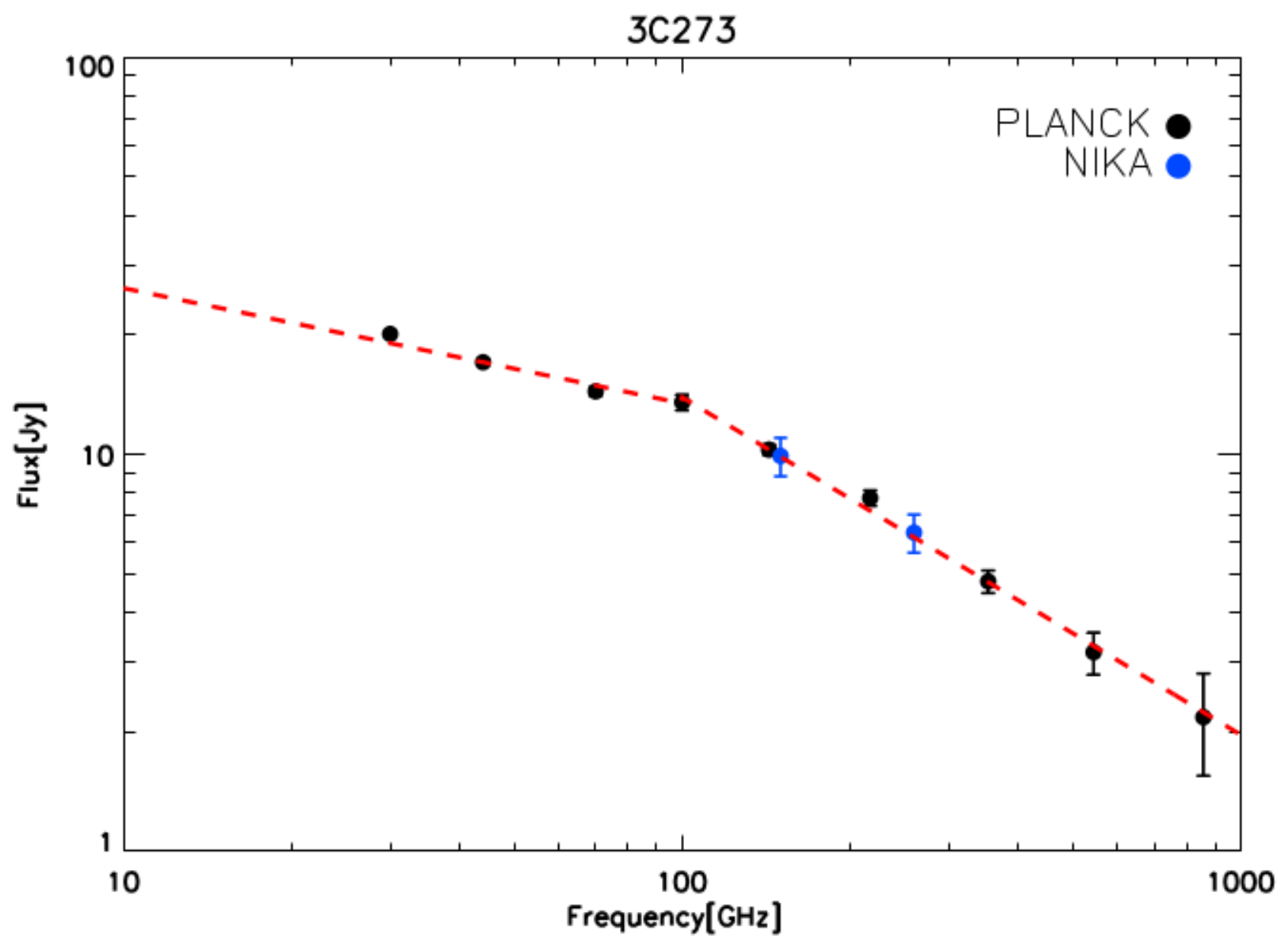}
   \caption{SED for the quasars 3C286 (left) and
     3C273 (right). We consider data from {\it Planck} \citep[black dot,
     ][]{planckcatalogue}; ALMA \citep[black diamond,][]{almacalib}; XPOl
     \citep[black triangle,][]{xpol} and \nika\ (blue dot, this
     paper). For the fit, a color correction of 5\% and 6\% at 260 GHz and 150 GHz, respectively, has been considered for both quasars. \label{sed}}
  \end{center}
   \end{figure*}

 The measurement of relatively stable quasars, such as 3C286 and 3C273, allows us to also
 to cross check the quality of the \nika\ photometry in
 intensity. Fig.~\ref{sed} presents the spectral energy density (SED)
 as a function of frequency in GHz for 3C286 (left) and 3C273 (right). The
 \nika\ intensity flux and uncertainties at 1.15 and 2.05 mm are represented in
 blue.  Results from other experiments including XPOL \citep{thum2008},
 {\it Planck} \citep{planckcatalogue}, and ALMA \citep{almacalib} are presented in
 black. We observe that the \nika\ data are consistent within error bars with
 other experimental results. We find that 3C286 data are consistent with a
 synchrotron spectrum in the form of a power law, $\propto$ $\nu^{\rm \beta}$,
 with spectral index $\beta$ $\simeq$ -1.007 $\pm$ 0.033 (dashed red line).  To
 explain the 3C273 data, we considered two power laws with spectral
   indices $\beta_1$ $\simeq$ -0.29$\pm$0.05 and $\beta_2$ $\simeq$
 -0.85$\pm$0.06 (dashed red line) and knee frequency of 100 GHz.
  
  \begin{figure*}
   \begin{center}
   \includegraphics[%
   width=0.324\linewidth,keepaspectratio]{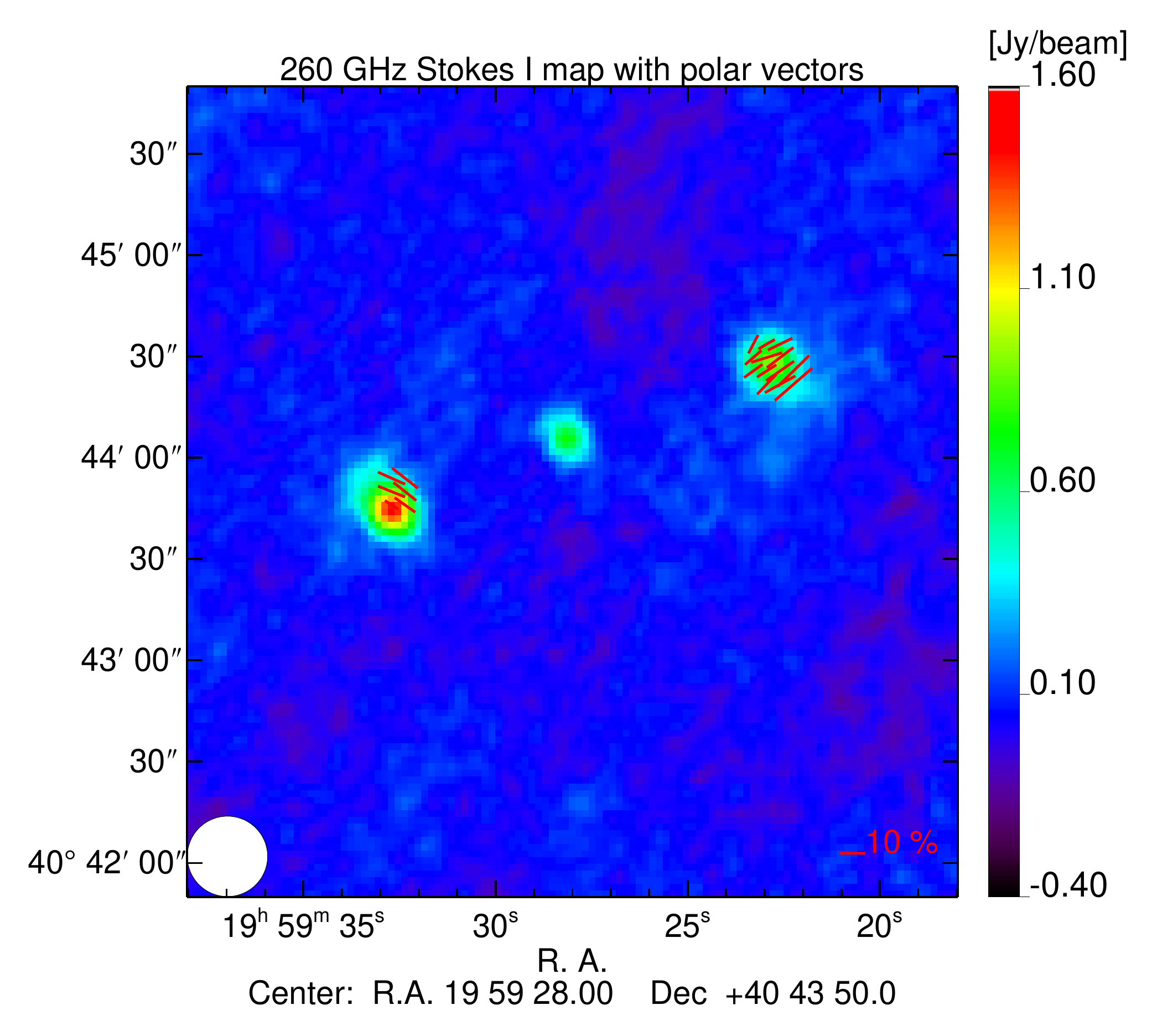}
    \includegraphics[%
   width=0.33\linewidth,keepaspectratio]{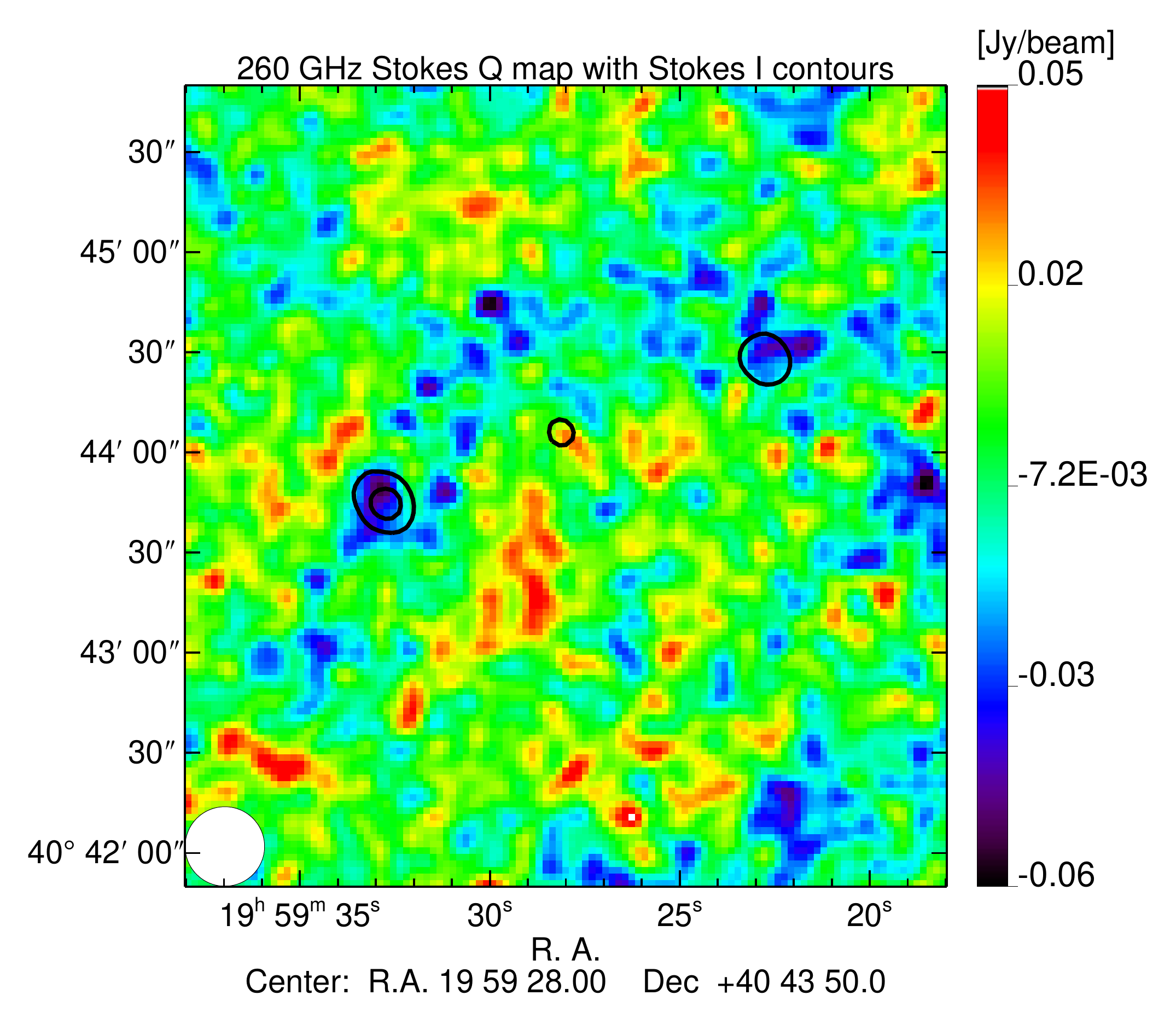}
    \includegraphics[%
   width=0.33\linewidth,keepaspectratio]{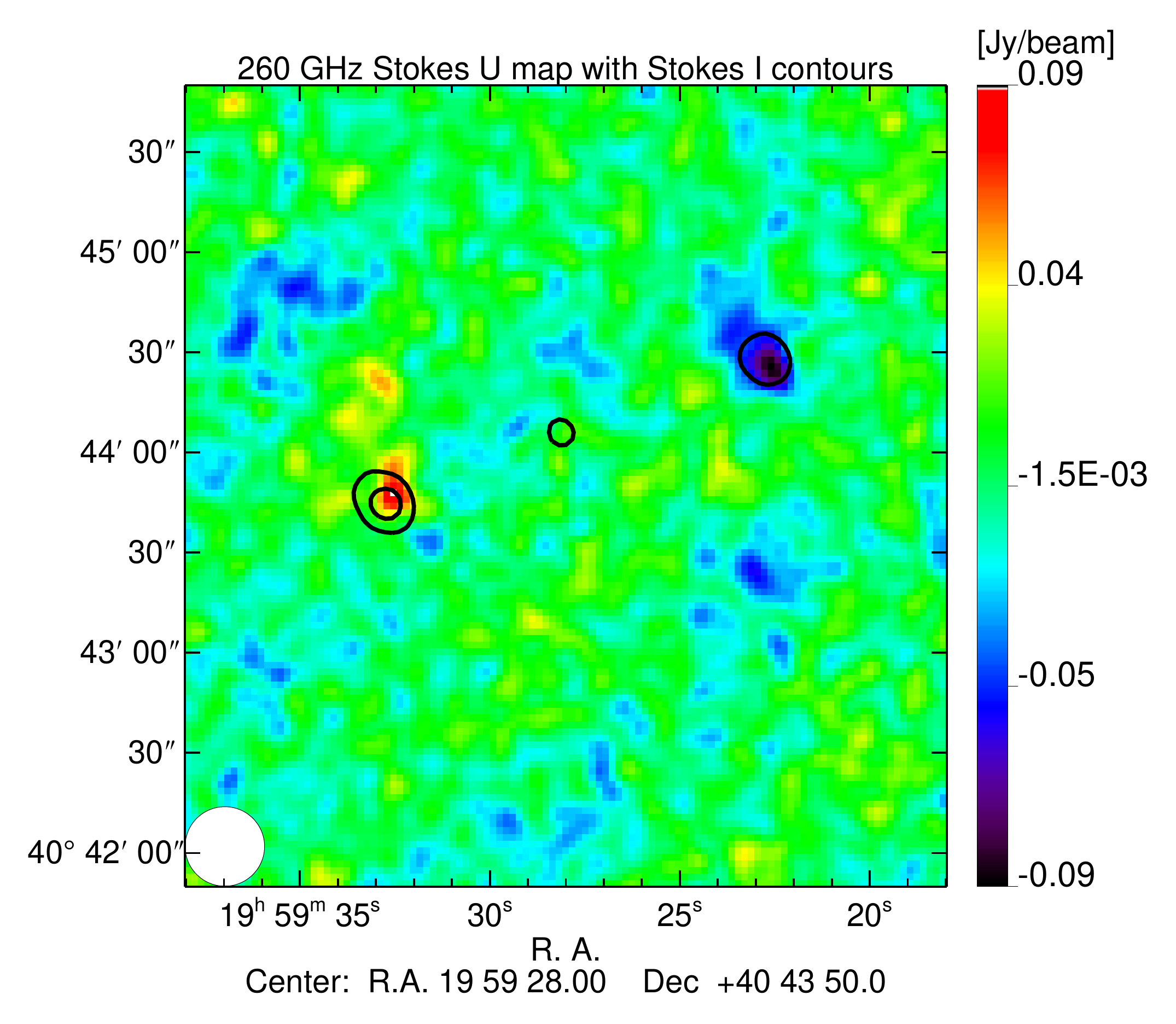}
    \includegraphics[%
   width=0.324\linewidth,keepaspectratio]{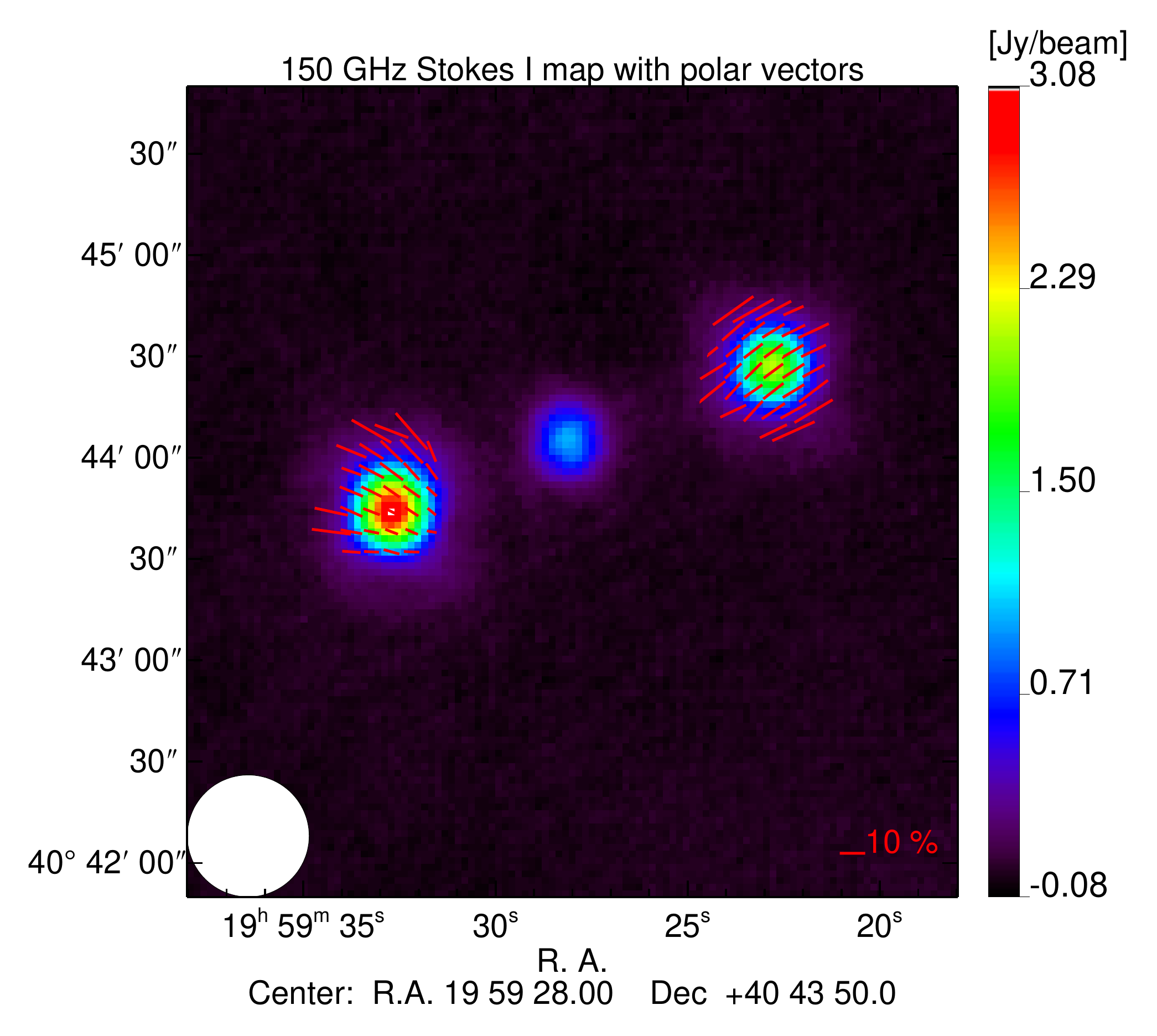}
    \includegraphics[%
   width=0.324\linewidth,keepaspectratio]{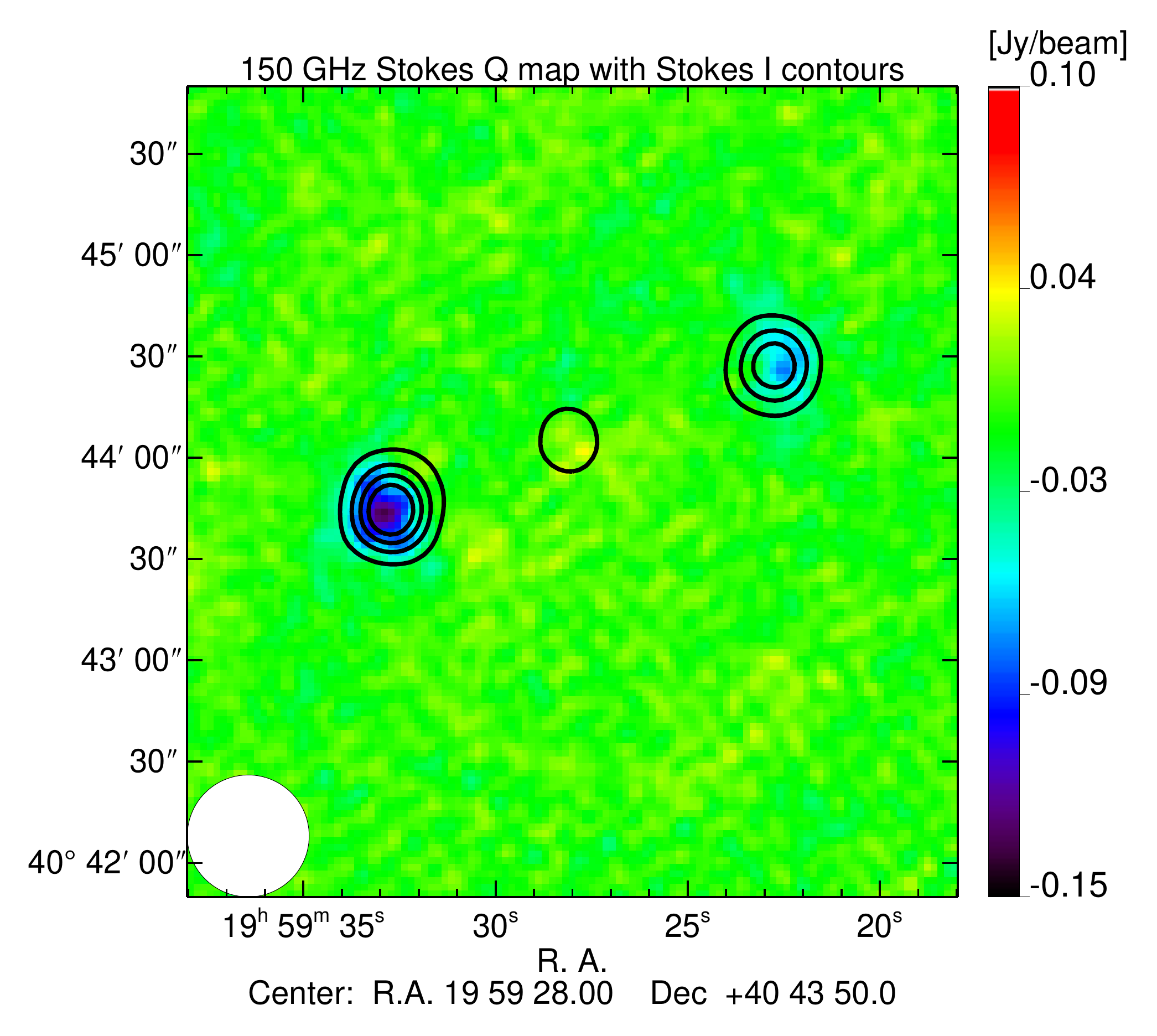}
    \includegraphics[%
   width=0.33\linewidth,keepaspectratio]{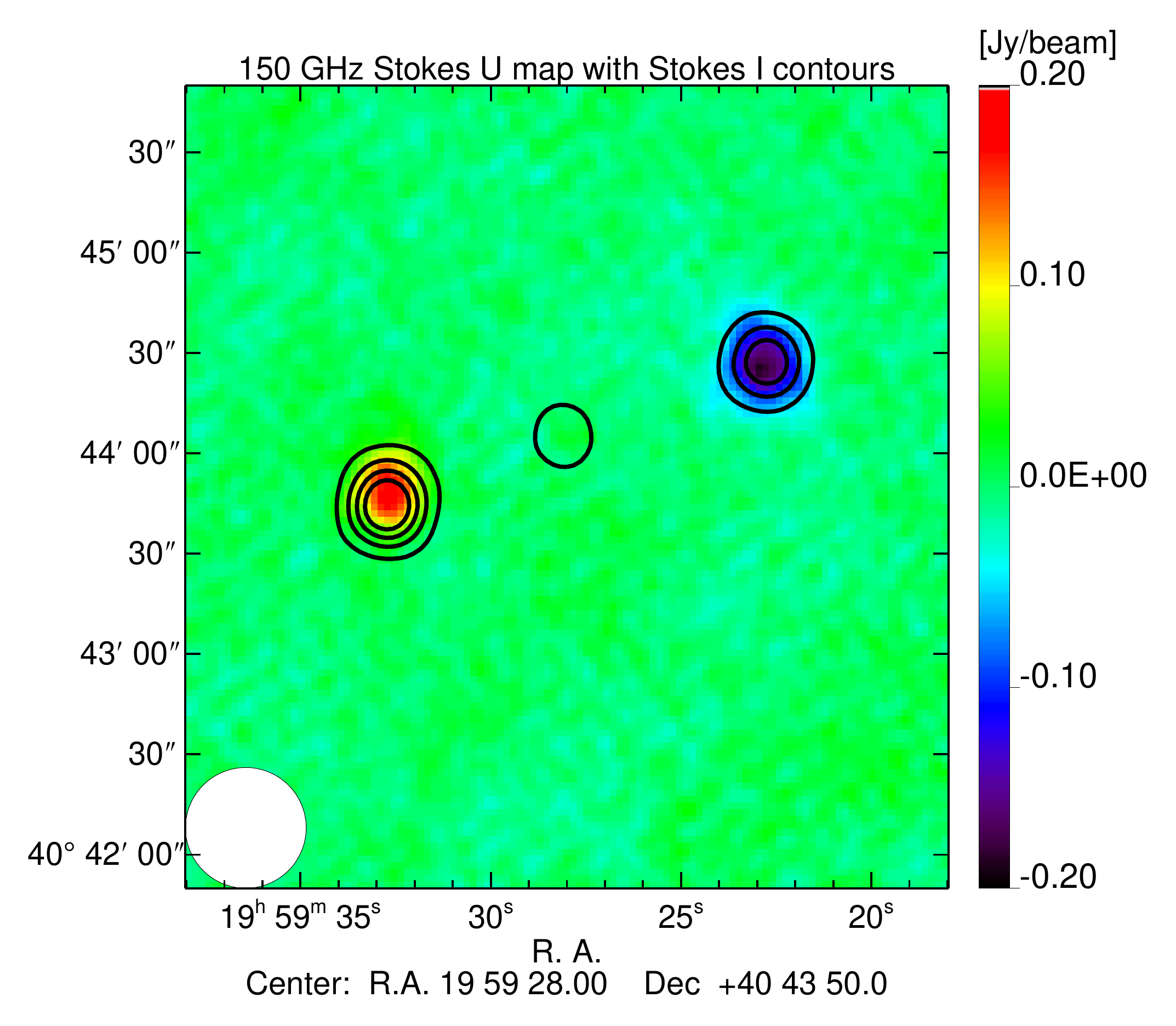}
    \caption{Cygnus~A Stokes  $I$, $Q$, and $U$ maps  at 260 GHz
      (top) and 150 GHz (bottom). The $Q$ and $U$ maps are smoothed with a Gaussian filter of 6 arcseconds. At 260 GHz the $I$ map is also smoothed 
      to 4 arcseconds for display purposes. The effective beam FWHM is shown as a white circle
      in the bottom left corner of each panel. The contours in $Q$ and $U$ represent the intensity map for
      each frequency. They start from 0.5 Jy/beam with steps of 0.5
      Jy/beam. Polarization vectors are plotted in red in the intensity image when $I$ $\textgreater$ 0 and $P$ $\textgreater$ $2\sigma_P$.}
     \label{cygnusa_maps}
   \end{center}
   \end{figure*}

 \begin{figure*}
   \begin{center}
   \includegraphics[%
   width=0.33\linewidth,keepaspectratio]{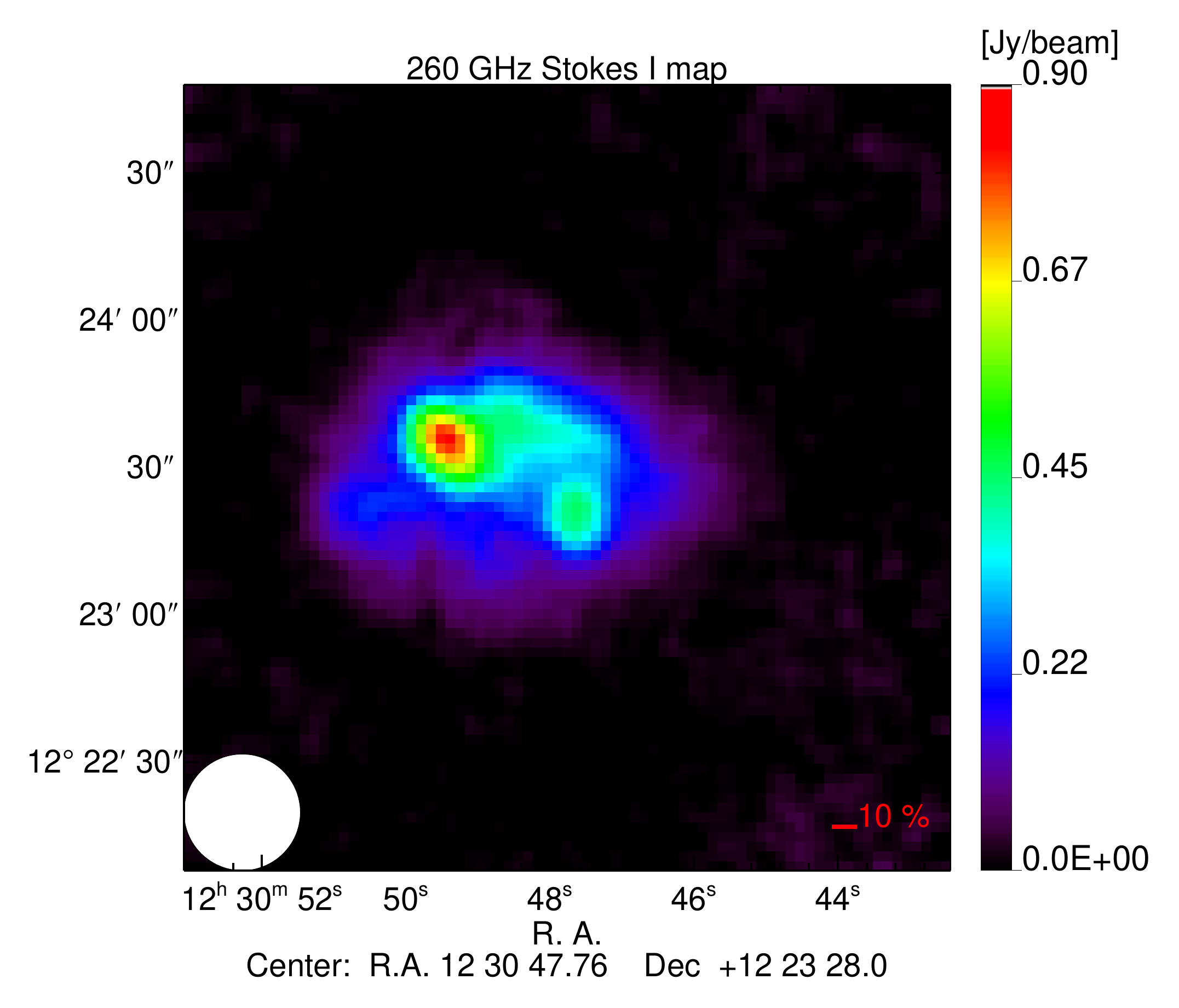}
    \includegraphics[%
   width=0.33\linewidth,keepaspectratio]{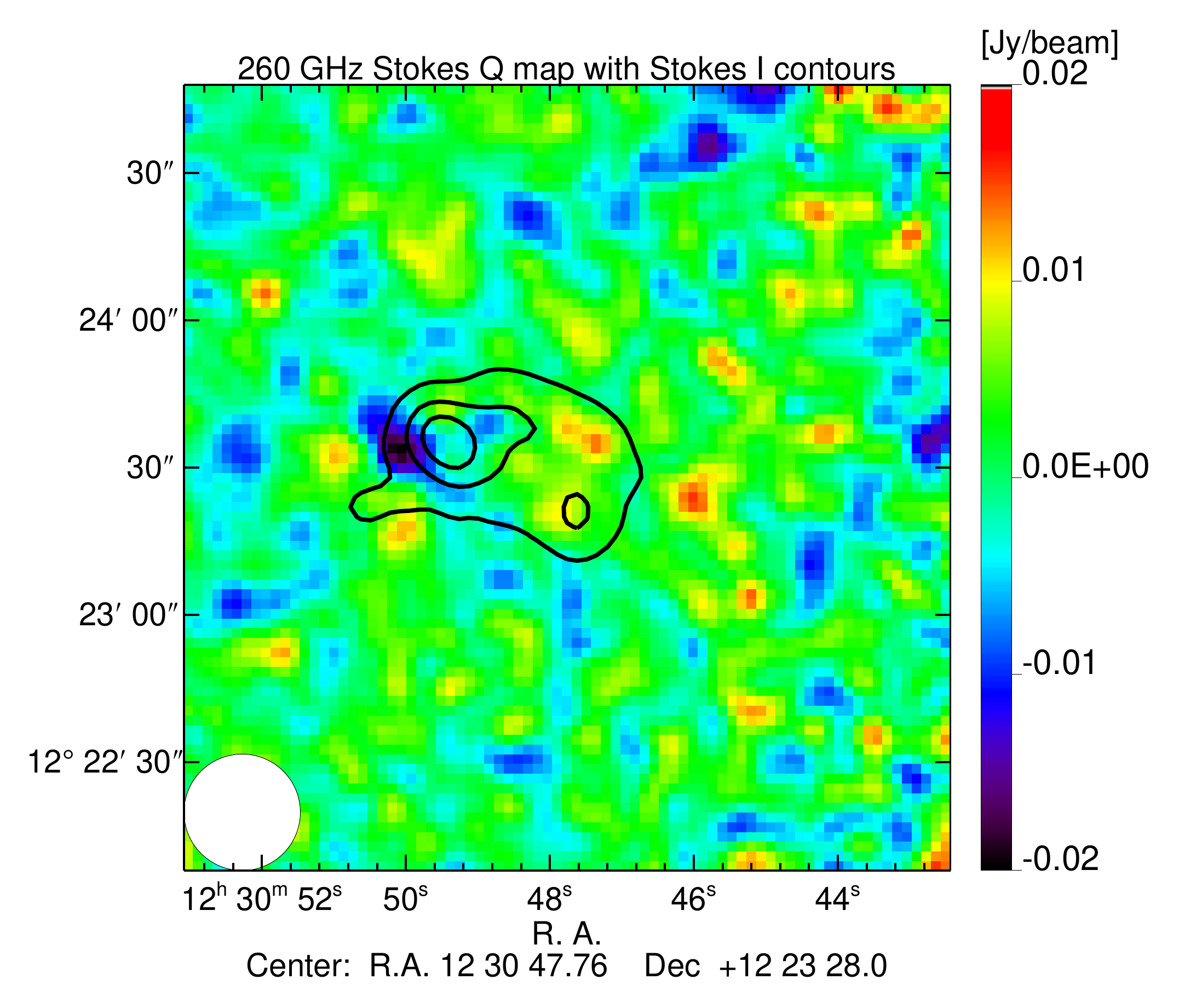}
    \includegraphics[%
   width=0.33\linewidth,keepaspectratio]{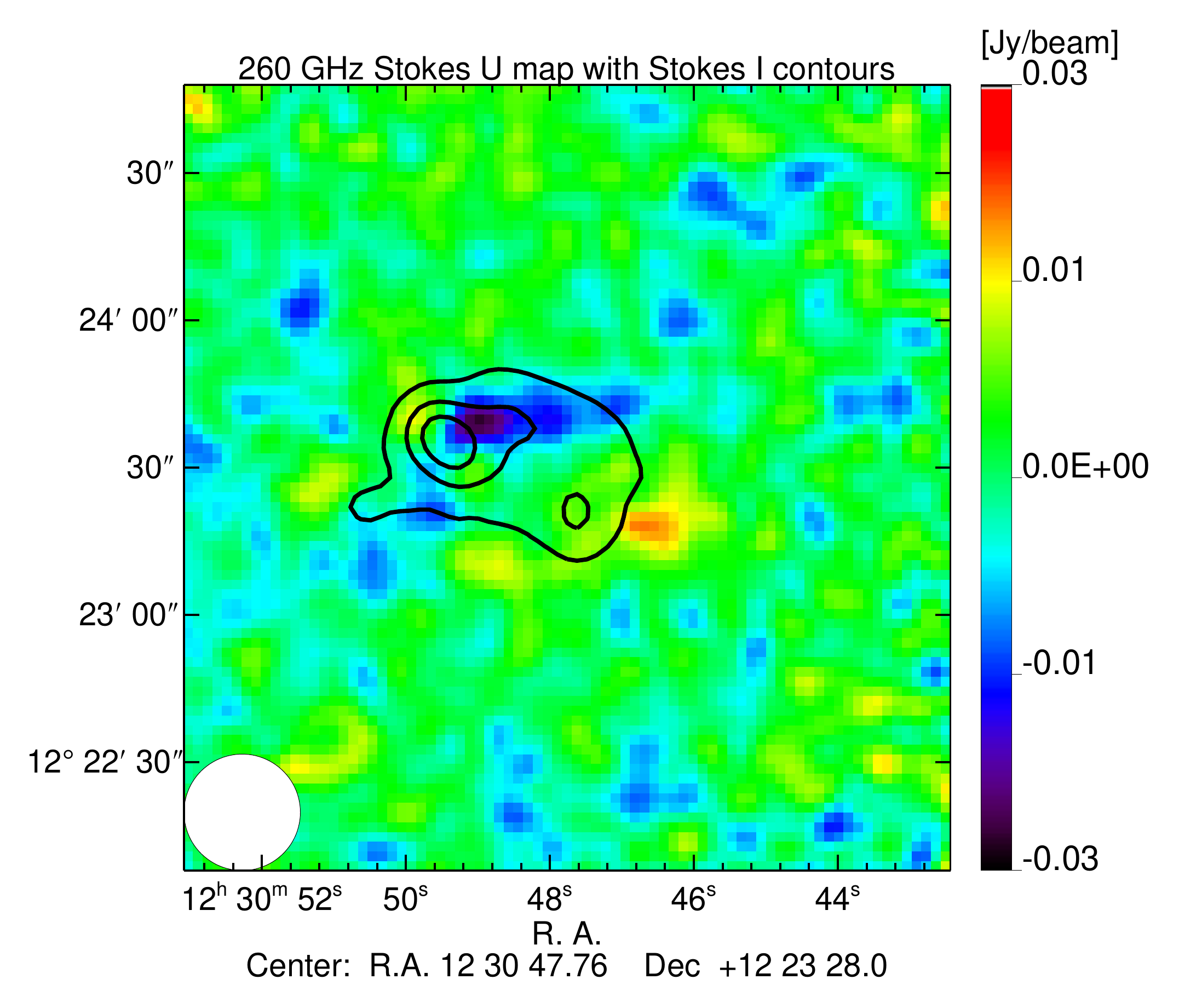}
    \includegraphics[%
   width=0.33\linewidth,keepaspectratio]{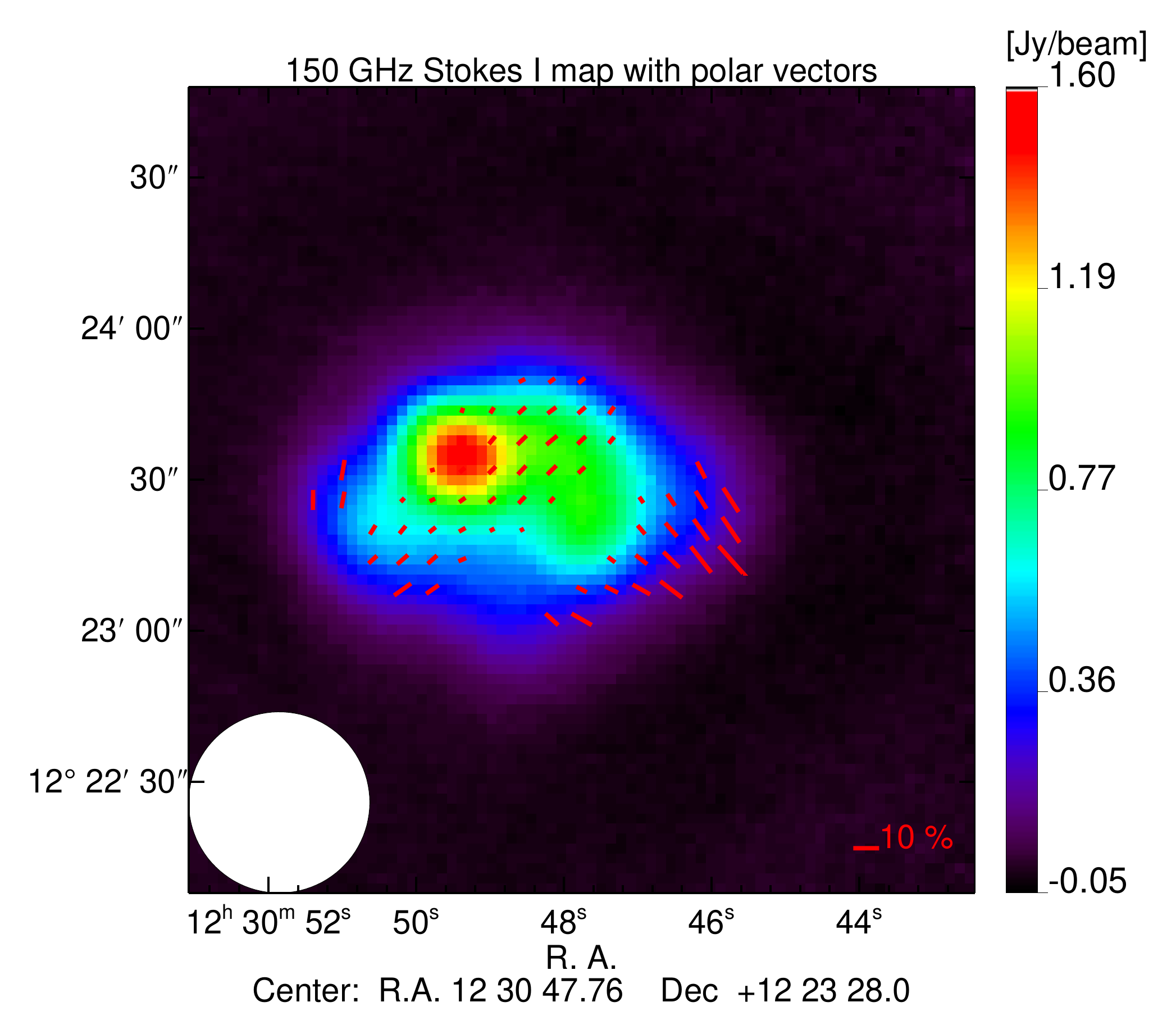}
    \includegraphics[%
   width=0.33\linewidth,keepaspectratio]{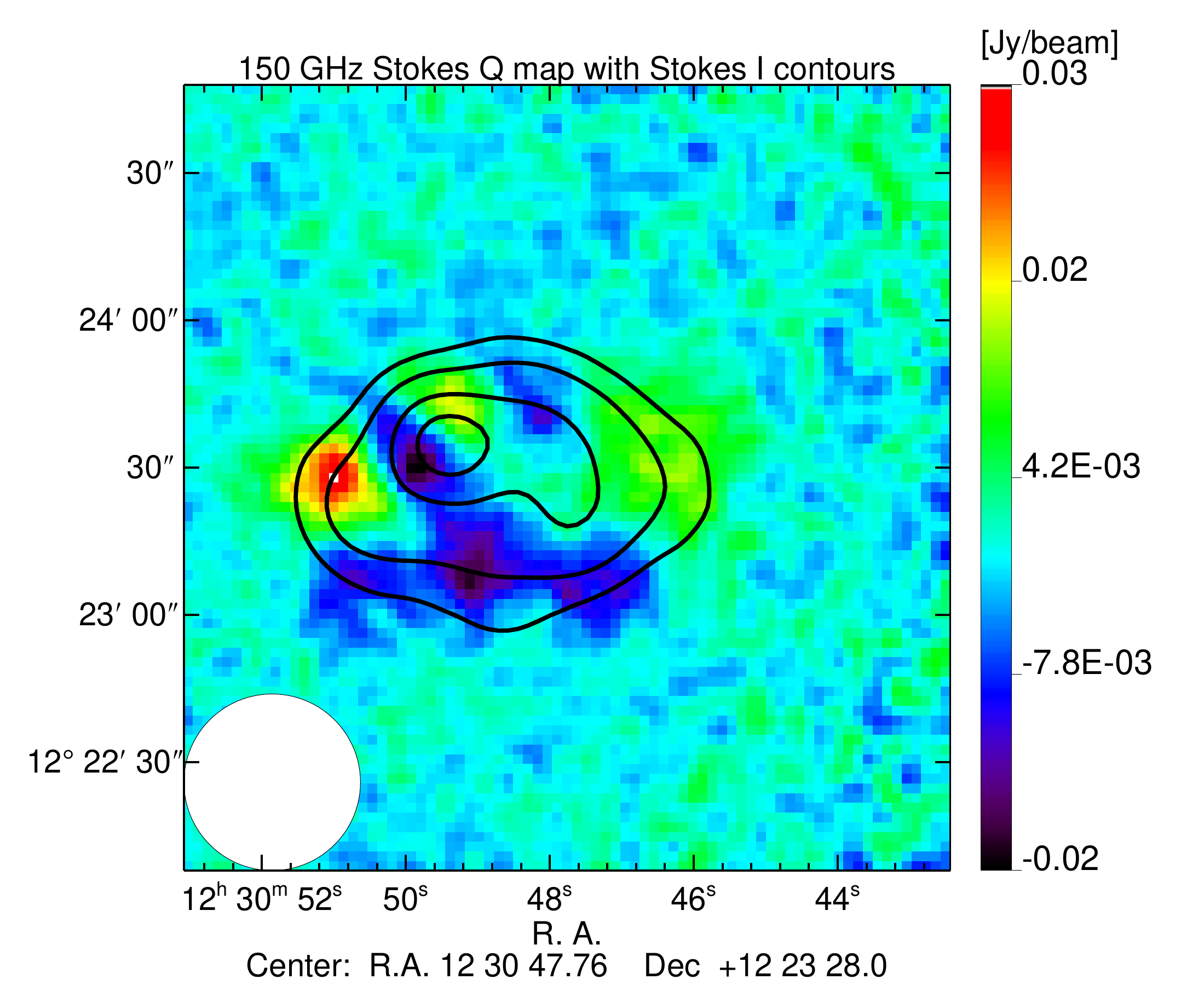}
    \includegraphics[%
   width=0.33\linewidth,keepaspectratio]{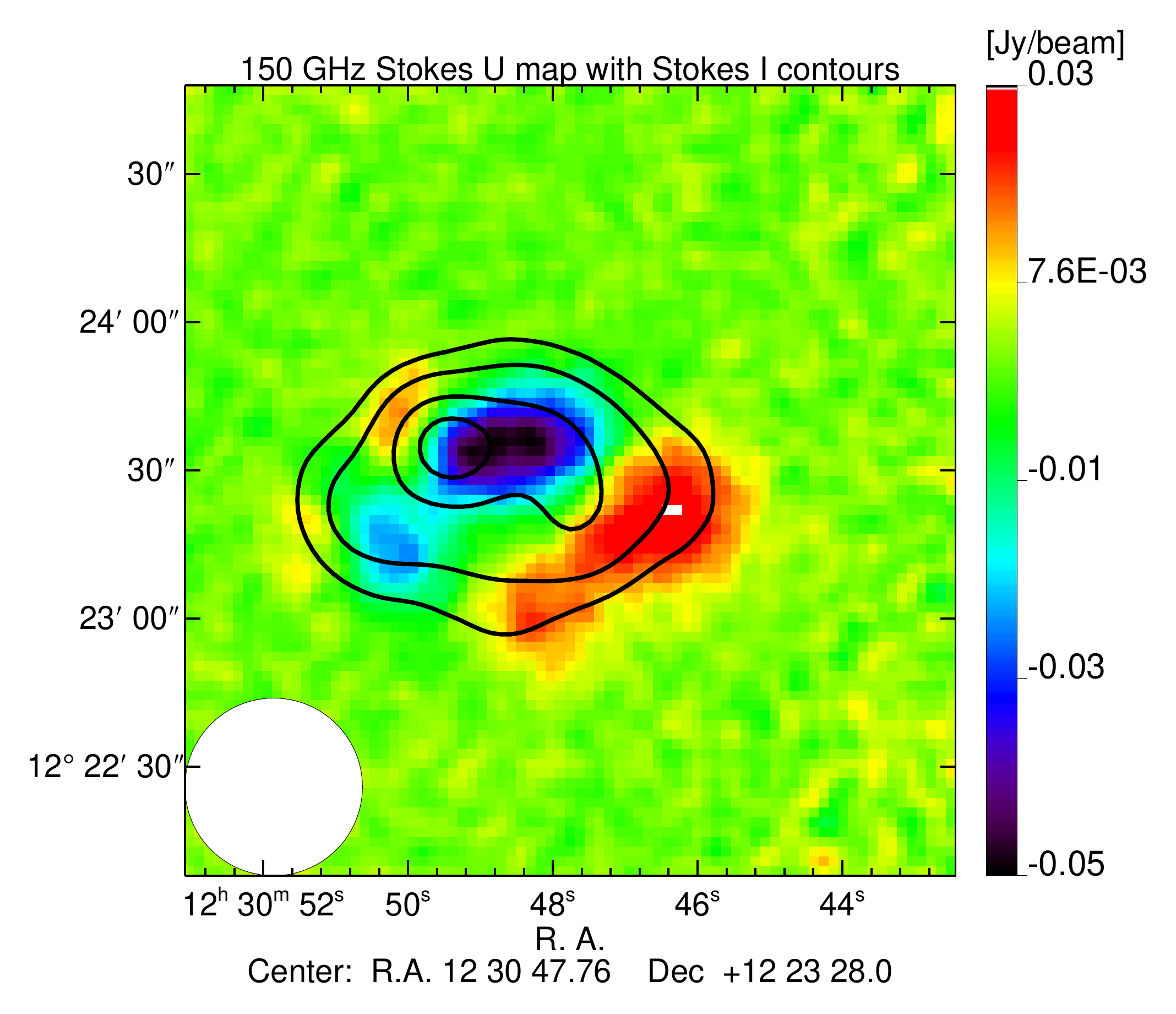}
    \caption{ \nika\ M87 Stokes  $I$, $Q$, and $U$ maps  at 260 GHz (top) and 150 GHz (bottom). 
    For display purposes, at 260 GHz, the $Q$ and $U$ maps are smoothed with a Gaussian filter of 6 arcseconds and the
      $I$ map of 4 arcseconds. At 150 GHz the $Q$ and $U$ maps are also smoothed to 4 arcseconds, while the $I$ map is not smoothed. The contours in $Q$ and $U$ represent the intensity map for each frequency, starting from 0.2 Jy/beam with steps of 0.2 Jy/beam. Polarization vectors are plotted in red in the intensity image when $I$ $\textgreater$ 0 and $P$ $\textgreater$ 2$\sigma_P$.}
     \label{M87_maps}
   \end{center}
   \end{figure*}
   
    \begin{figure*}
   \begin{center}
     \includegraphics[%
   width=0.33\linewidth,keepaspectratio]{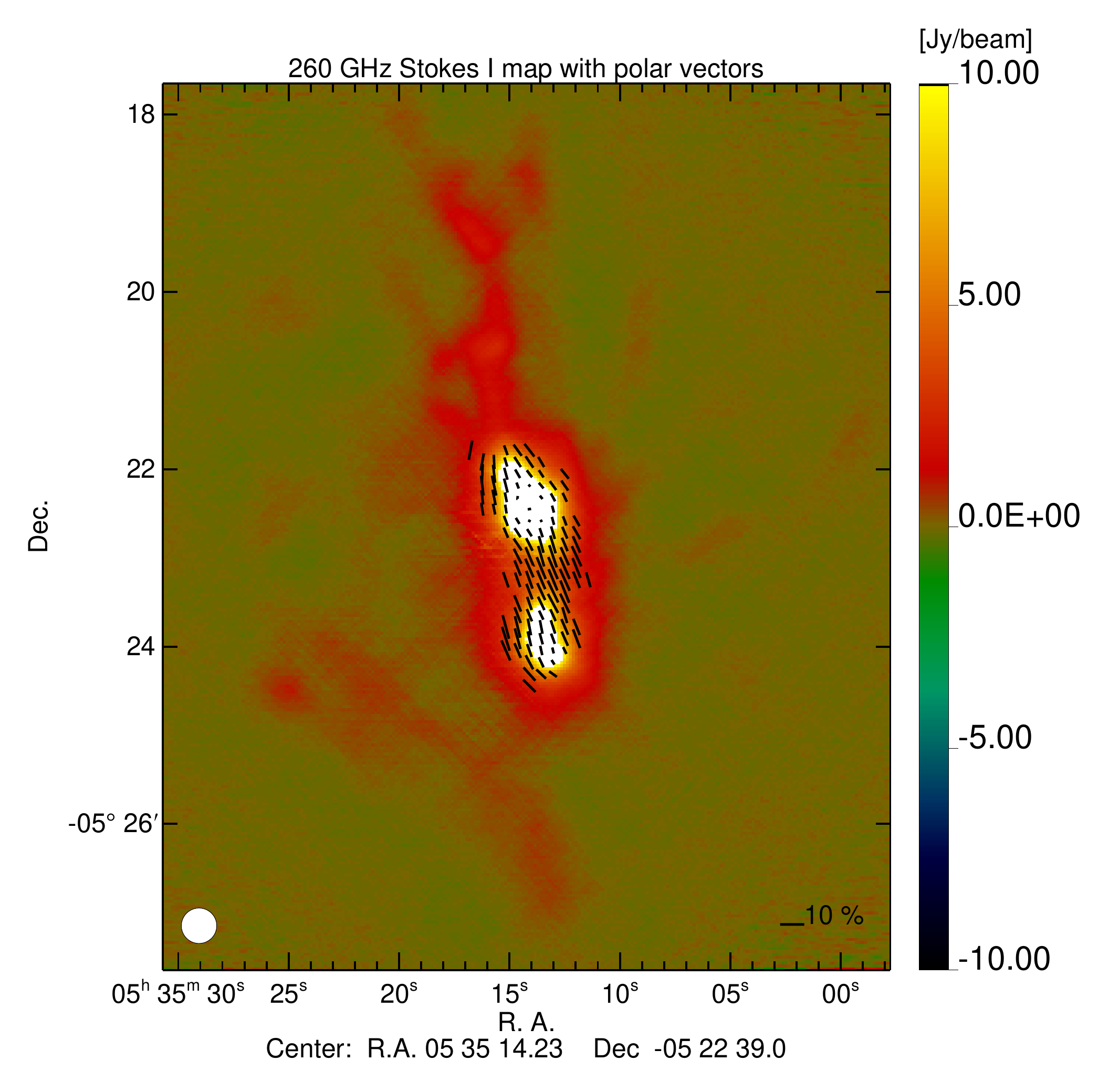}
     \includegraphics[%
   width=0.315\linewidth,keepaspectratio]{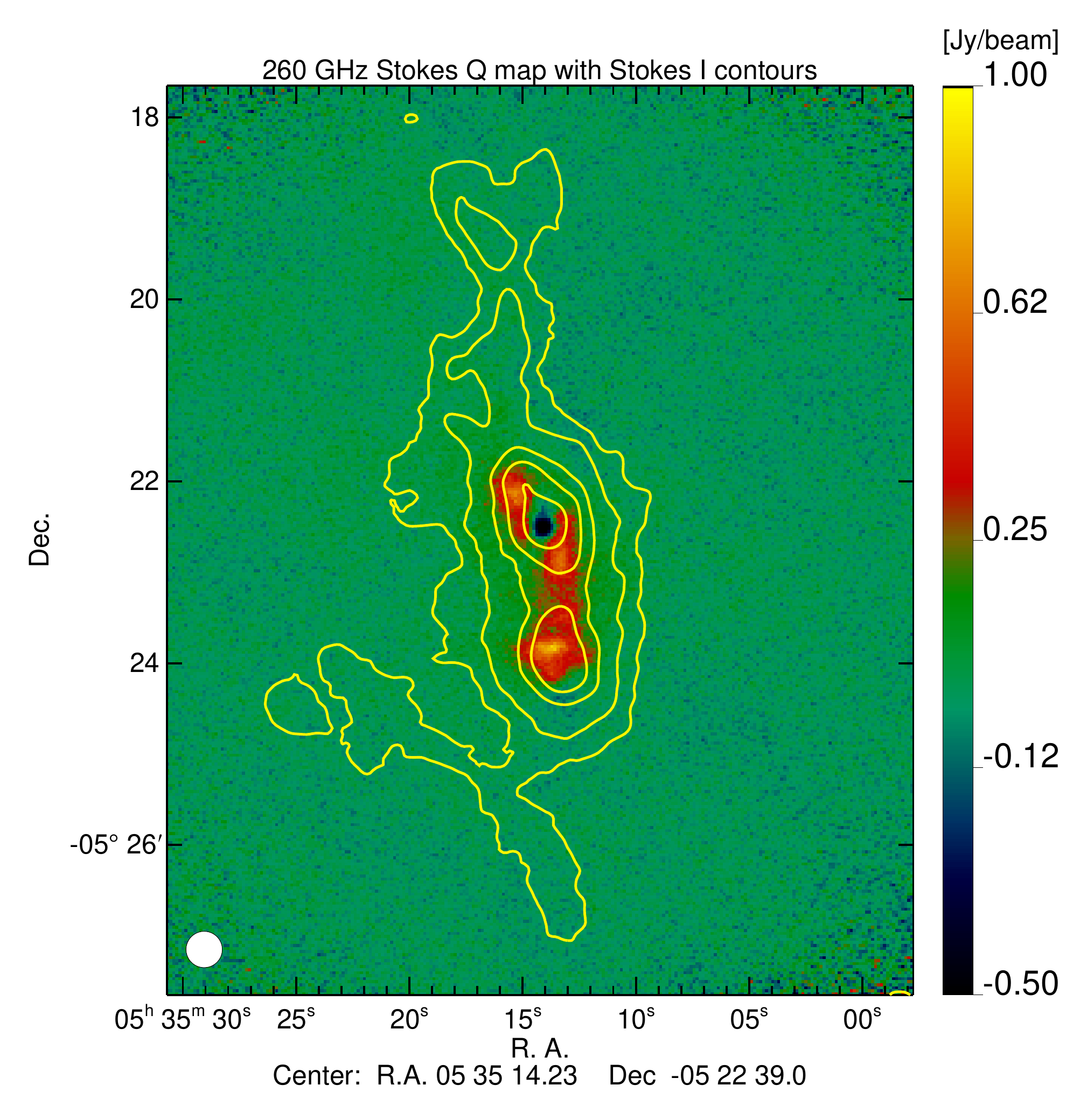}
    \includegraphics[%
   width=0.315\linewidth,keepaspectratio]{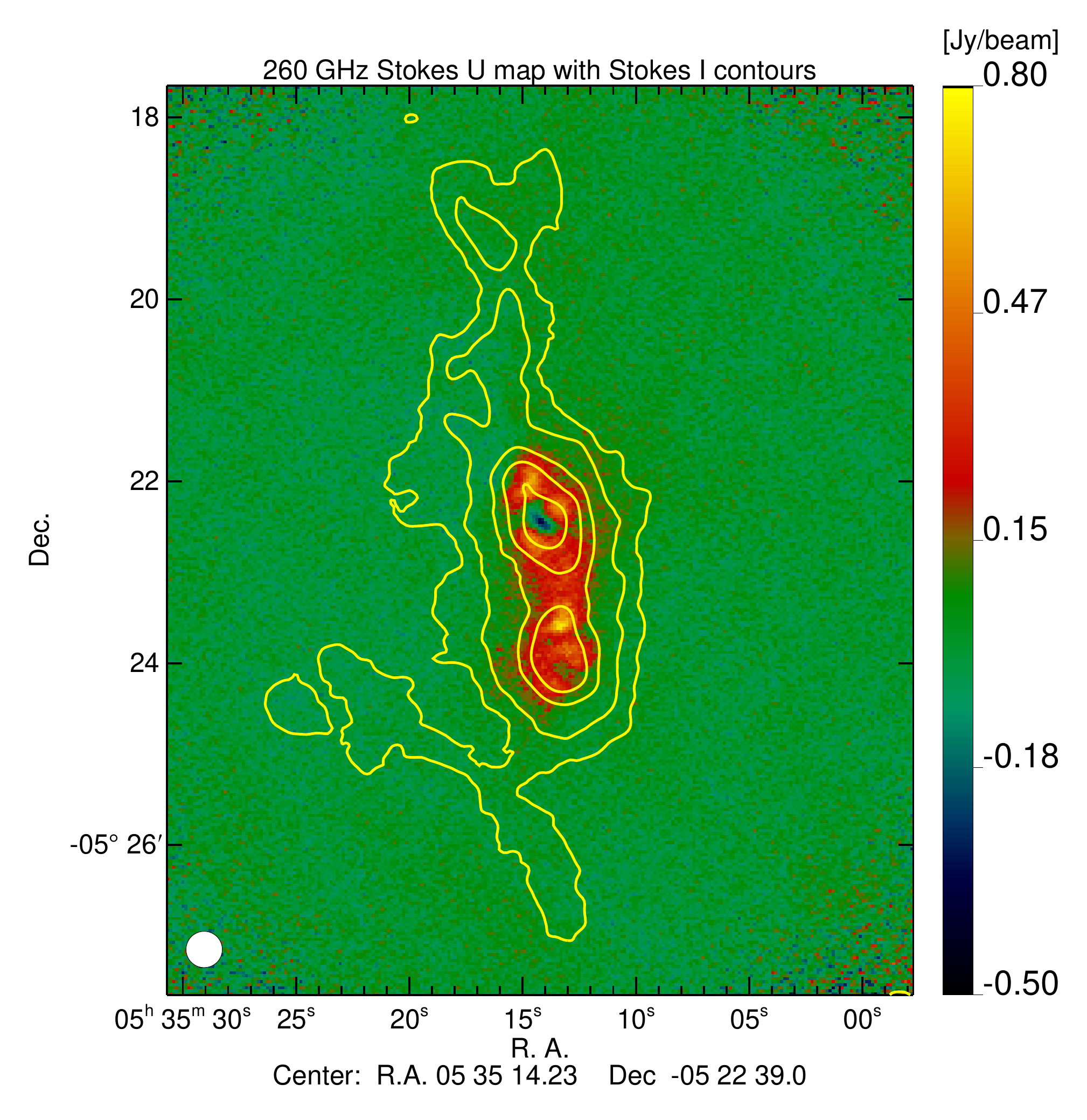}
      \includegraphics[%
   width=0.33\linewidth,keepaspectratio]{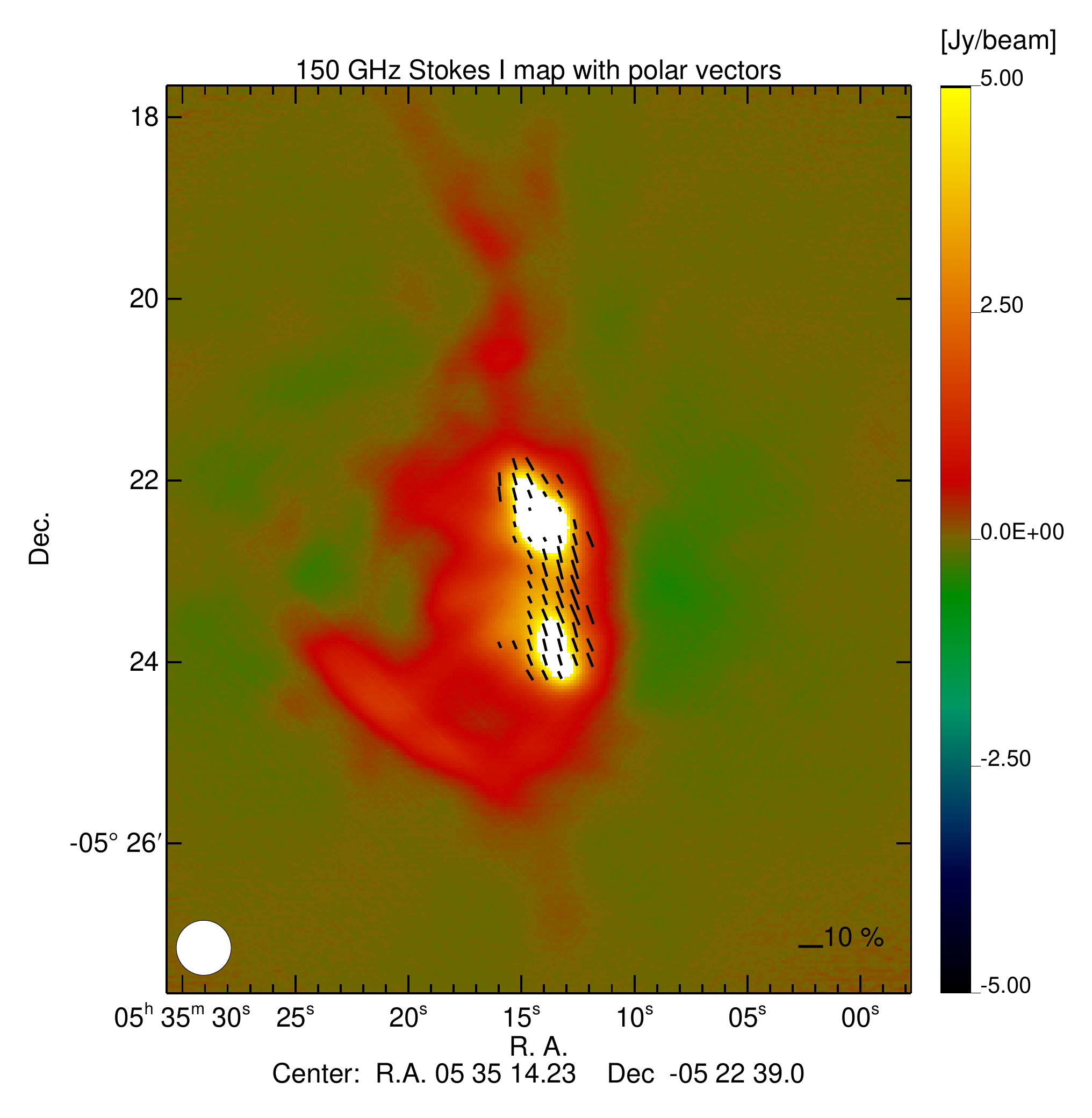}
     \includegraphics[%
   width=0.33\linewidth,keepaspectratio]{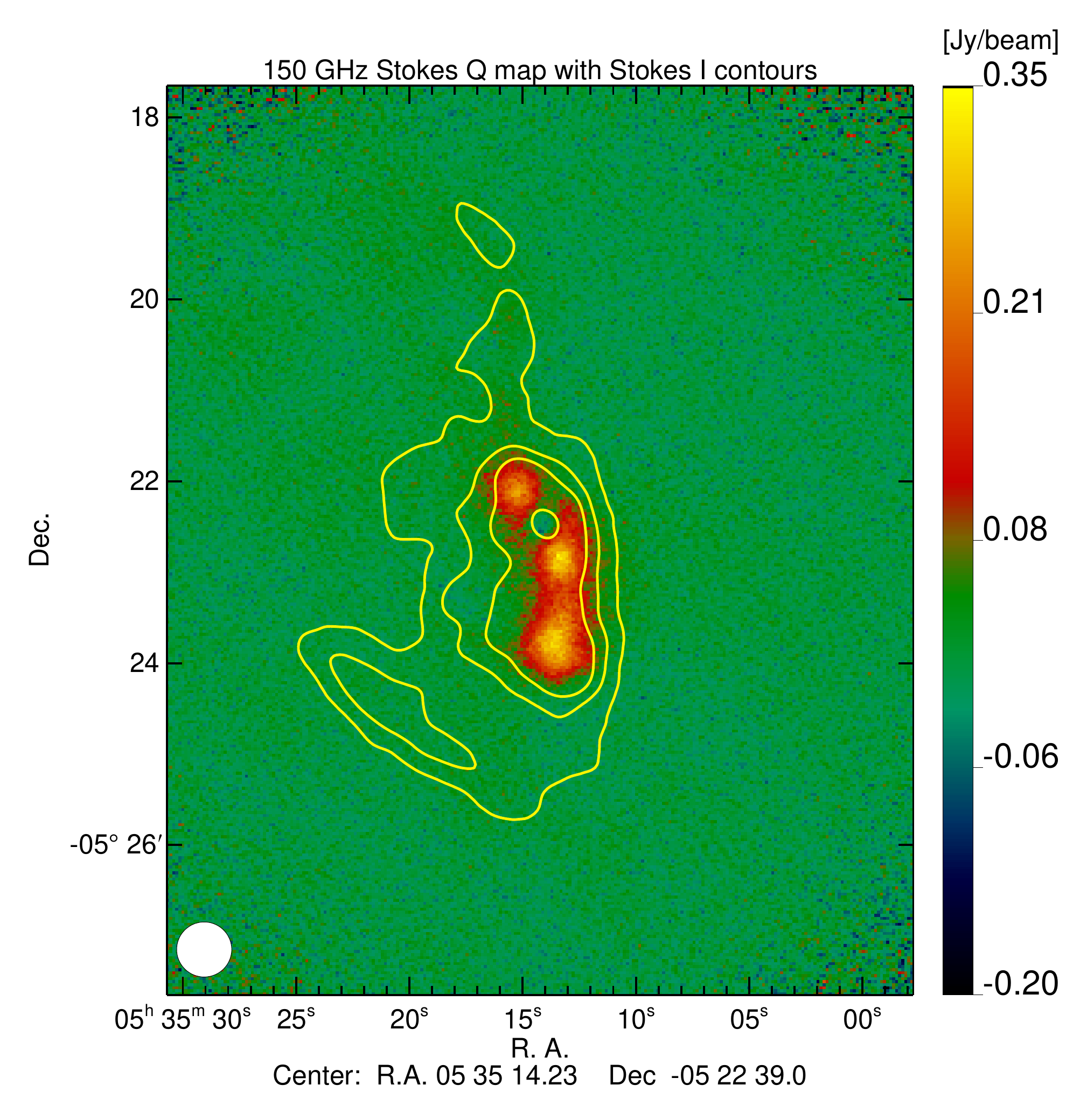}
    \includegraphics[%
   width=0.33\linewidth,keepaspectratio]{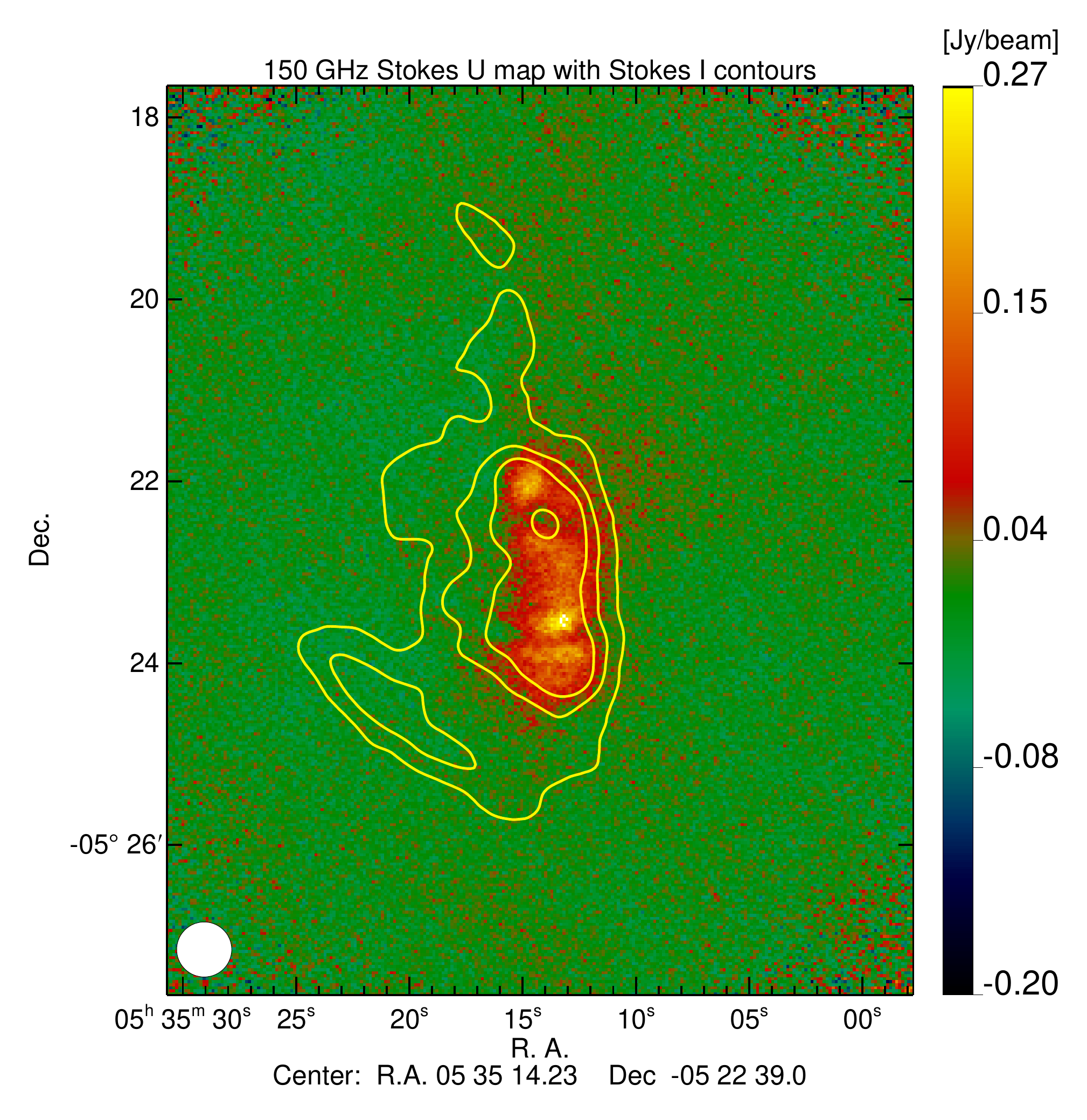}
 \caption{ {\it NIKA} Stokes $I$, $Q$, and $U$ maps of Orion
   OMC-1 at 260 GHz (top) and 150 GHz (bottom). The intensity contours over-plotted in the
   $Q$ and $U$ maps correspond to (0.3, 1, 3, 6, 15, and 48) Jy/beam at 260 GHz and
   (0.3, 1, 2, 10, 14) Jy/beam at 150 GHz. Polarization vectors are plotted in black in the intensity image when $I$ $\textgreater$ 0 and $P$ $\textgreater$ 2$\sigma_P$.}
   \label{Orion}
   \end{center}
   \end{figure*}
    \begin{figure*}[h!]
    \begin{center}
     \includegraphics[%
   width=0.8\linewidth,keepaspectratio]{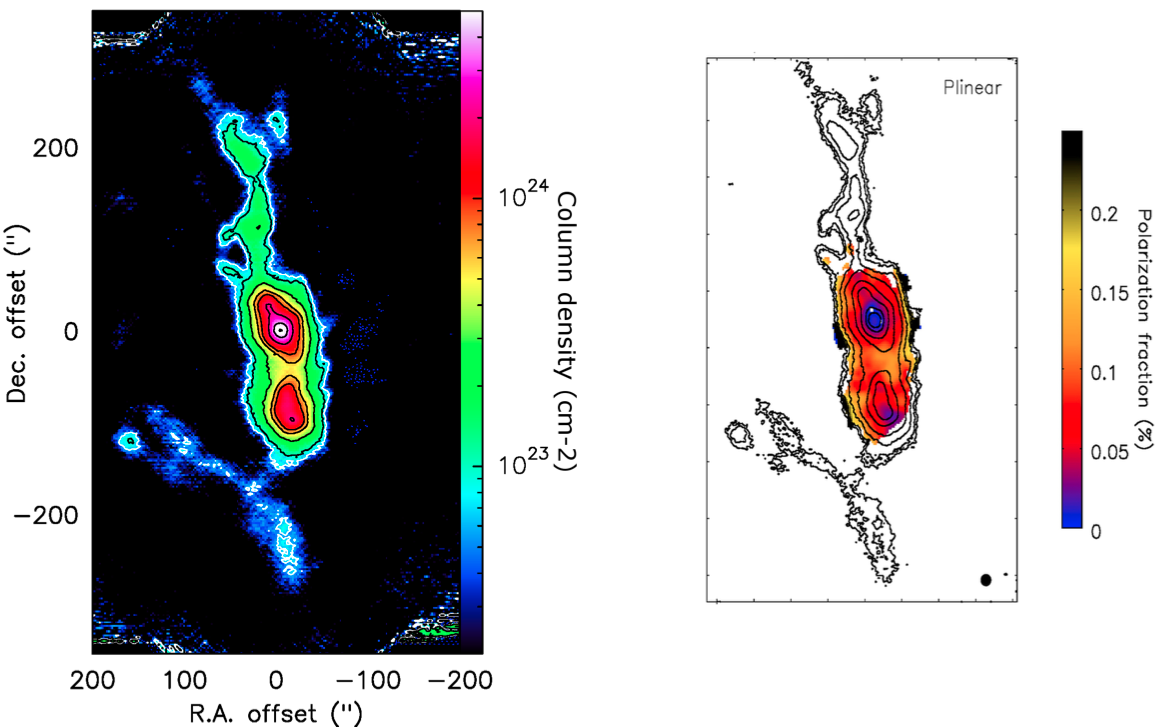}
    \caption{Column density map (left) obtained from the intensity map $I$ at 1.15 mm. For comparison the polarization fraction is reported on the right panel of the figure.}
    \label{column_density}
\end{center}
  \end{figure*}
  
      \begin{figure}
  \caption{ Polarization intensity (a), degree (b) and angle (c) maps of Orion Molecular Cloud (OMC-1) at 260 GHz (left) and 150 GHz (right) with intensity contours over-plotted. Only the regions with P $\textgreater$ 2 $\sigma_P$ are plotted on (b) and (c). At both frequencies, for display purposes, the polarization intensity maps are smoothed with a Gaussian filter of 6 arcseconds; while the polarization degree and angle maps are smoothed to 2 arcseconds.}
     \begin{center}
  	\setlength{\unitlength}{\columnwidth} 
\begin{picture}(2,2.5)
 \put(0,2.5){(a)}
           \put(0.05,2.05) { \includegraphics[%
   width=0.4\linewidth,keepaspectratio]{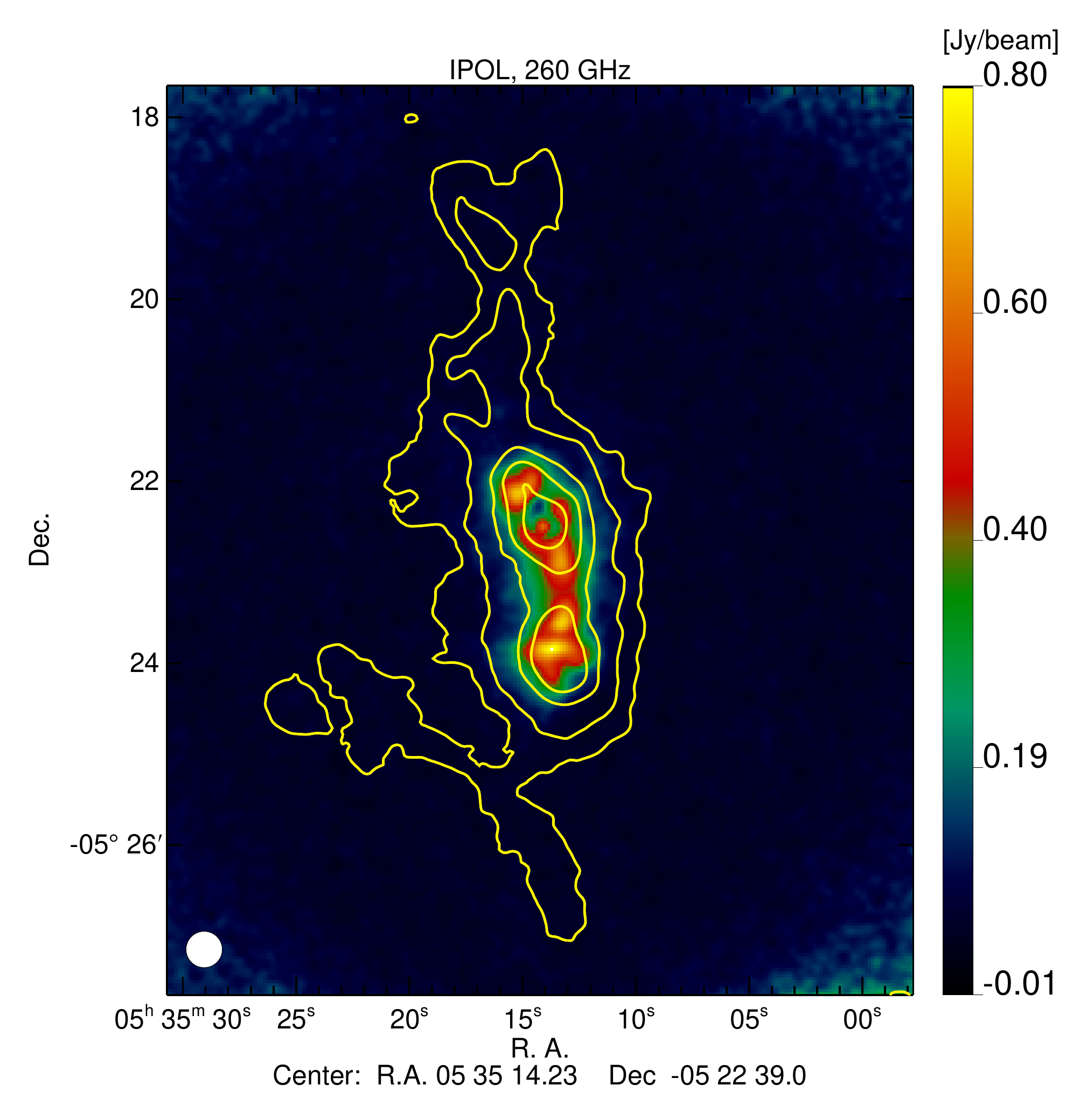}}
      \put(0.5,2.05){ \includegraphics[%
   width=0.4\linewidth,keepaspectratio]{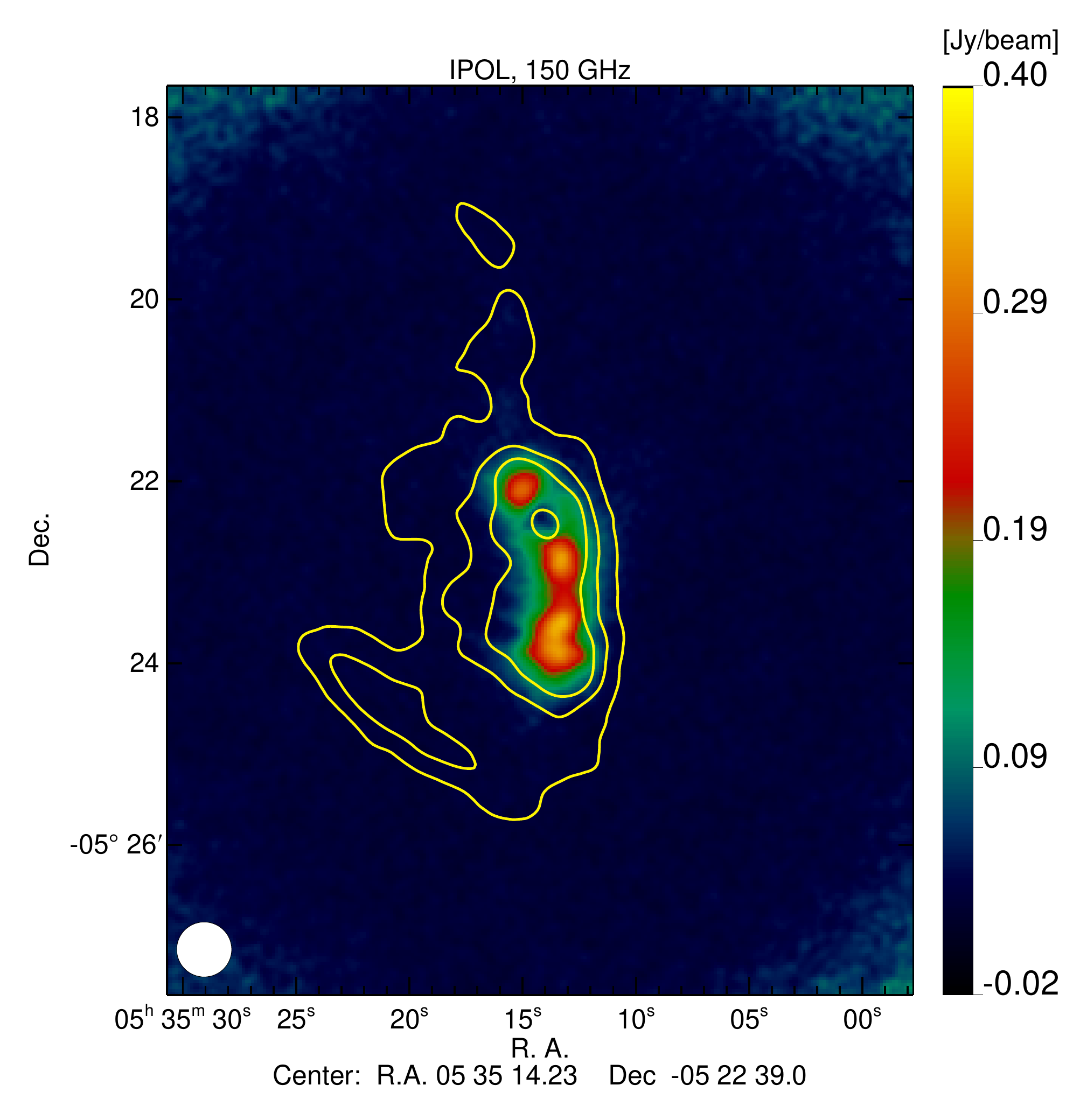}}

    \put(0,2.0){(b)}
   
      \put(0.05,1.55){  \includegraphics[width=0.4\linewidth,keepaspectratio]{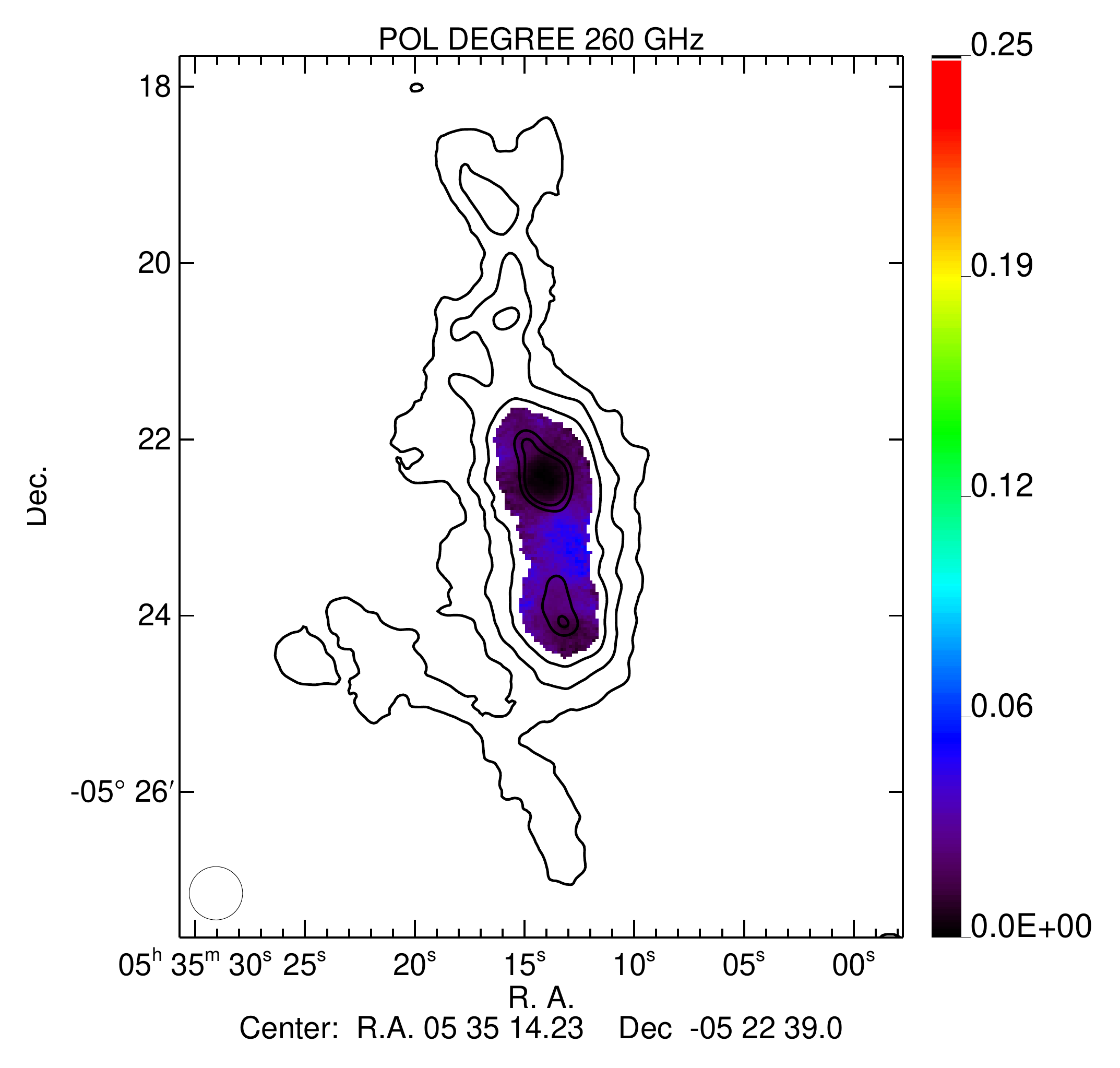}}
	\put(0.5,1.55){\includegraphics[width=0.4\linewidth,keepaspectratio]{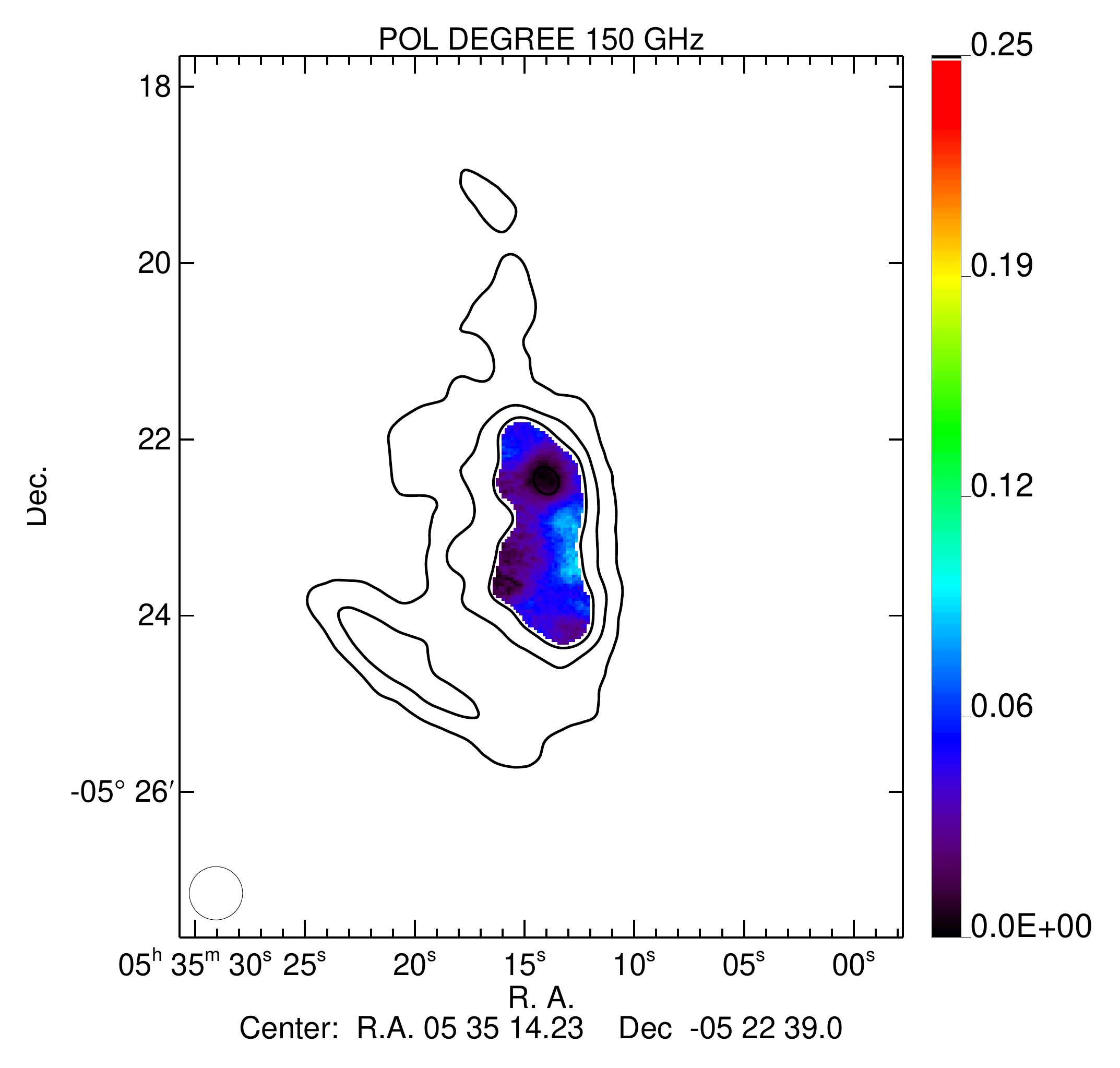}}
	
     \put(0,1.5){(c) }
   
    \put(0.05,1.05){\includegraphics[%
   width=0.4\linewidth,keepaspectratio]{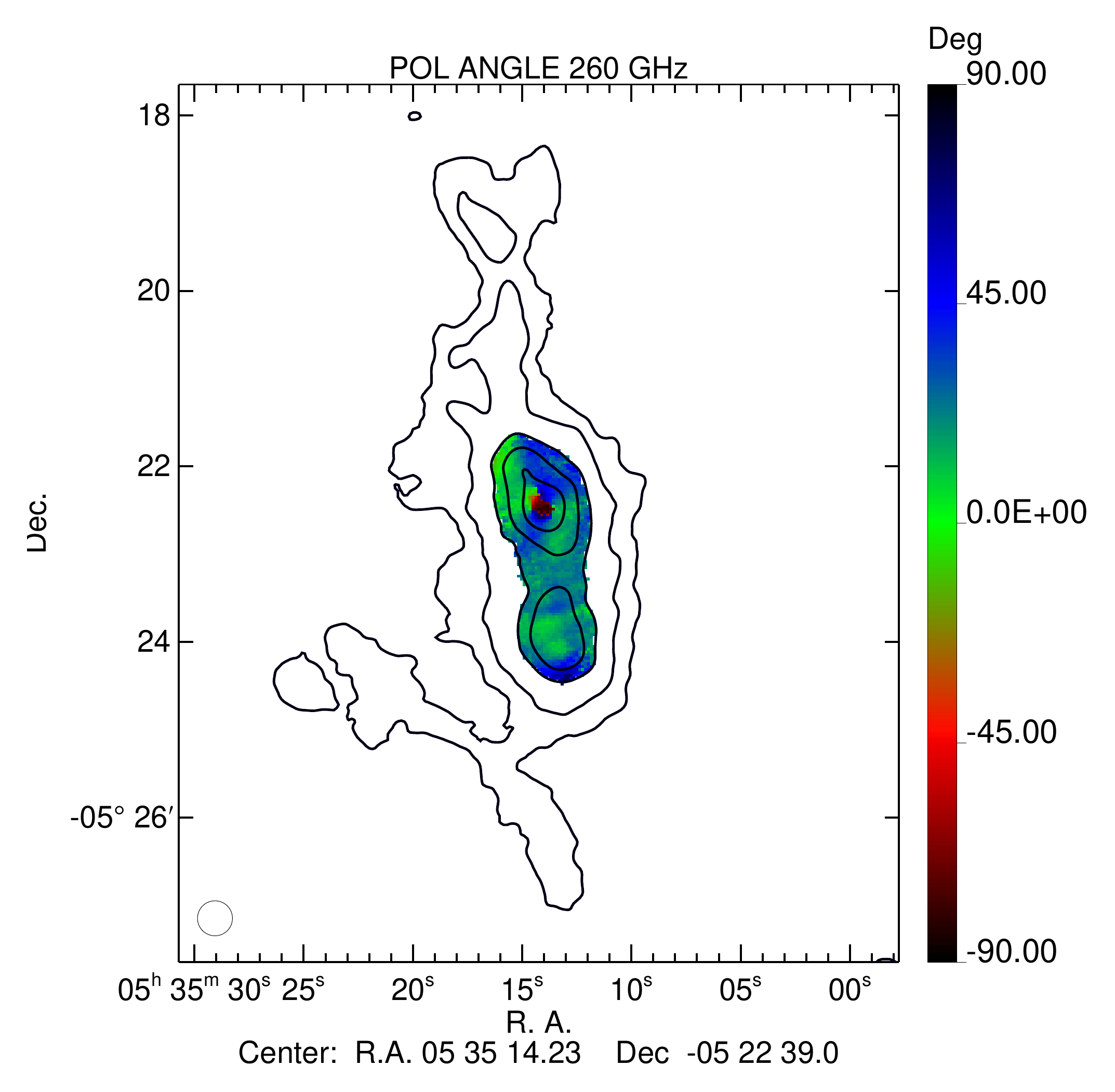}}
    \put(0.5,1.05){\includegraphics[%
   width=0.4\linewidth,keepaspectratio]{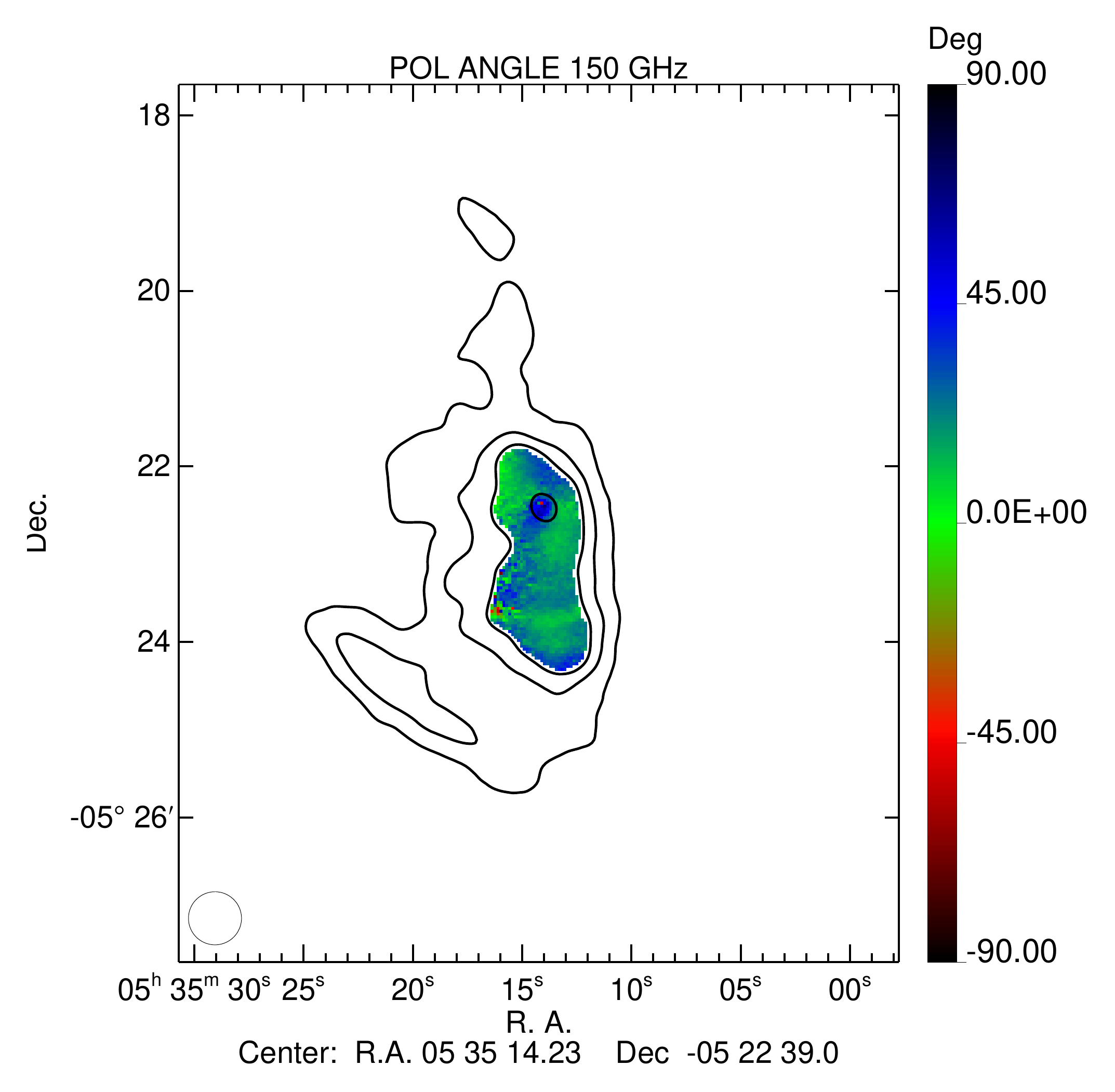}}
	\end{picture}
 \end{center}
\label{Orion_poldeg}
  \end{figure}
  
 \section{\nika\ polarized observations of compact and extended sources}
 \label{sec:extended}
 We discuss here observations of compact and extended polarized sources, which
 allow us to further validate the quality of the reconstruction of the polarized
 sky signal with the \nika\ camera. Special care is taken on the verification of
 the validity of the leakage correction algorithm, which can affect the
 reconstruction of the direction of polarization across the source. We have
 performed observations of different types of sources: Cygnus~A, a radio galaxy
 with diffuse emission between the two radio lobes; M87, an external galaxy; and Orion OMC1, a nearby highly polarized galactic cloud. Observations of the Crab nebula have been performed several times
 during the \NIKA\ polarization runs. 
 A preliminary report on these observations has been given in \citet{Ritacco2015}.
 A complete study of the polarized continuum emission observed at mm wavelengths for the Crab nebula is 
 beyond the scope of this paper and will be included in a forthcoming publication ({\it in prep.}).

\subsection{Cygnus~A}
Cygnus~A is a typical radio galaxy with twin jets of plasma emanating from its
nucleus and forming two extended radio lobes. Cygnus~A is the most
powerful Fanaroff- Riley II (FRII) radio galaxy in the local environment. It has
been well studied in terms of spatial resolution as it lies at a distance of 227
Mpc. At low radio frequencies, the synchrotron emission from the two giant lobes
dominates \citep{1974MNRAS.166..305H}. At higher frequencies, the hotspots
(working surfaces in the lobes) and the galaxy core become more prominent. The
southern and northern hotspots are at 50 and 70 arcsec from the core,
respectively.  The complex structure of Cygnus~A can be well explained by
assuming that it consists of two components polarized in opposite
directions and of an unpolarized core \citep{1967ApJ...150L..15S, 1966SvA....10..214S, 1968ApJ...151...53M}.

The \nika\ Stokes $I$, $Q$, and $U$ maps of Cygnus~A at 260
(top row) and 150 (bottom row) GHz are presented in Fig.~\ref{cygnusa_maps}.
On the intensity maps, the polarization vectors are over-plotted in red. On the
$Q$ and $U$ maps, intensity contours are also represented in black.  As expected,
on the intensity maps at 1.15 and 2.05 mm, we clearly observe three compact
sources that correspond to the core and the two hotspots. By contrast, the
polarization maps show only two polarized regions that correspond to the hotspots and nothing on the core. This is an unambiguous confirmation of the
astrophysical origin of the observed polarization: if it was due to the instrumental
polarization, it would be proportional to the total intensity and therefore show up
on the core. Such considerations
are further confirmed on other extended sources as presented in the following sections.
\subsection{M87}
 M87, also designated as 3C274, or NGC 4486, or Virgo A, is a giant elliptical
 galaxy \citep{dev76} located near the core of the Virgo cluster.  Its nucleus is
 a radio and X-ray source from which emanates an optical jet.  M87 is
 estimated to be approximately 16 Mpc from Earth \citep{mou80}. The core and
 the jet can be seen at all wavelengths from radio to X-rays.
  Fig.~\ref{M87_maps} presents the \nika\ Stokes $I$, $Q$, and
 $U$ maps at 260 (top) and 150 (bottom) GHz. The polarization vectors are
 over-plotted in red on the intensity maps representing both the degree and
 orientation of polarization. The peak surface brightness of the $I$ map is $\sim$ 0.6 Jy/beam
and $\sim$ 1.5 Jy/beam at 260 GHz and 150 GHz, respectively. 
Such a spectral index indicates that mm continuum emission in M87 is indeed dominated by synchrotron emission.
 At 260 GHz we do not have a significant detection of the polarization and the expected signal in $Q$ and $U$ maps is limited by the noise, as shown in the
 top panels of Fig.~\ref{M87_maps}.
 However, the detection at 150 GHz shows the polarization vectors 
 well aligned following the intensity contours. They suggest the
 existence of a large-scale, ordered magnetic field in the radio lobes of M87.
  
\subsection{Orion OMC-1}
The Orion molecular cloud (OMC-1) is the closest site of OB star formation. The
Nebula (KL) is the flux peak from far infrared to millimeter wavelengths on the
OMC-1 ``ridge" \citep{schle1998}.  A sub-millimeter peak (KHW) with equal mass
but lower dust temperature, is found 90 arcsec south of KL along the ridge
\citep{KHW1982}. At KL and KHW, the polarization fraction increases with the
wavelength \citep{schle1998}.

Fig.~\ref{Orion} presents the \nika\ Stokes $I$, $Q$, and $U$ maps of Orion OMC-1 at 260 (top) and 150
(bottom) GHz.  Polarization vectors are over-plotted on the intensity maps,
showing both the polarization degree and orientation.
 The peak surface brightness of the OMC-1
emission is approximately 45.8 Jy/beam and 14 Jy/beam at 260 GHz and 150 GHz,
respectively. The size of the map is 8 $\times$ 8 arcminutes and is obtained by the
co-addition of 18 maps for a total observational time of approximately 5h. 
The left panel of Fig.~\ref{column_density} shows the column density map obtained from the continuum emission of Orion OMC-1 observed at 1.15 mm assuming an
average homogeneous dust temperature of 30 K, see \citep{2014A&A...566A..45L}. Mostly driven by our sensitivity limit, polarization is only detected at column densities $\textgreater$ 3 $\times$ 10$^{23}$ cm$^{-2}$ in the map.
Depolarization is observed at column densities $\textgreater$ 4 $\times$ 10$^{24}$ cm$^{-2}$ at 1.15 mm, while the 2.05 mm polarized fluxes seems less sensitive to depolarization. 

The orientation of the polarization vectors is consistent between the 260 and 150 GHz maps, confirming the same physical origin of the observed
polarization. 
If we trust that the magnetic field is orthogonal to the direction of polarization, we observe a very organized magnetic field topology with field lines mostly oriented to the integral-shaped filament, and suggesting some bending of the field lines along the major axis of the filament towards the high-column-density cores. 
This magnetic field lines morphology could be due to the \vec{B} dragged by large-scale converging material accreting along the filament onto the core, or to 
the \vec{B} being pushed by the powerful winds of the Orion nebula.
These structures and filaments are consistent with OMC-1
observations performed with SCUPOL \citep{scubapol} at 850 $\mu$m.  
Previous observations at 1.3 mm performed by \cite{leach1991} show that at the KL
position, the average polarization rises to 4\%-5\% while at KL, the
polarization drops to 0.6\%. 

The polarization intensity maps $P$ are reported on the top panel of Fig.~\ref{Orion_poldeg} (a), showing the expected ``polarization hole'' across the KL nebula, already observed by \cite{schle1998}. The polarization fraction maps $p$ as observed at both \NIKA\ frequencies are reported on the central panel of Fig.~\ref{Orion_poldeg} (b).
On both maps, we observe a polarization fraction that reaches a level of approximately 10\% of the total intensity in regions where the diffuse intensity emission is observed. This polarization fraction decreases greatly near the KL nebula.
The polarization angle $\psi$ is shown on the bottom panel of Fig.~\ref{Orion_poldeg} (c).  Averaging across the KL nebula on 10 arcseconds with central position $\alpha_{J2000}$:05:35:14.098, $\delta_{J2000}$:-05:22:31.00  on the \nika\ Orion maps at 260 GHz (1.15 mm) we find that the angle and degree of polarization are
$\psi$ = (37.74 $\pm$ 3.56)$^\circ$ and $p$ = (0.6 $\pm$ 0.2) $\%$.  
The uncertainties reported here are purely statistical. We have also to consider the systematic
uncertainties due to HWP zero position 1.8$^{\circ}$ as well as the absolute calibration error calculated on Uranus, approximately $\sim$14\% at 260 GHz and 5\% at 150 GHz.
In Tab.~\ref{tab:tab_orion_polka_scupol}, we summarize the results obtained on the KL region
by SCUPOL and POLKA \citep{polka_apex} experiments
for comparison with \nika\ results.
\begin{table*}
 \centering
 \caption{Summary of KL nebula polarization degree and angle results obtained by previous experiments and \nika. An absolute uncertainty of 1.8$^{\circ}$ has to be added to the statistical angle uncertainties reported here.}
 \footnotesize
 \begin{tabular}{cccccccccccc}
 \hline
 \hline
  \multicolumn {3} {c} {p [$\%$]} & \multicolumn {3} {c} {$\psi$ [$^\circ$]} \\
 \hline
 POLKA  & SCUPOL  &  \nika\ & \nika\ & POLKA  & SCUPOL   &  \nika\ & \nika\ \\
  \hline
   [870 $\mu$ m] &  [850 $\mu$ m]  &  [1.15 mm] & [2.05 mm]  &  [870 $\mu$ m] &  [850 $\mu$ m]  &  [1.15 mm] & [2.05 mm] \\
  \hline
  0.7 $\pm$ 0.2 & 0.7 $\pm$ 0.1 & (0.6 $\pm$ 0.2) & (1.0 $\pm$ 0.2) &  32.8 $\pm$ 7.6 & 40.8 $\pm$ 5.4 & [37.73 $\pm$ 3.56] & [25.35 $\pm$ 2.15] \\ 
 \hline
 \hline
 \end{tabular}
 \label{tab:tab_orion_polka_scupol}
 \end{table*}    
 
 \section{Conclusions}\label{conclusions}
This paper presents the first astrophysical polarization measurements with KIDs. 
For these measurements, we have adopted a simplified
polarization system consisting of an achromatic, continuously rotating HWP at
approximately 3~Hz, an analyzer, and arrays of KIDs not sensitive to polarization.  The
fast modulation of the input polarization signal with the HWP allowed us to
significantly reduce the atmospheric emission in the polarized
signal. Instrumental polarization in the form of intensity to polarization
leakage with a non-trivial-point-spread function has been observed at the level of 2\% to 3\% peak to peak for point- like
and extended sources, respectively.  We have successfully developed an algorithm
to correct for this systematic effect. We are then left another kind of
instrumental polarization that generates a polarized signal directly
proportional to intensity at the level of 0.7\% and 0.6\% at 1.15 and 2.05~mm,
respectively, that can be corrected.
We have observed 3C286, a quasar used as a standard polarization calibrator in the
literature and have found a total flux, a polarization degree, and orientation in
agreement with existing data. These results confirms findings for other quasars, such as 3C279, 3C273, and 0923+392, for which we either comparable results in the literature or performed simultaneous measurements with XPOL. We have also observed
compact and extended sources such as Cygnus~A, OMC-1, and M87, and, again, found
consistent results with existing polarization maps at approximately the same wavelength
({\it e.g.} OMC-1 \cite{scubapol}). All these observations establish the accuracy of
our system and analysis on astronomical sources with fluxes of approximately one Jansky
and degrees of polarization as low as 3\%. On extended sources such as OMC-1, \nika\ observations confirm
that polarization vectors align well with the intensity structures indicating the
presence of well ordered magnetic fields. To our knowledge, our observations of
Cygnus~A and M87 are the first ones in polarization at millimetric
wavelengths.
\nika\ has been a successful test-bench for the \nikad \footnote{http://ipag.osug.fr/nika2} camera, which shares the
same polarization system, although limited to the 260~GHz
channel. \nikad\ will observe the sky at the
same frequencies with ten times more detectors and a FOV of 6.5
arcminutes. The \nikad\ camera has been installed at the IRAM 30 meter telescope
in Spain on October 2015 to start its commissioning phase for unpolarized
observations. A polarization dedicated commissioning will follow, during which
we will improve our understanding of the observed instrumental polarization and
our ability to correct for it. 

This paper shows the potentialities of an instrument based on
KIDS and a fast and continuously rotating HWP to measure polarization, especially
from the ground, where atmosphere is a nuisance, even more at low temporal
frequencies and large angular scales. It opens the way to forthcoming
observations with \nikad\ that will undoubtedly provide advances in the field of
Galactic emission and interactions with the magnetic field.
 
 \bibliography{biblio}   
 \begin{acknowledgements}
We would like to thank the IRAM staff for their support during the campaigns. 
The NIKA dilution cryostat was designed and built at the Institut N\'eel. 
In particular, we acknowledge the crucial contribution of the Cryogenics Group, and 
in particular Gregory Garde, Henri Rodenas, Jean Paul Leggeri, and Philippe Camus. 
This work has been partially funded by the Foundation Nanoscience Grenoble, the LabEx FOCUS ANR-11-LABX-0013 and 
the ANR under the contracts "MKIDS", "NIKA" and ANR-15-CE31-0017. 
This work has benefited from the support of the European Research Council Advanced Grant ORISTARS 
under the European Union's Seventh Framework Programme (Grant Agreement no. 291294).
We acknowledge fundings from the ENIGMASS French LabEx (R. A. and F. R.), 
the CNES post-doctoral fellowship program (R. A.),  the CNES doctoral fellowship program (A. R.) and 
the FOCUS French LabEx doctoral fellowship program (A. R.).
\end{acknowledgements}
\end{document}